\documentclass[fleqn,usenatbib]{mnras}

\usepackage{newtxtext,newtxmath,xfrac,mathtools,booktabs}

\usepackage[T1]{fontenc}

\DeclareRobustCommand{\VAN}[3]{#2}
\let\VANthebibliography\thebibliography
\def\thebibliography{\DeclareRobustCommand{\VAN}[3]{##3}\VANthebibliography}

\usepackage{graphicx}	
\usepackage{amsmath}	
\usepackage{color}
\usepackage{orcidlink}

\definecolor{Red}{rgb}{0.65,0.08,0.05}
\definecolor{Green}{rgb}{0.05, 0.5, 0.25}
\definecolor{LGreen}{rgb}{0.17, 0.84, 0.57}

\title[The phase-space distribution of ICL and DM]{Intracluster light as a dark matter tracer: how their spatial and kinematic relationship is shaped by satellite demographics}

\author[G. Martin et al.]{
G. Martin \orcidlink{0000-0003-2939-8668},$^{1}$\thanks{E-mail: garreth.martin@nottingham.ac.uk} F. R. Pearce \orcidlink{0000-0002-2383-9250},$^{1}$ N. A. Hatch \orcidlink{0000-0001-5600-0534},$^{1}$ H. J. Brown \orcidlink{0009-0004-7521-8204},$^{1}$ M. Butler \orcidlink{0009-0004-9273-4630},$^{1}$ Y. M. Bah\'{e} \orcidlink{0000-0002-3196-5126},$^{1,2}$\newauthor
~W. Cui \orcidlink{0000-0002-2113-4863},$^{3,4,5}$ Y. Dubois \orcidlink{0000-0003-0225-6387}$^{6}$, and A. Knebe \orcidlink{0000-0003-4066-8307}$^{3,4,7}${
}
\\
$^{1}$School of Physics \& Astronomy, University of Nottingham, University Park, Nottingham NG7 2RD, UK\\
$^{2}$Laboratoire d'Astrophysique, \'{E}cole Polytechnique F\'{e}d\'{e}rale de Lausanne (EPFL), Observatoire de Sauverny, 1290 Versoix, Switzerland\\
$^{3}$Departamento de Física Te\'orica, M\'odulo 15, Facultad de Ciencias, Universidad Aut\'onoma de Madrid, 28049 Madrid, Spain\\
$^{4}$Centro de Investigaci\'on Avanzada en F\'isica Fundamental (CIAFF), Facultad de Ciencias, Universidad Aut\'onoma de Madrid, 28049 Madrid, Spain\\
$^{5}$Institute for Astronomy, Royal Observatory, Edinburgh EH9 3HJ, UK\\
$^{6}$Institut d’Astrophysique de Paris, UMR 7095, CNRS, Sorbonne Universit\'e, 98 bis boulevard Arago, 75014 Paris, France\\
$^{7}$International Centre for Radio Astronomy Research, University of Western Australia, 35 Stirling Highway, Crawley, Western Australia 6009, Australia
}

\date{Accepted XXX. Received YYY; in original form ZZZ}

\pubyear{\the\year{}}

\begin{document}
\label{firstpage}
\pagerange{\pageref{firstpage}--\pageref{lastpage}}
\maketitle

\begin{abstract}
We investigate how the orbital evolution and mass distribution of infalling satellite galaxies shape the phase-space and radial distributions of intracluster light (ICL) relative to the underlying cluster dark matter (DM) halo. Using controlled, self-consistent $N$-body simulations, we follow the tidal stripping and orbital evolution of satellite galaxies as they are accreted into a live cluster halo, systematically varying satellite--to--host mass ratio and orbital circularity. From these experiments, we measure the specific orbital energy and angular momentum of stripped stellar and DM material, finding that stripped stars consistently occupy lower-energy and lower-angular momentum regions of phase-space than stripped DM. The magnitude of this difference increases strongly towards more equal satellite--to--host mass ratios, while dependence on circularity is weak.
We construct a predictive model for the phase-space properties of stripped stars and DM from a given infalling satellite population and find that phase-space differences are driven primarily by the characteristic mass of the satellite stellar mass function. The ICL is always more centrally concentrated than the DM, with offset magnitude increasing towards higher characteristic masses. Comparisons with four cosmological hydrodynamical simulations show that, once the satellite stellar mass function is matched, the model reproduces the radial stellar-to-DM density profile offsets to better than inter-simulation scatter. This demonstrates that the radial ICL-DM relationship is largely governed by satellite demographics. With adequate constraints on the infalling satellite population, ICL density profiles can therefore be used as informative tracers of the underlying radial DM distribution in clusters.

\end{abstract}

\begin{keywords}
galaxies: clusters: general -- galaxies: interactions -- methods: numerical
\end{keywords}




\section{Introduction}

Galaxy clusters are key laboratories for cosmology because they are the most massive virialised structures expected to form through hierarchical growth in a $\Lambda$CDM Universe. Their mass, spatial structure, and assembly history encode both the underlying cosmology and the nonlinear physics of structure formation. For example, cluster-scale halo shapes and density profiles provide constraints on the nature of dark matter (DM), while their abundances and scaling relations test cosmological parameters \citep[e.g.][]{Allen2011, Kravtsov2012}. Mapping these structures requires tracers of the dominant DM halo, which is not directly observable \citep[e.g.][]{West1995,Borgani2001,Arnaboldi2004,Gifford2013,Kluge2025}.

Clusters are also permeated by intracluster light (ICL), the combined light emitted by the diffuse component of stars not bound to any individual galaxy. These stars originate predominantly from tidal stripping and merging of satellites during cluster assembly, with some additional contribution from pre-processed stars already liberated from their parent galaxies upon cluster infall \citep[e.g.][]{Byrd1990, Murante2007, Contini2014, Mihos2017, Contini2019, Chun2023, Ragusa2023, Brown2024}. The ICL formed through this process is expected to be dominated by a small number of progenitors and to retain memory of the timing and orbital properties of individual accretion events \citep{Bullock2005, Johnston2008}. The ICL constitutes a significant fraction \footnote{The ICL fraction is definition- and method-dependent; observational and simulation studies suggest it typically contributes 5--40 per cent of the total stellar light in clusters, with the exact value varying with measurement methodology \citep[][]{Brough2024}.} of the stellar mass budget of clusters and provides a complementary baryonic tracer of cluster growth and dynamics \citep{Cui2014}. Because its spatial and kinematic distributions are set primarily by the cluster potential and the cumulative accretion history of satellites, the ICL encodes information about both assembly processes and the properties of the underlying DM halo that dominates the matter content of the cluster \citep{Gonzalez2007, Gonzalez2013, Montes2019b, Golden-Marx2023, Golden-Marx2025, Montenegro-Taborda2025, Contreras-Santos2025}.

Observationally, next-generation surveys such as \textit{Euclid} \citep{Laureijs2011,Scaramella2022,Mellier2025} and LSST \citep{Ivezic2019} will enable systematic ICL studies across large cluster samples to higher redshifts \citep[e.g.][]{Bellhouse2025}. \textit{Euclid} is forecast to detect ICL in tens of thousands of clusters up to $z=1.5$, while LSST will reach $\mu=30$--$31~\mathrm{mag\,arcsec^{-2}}$ ($3\sigma$, $10\arcsec\times10\arcsec$) \citep{Martin2022,Brough2024,Englert2025}. These datasets will enable direct comparison of ICL with weak lensing and X-ray observations, allowing ICL to trace halo properties and assembly histories. However, this requires careful understanding of how stars and DM are distributed during cluster assembly.

A growing body of observational and theoretical work demonstrates a strong correspondence between the large-scale morphology of the ICL and that of the cluster DM halo. For example, \citet{Montes2019b} showed that the projected shape of the ICL closely follows that of the lensing-derived total mass distribution in \textit{Hubble Space Telescope} Frontier Fields clusters, while \citet{Ellien2025} shows similar results using \textit{Euclid} Early Release Observations. Cosmological simulations also report alignments between ICL and DM on large scales \citep[e.g.][]{Yoo2024, Yoo2025, Fernandez2026}. This correspondence indicates that the ICL traces at least some of the cluster’s dynamical and spatial properties. However, detailed studies reveal that the relation is not one-to-one. Numerical simulations show that the ICL's kinematic properties differ systematically from those of both cluster galaxies and the underlying DM halo \citep{Rudick2006, Rudick2011, Dolag2010} and occupies systematically lower orbital energies and higher anisotropies than the background DM halo, producing phase-space offsets that are sensitive to a cluster’s accretion history \citep{Butler2025}. These kinematic differences directly result in the ICL having significantly more compact radial density profiles than the DM, as seen in both observations and simulations \citep{Pillepich2018,Contini2020,Alonso-Asensio2020,Chen2022, Diego2023,Diego2024, Contreras-Santos2024,Contreras-Santos2025,Manuwal2025}. While simulation-calibrated scaling relations can recover DM profiles consistent with independent mass estimates within current uncertainties \citep[e.g.][]{Alonso-Asensio2025}, the scatter in such relations across different simulations and cluster populations remains to be fully quantified, and the systematic differences identified above imply that the underlying DM distribution cannot be precisely inferred from na\"ive one-to-one stellar-to-DM mappings.

Understanding why the ICL exhibits these distinctive phase-space properties requires examining the physical mechanisms that govern stellar stripping. \citet{Butler2025} proposed physical explanations for these systematic differences, two of which we test in this work. First, the more extended DM component of infalling satellites is stripped more efficiently than the centrally concentrated stellar component \citep[e.g.][]{Smith2016,Haggar2021}. This naturally suggests that intracluster stars should occupy lower orbital energies and angular momenta than the DM stripped from the same systems. Second, dynamical friction and tidal torques cause massive satellites to lose orbital energy  \citep{Chandrasekhar1943}. While DM is preferentially removed at earlier times, the centrally concentrated stars can survive until later and are therefore deposited onto more tightly bound orbits. The resulting phase-space offsets between stripped stars and DM are therefore expected to vary with the relative efficiency of tidal stripping, dynamical friction and tidal torques in addition to the mass, internal structure and orbital configuration of the satellites that contribute to the ICL.

These dependencies imply that using the ICL as a tracer of halo mass, shape, or dynamics requires more than a simple mapping between stars and DM. Quantifying the strength and physical origin of the resulting spatial and phase-space offsets is a prerequisite for employing diffuse starlight as a precision probe of cluster DM haloes with forthcoming \textit{Euclid} and LSST data, though practical application will also require careful treatment of observational effects including projection, PSF and noise, spatially varying mass-to-light ratios, and the separation of ICL from the BCG \citep{Brough2024, Manuwal2025}.

The paper proceeds as follows. In Section \ref{sec:method} we describe the controlled $N$-body simulations of satellites accreted into a cluster halo. Section \ref{sec:results} presents the results of these experiments and examines how the phase-space distributions and the resulting offsets in specific energy and angular momentum depend on satellite mass ratio and orbital configuration. Then in Section \ref{sec:population_modeling}, we introduce our methodology to integrate the single-satellite results over plausible satellite populations and in Section \ref{sec:population_offsets}, we quantify how the population-averaged offsets depend on satellite demographics. Throughout this analysis, we provide interactive figures that allow the effects of varying the assumed satellite population parameters to be explored directly; an index of these interactive plots is available at \href{https://garrethmartin.github.io/interactive-profiles-ICL/}{garrethmartin.github.io/interactive-profiles-ICL}. In Section \ref{sec:cosmo_comparison}, we compare these predictions to cosmological simulations to identify any additional effects not captured by our methodology. Finally, in Section \ref{sec:discussion} we discuss the physical interpretation of our results, their limitations, and the implications for using the ICL as a tracer of cluster DM haloes. Section \ref{sec:conclusions} summarises our main conclusions.

\section{Controlled simulations and analysis}
\label{sec:method}

We begin by generating model satellite galaxies, each comprising both DM and stellar components with spatial distributions described by \citet{Hernquist1990} profiles. These satellites are subsequently injected with varying orbital configurations into a live cluster halo with an associated central galaxy.

\subsection{\textsc{Gadget-4}}
We make use of \textsc{Gadget-4}, an open-source cosmological $N$-body and smoothed particle hydrodynamics code \citep{Springel2021}. \textsc{Gadget-4} computes gravitational forces using a hierarchical tree algorithm \citep{Barnes1986}. The tree component employs a multipole expansion with an adaptive opening criterion. Gravitational softening is applied using a cubic spline kernel. Although \textsc{Gadget-4} also includes a particle-mesh scheme for long-range forces, we make use only of the tree-based gravity solver, as the relatively small size of our simulation domain does not benefit from a hybrid TreePM approach. We adopt force softening lengths of 200~pc and 50~pc for DM and stellar particles, respectively. As shown below, this choice maintains the stability of both satellite and cluster models at all resolution levels. Moreover, as demonstrated in \citet{Martin2024}, provided the smoothing lengths are sufficiently small, the precise softening scale has negligible influence on stripping efficiency.

A detailed description of the gravity solver, SPH formulation, time integration scheme, domain decomposition, and parallelisation strategy implemented in \textsc{Gadget-4} is given in \citet{Springel2021}.

\subsection{Simulation setup}
\label{sec:setup}

We generate initial conditions in which a single satellite is injected at a radius of $R_{200}$ into a cluster halo that will attain a mass of $10^{14.5}\,{\rm M_{\odot}}$ at $z=0$, characteristic of nearby intermediate-mass clusters such as the Virgo Cluster \citep[e.g.][]{McLaughlin1999}. Details of the cluster and satellite models are given in Sections \ref{sec:method:models:cluster} and \ref{sec:method:models:satellites}.

We adopt a cluster halo of fixed global mass, which responds self-consistently to the infalling satellite. This assumes the cluster potential evolves adiabatically, on long timescales compared to satellite orbital periods, a reasonable approximation for cluster-scale haloes whose late-time growth is dominated by gradual accretion rather than major mergers \citep{Wechsler2002,Zhao2003,McBride2009}.

In such adiabatically evolving systems, orbital actions remain approximately conserved \citep{Binney2008}. While the absolute energies and angular momenta of individual particles would evolve in a growing halo, the ensemble-averaged ratios between stellar and DM components, $\langle \varepsilon \rangle_\star / \langle \varepsilon \rangle_{\rm DM}$ and $\langle h \rangle_\star / \langle h \rangle_{\rm DM}$, should remain approximately invariant \citep{Blumenthal1986,Gnedin2004}. We adopt this adiabatic approximation as our baseline description of the differential response between stripped stars and DM. Potential violations and their impact on the inferred ratios are discussed in Section~\ref{sec:discussion}.

Each simulation is evolved for 10~Gyr with snapshots output every 100~Myrs. For the majority of the analysis presented in this work, we restrict measurements to the first 6~Gyr of evolution, corresponding to the time elapsed between the cluster formation redshift, defined as the epoch at which the halo has assembled half of its $z=0$ mass ($z_{\rm form}=0.6$), and $z=0$. At each output snapshot, we identify particles gravitationally bound to the satellite by evaluating the gravitational potential using a Barnes-Hut tree method \footnote{implemented as \url{https://github.com/garrethmartin/bh_potential}} applied to the previously identified bound set. A particle is considered unbound once its total specific energy satisfies
\begin{equation}
\frac{1}{2}|\textbf{v} - \textbf{v}_{\rm sat}|^{2} + \phi(\textbf{x})_{\rm sat} > 0,
\end{equation}
where $\phi(\textbf{x})_{\rm sat}$ is the gravitational potential at position, $\textbf{x}$, computed from the particles found to still be bound at the preceding snapshot and $\textbf{v}_{\rm sat}$ denotes the instantaneous velocity of the satellite. Once a particle becomes unbound, it is subsequently treated as unbound in all subsequent snapshots.

For all particles, we also compute the specific orbital energy and specific angular momentum relative to the centre of mass and bulk velocity of the entire system, defined as
\begin{equation}
\varepsilon = \frac{1}{2}|\mathbf{v} - \mathbf{v}_{\rm COM}|^{2} + \phi(\textbf{x}),
\end{equation}
\begin{equation}
h = |(\mathbf{x} - \mathbf{x}_{\rm COM}) \times (\mathbf{v} - \mathbf{v}_{\rm COM})|,
\end{equation}
where $\phi(\textbf{x})$ is the potential of the entire system measured at $\textbf{x}$ and where $\mathbf{x}_{\rm COM}$ and $\mathbf{v}_{\rm COM}$ denote the instantaneous centre-of-mass position and velocity of the combined cluster-satellite system. These quantities are subsequently used to track the orbital evolution of the bound remnant and the kinematic properties of the stripped debris.

All quantities analysed below refer to the phase-space distributions of stripped material measured at the end of the simulated evolution, after the debris has evolved within the live cluster potential.

\subsection{Cluster and satellite models}

Below, we describe the methodology used to create the cluster and satellite models, along with the rationale behind our selected properties for both clusters and satellites. We consider a range of numerical resolutions, summarised in Appendix~\ref{sec:resolution_test}, with the adopted resolution for each simulation scaled to the satellite mass in order to ensure that satellites and the cluster halo are consistently well resolved across the full mass range studied. We further verify that the cluster halo and central galaxy remain numerically stable over the timescales of interest at all adopted resolutions. These choices are informed by previous work highlighting the importance of mass resolution and numerical stability for recovering reliable stripping rates and phase-space evolution \citep{Martin2024}.

\subsubsection{Cluster}
\label{sec:method:models:cluster}

The cluster is modelled at its expected formation redshift of $z_{\rm form} = 0.6$ \citep{Harker2006} as an NFW-like \citep{Navarro1997} halo with a mass of $M_{200} = 10^{14.2}\,{\rm M_{\odot}}$, a concentration of $c_{200} = 4.96$ following relations from \citet{Prada2012}, and a spin parameter of $\lambda = 0.035$, following \citet{Bullock2001}. For the stellar component, we adopt a mass of $M_{\star} = 10^{12.2}\,{\rm M_{\odot}}$ based on the stellar-to-halo mass relation from \citet{Moster2013}, and a half-mass radius of 12.5~kpc from the $z=1$ relation of \citet{Mowla2019}.

We generate the initial conditions of the cluster halo and central galaxy using \textsc{GalIC} \citep[][]{Yurin2014}. We construct an $N$-body system comprising a stellar bulge embedded in a DM halo using the parameters discussed above and summarised in bold in Table~\ref{tab:models}.
\textsc{GalIC} represents both the DM halo and stellar component with \citet{Hernquist1990} profiles; for the DM halo, the scale length is chosen to match the inner density profile of an NFW halo of the same concentration, while for the stellar component, the scale length is set to reproduce the adopted half-mass radius. Initial conditions for the satellites are generated in the same way, with Hernquist profiles for both components and parameters as listed in Table~\ref{tab:models}. In Appendix \ref{sec:resolution_test}, we demonstrate that the cluster DM and stellar radial density profiles do not deviate from these initial conditions outside of the central kpc.

\subsubsection{Satellites}
\label{sec:method:models:satellites}
\begin{figure*}
    \centering
    \includegraphics[width=0.95\textwidth]{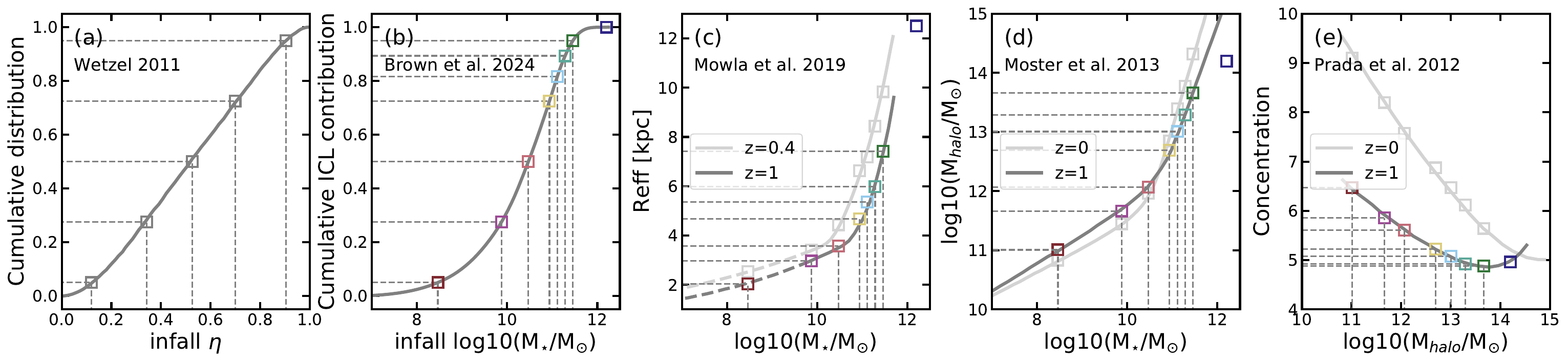}
    \caption{Scaling relations used to assign structural properties to the cluster and satellite models. Panels show: (a) the infall orbital circularity distribution ($\eta$) adopted from \citet{Wetzel2011}, with symbols indicating the 5 circularities selected for the simulations spanning the 5th to 95th percentiles of the distribution, (b) the cumulative fraction of ICL stellar mass contributed by satellites below a given infall stellar mass from \citet{Brown2024}, (c) the stellar mass--size relation from \citet{Mowla2019}, (d) the stellar--to--halo mass relation from \citet{Moster2013}, and (e) the halo concentration--mass relation from \citet{Prada2012}. Symbols in panels (b)--(e) indicate the 8 satellite models listed in Table \ref{tab:models}. Relations are shown at both $z\sim1$ and $z\sim0$ to illustrate redshift evolution. Satellite properties are assigned using the $z=1$ relations, while the cluster is modelled at its formation redshift $z=0.6$, leading to a small offset relative to the plotted scaling relations.}
    \label{fig:satellite_props}
\end{figure*}

Initial conditions for each satellite are also generated using \textsc{GalIC}. Each satellite is modelled as a stellar bulge embedded within an isotropic DM halo, consistent with the setup used for the cluster halo and central galaxy. Satellite stellar masses are selected to sample evenly from the range of systems expected to contribute to the ICL by $z=0$, as inferred from the cumulative ICL contribution of \citet{Brown2024} (shown in Figure~\ref{fig:satellite_props}b). Additional higher-mass models are included to better sample the mass ratio regime in which the DM--ICL orbital energy and angular momentum offset varies most rapidly (see Figure~\ref{fig:energy_am_ratio_vs_mratio}).

Additional satellite properties corresponding to each selected stellar mass are selected using the same relations as for the cluster: the size--mass relation ($M_{\star}$--$R_{\rm eff}$), the stellar--to--halo mass relation ($M_{200}$--$M_{\star}$), and the concentration--mass relation ($M_{200}$--$c_{200}$) are assigned using empirical relations from the literature. We adopt relations calibrated at $z=1$, the closest common redshift for which consistent data exist. These $z=1$ relations are shown in Figure~\ref{fig:satellite_props}, together with the closest relations corresponding $z\sim0$ to illustrate the degree of redshift dependence.

Stellar half-mass radii, halo masses and concentrations are assigned according to the relations found in \citet{Mowla2019}, \citet{Moster2013} and \citet{Prada2012} and shown in panels c, d, and e of Figure~\ref{fig:satellite_props} respectively. We assign a fixed spin parameter of $\lambda=0.035$ to each satellite, consistent with the cluster halo. As discussed in Section \ref{sec:discussion}, our choices for these parameters are likely to have some influence on our results, particularly as the extent of the satellite's stellar component relative to their halo scale radii will influence the relative binding of DM and stellar components.

As with the cluster halo model (see Appendix \ref{sec:resolution_test}), we perform similar resolution tests, again finding no significant deviations from the initial conditions until well within the stellar effective radius of each satellite.

For the satellite orbits, we choose a range of orbital configurations sampled across the expected circularity distribution spanning from the 5th to the 95th percentile described by \citet{Wetzel2011} and indicated in Figure~\ref{fig:satellite_props}. Satellites are injected at the apocentre of their respective orbits, which are fixed at $R_{200}$ of the cluster. For the least massive satellite, which is simulated at the highest mass resolution and is therefore the most computationally expensive, we restrict the orbital sampling to the median circularity and the 5th and 95th percentiles of the distribution. As shown in Section~\ref{sec:phase_space_evolution}, mergers of such disparate mass ratios have a negligible impact on the cluster phase-space structure.

\section{Stellar and DM phase-space properties from individual satellites}
\label{sec:results}

\subsection{Spatial distribution of stripped stars and DM}
\label{sec:spatial_dist}

\begin{figure*}
    \centering
    \includegraphics[width=0.95\textwidth]{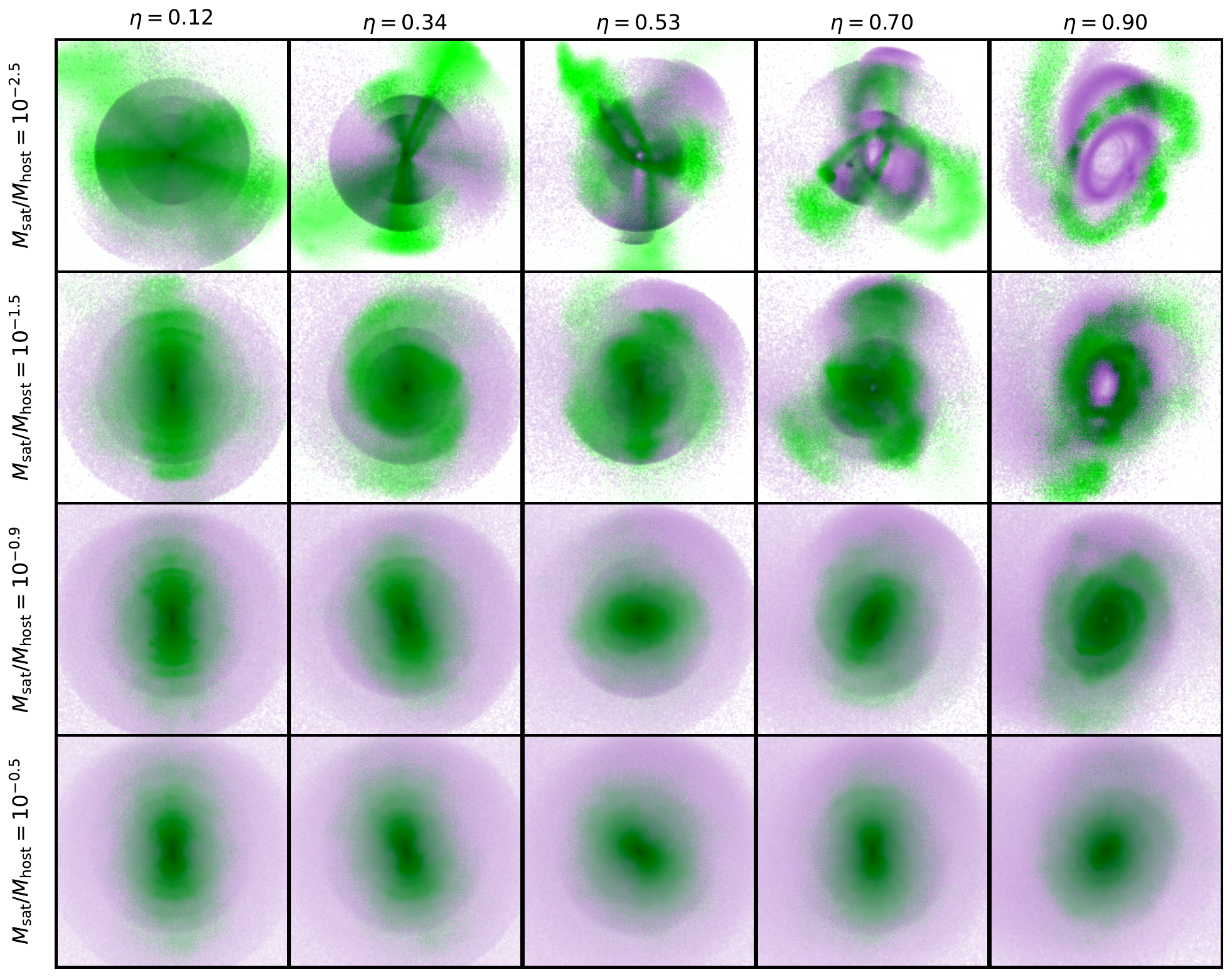}
    \caption{Final projected distributions of stripped stars (green) and DM (purple) for a grid of satellite mass ratios $M_{\rm sat} / M_{\rm host}$ and orbital circularities $\eta$. Each panel spans $2\times2$~Mpc. Columns vary orbital circularity; rows vary mass ratio. An interactive version of these plots showing the distribution for all combinations of mass ratio and orbital circularity can be found here: \href{https://garrethmartin.github.io/interactive-profiles-ICL/index.html\#energy-am}{garrethmartin.github.io/interactive-profiles-ICL/index.html\#energy-am}.}
    \label{fig:satellites_grid}
\end{figure*}

We begin by examining the spatial distribution of stripped stars and DM produced by our controlled simulations. Figure \ref{fig:satellites_grid} shows the final configuration of material stripped from satellites of different masses and orbital configurations, with stars indicated in green and DM in purple. Stellar debris becomes visibly more centrally concentrated towards more similar mass ratios and towards decreasing circularity. At high circularity, both components form coherent streams (especially at disparate mass ratios); at low circularity the debris form shell-like structures. In more equal mass ratio mergers the stellar distribution is markedly more elliptical than the DM distribution, which remains comparatively round, particularly in low-circularity mergers \citep[see also][]{Karademir2019}. For the most disparate mass ratios on the most circular orbits, there is very little visible stellar debris owing to more inefficient stellar stripping, which is the result of the more compact stellar distributions and relatively large DM mass in these systems.

These trends hint at systematic differences in how the stellar and DM components respond to tidal stripping. The DM forms smoother, rounder debris, consistent with material drawn from a wide range of initial radii and therefore a broad spread in orbital energies, leading to more rapid phase-mixing. The stellar debris, originating from the tightly bound central regions of the satellite, appear more centrally concentrated and often retain coherent morphology (streams or shells), suggestive of slower mixing. In similar mass ratio cases, the stellar debris also show higher ellipticity and concentration than the corresponding DM.

These observations suggest that the contrast between DM and stellar debris is closely tied to the satellite’s orbital evolution. Variations in orbital energy and angular momentum during the merger, which are driven by the satellite's infall mass ratio and circularity, determine both the degree of stripping and the subsequent distribution of debris.

\subsection{Evolution of satellite orbits}
\label{sec:satellite_evo}

To quantitatively explain these spatial differences, we examine satellite orbital evolution to establish the range of orbital energies and angular momenta accessible to stripped material, and how these depend on satellite mass ratio and infall circularity.

We begin by studying how satellite orbits evolve as a function of their initial infall properties. This is important because stars and DM are stripped from satellites with energies and angular momenta approximately symmetrically offset about that of the satellite’s orbit \citep[e.g.][]{Hendel2015}. Since DM is typically stripped earlier than stars, any subsequent orbital decay is expected to generate a systematic offset between the mean orbital energies of the stripped stellar and DM components.

\begin{figure}
    \centering
    \includegraphics[width=0.45\textwidth]{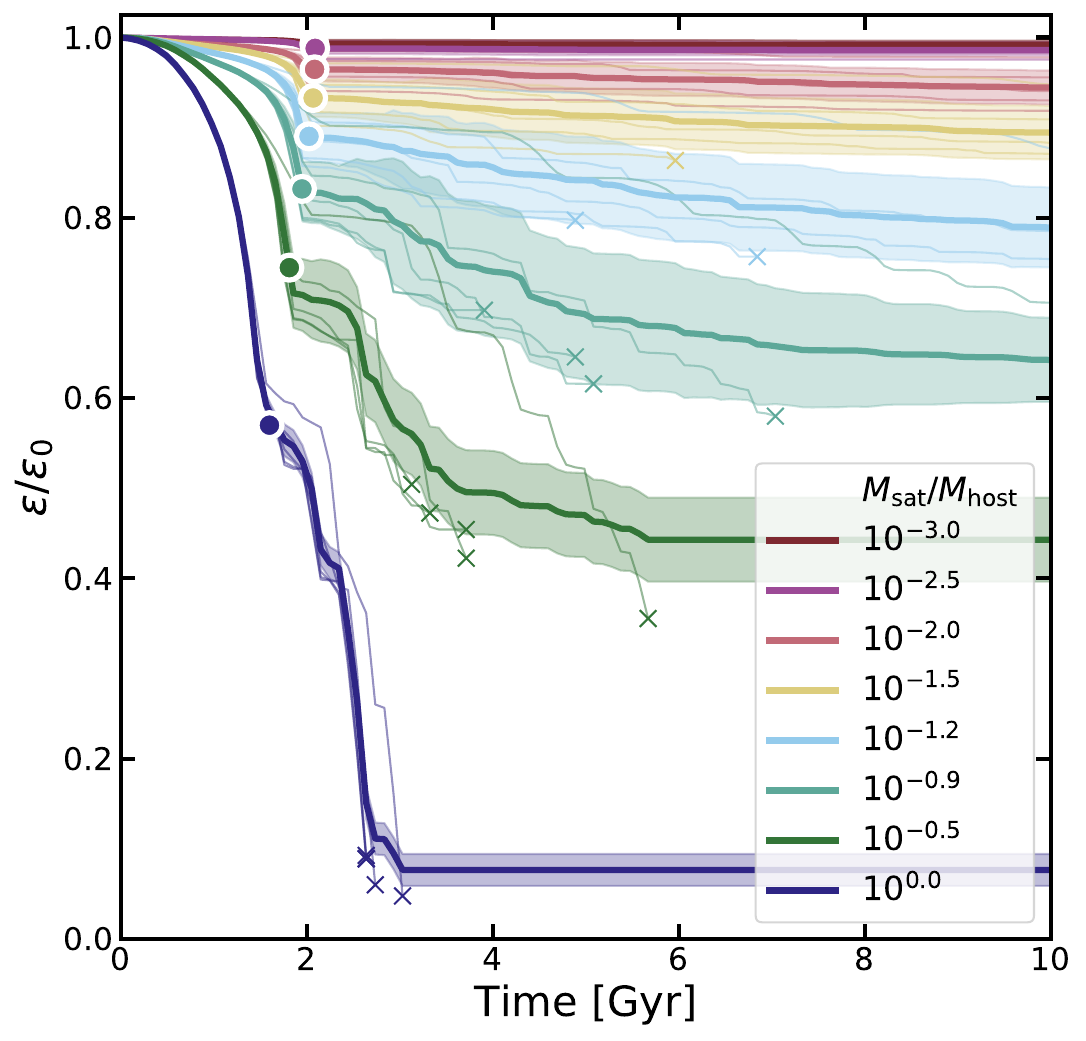}
    \caption{The cluster-centric specific orbital energy evolution of different mass ratio satellites. Thin lines show individual orbits, terminated by a cross where the satellite is completely destroyed, and thick lines show the weighted average orbital energy over all orbital configurations. For the purpose of averaging the profiles, all satellites are included, and for those that are disrupted, the last available orbital energy is assumed to persist. The time when each satellite reaches its first pericentric passage is marked by a filled circle.}
    \label{fig:sat_orbit}
\end{figure}

Figure~\ref{fig:sat_orbit} shows the evolution of satellite orbital energy normalised by its initial value $\varepsilon/\varepsilon_{0}$ as a function of time for a range of orbits and satellite--to--host mass ratios. Individual thin curves correspond to single infall orbits, colour-coded by mass ratio. For each mass ratio, the thick solid line indicates the weighted average over orbital circularities, using the relative frequencies from \citet{Wetzel2011}, with shaded regions denoting the weighted $1\sigma$ scatter. In all cases, $\varepsilon/\varepsilon_{0}$ decreases monotonically with time. The rate of decay depends strongly on satellite mass ratio: more massive satellites lose orbital energy more rapidly owing to stronger dynamical friction and tidal torques.

All satellites exhibit an initially steeper decline in orbital energy prior to first pericentric passage (indicated by filled circles along each satellite's track), when they retain most of their DM haloes. After the bulk of the halo is removed following first pericentre \citep{Smith2016,Martin2024}, orbital decay proceeds more slowly. Equal-mass mergers coalesce almost immediately, while 1:3 mergers typically merge with the central galaxy within one or two pericentric passages. At more disparate mass ratios, orbital energy loss becomes progressively weaker. Even for highly radial orbits reaching pericentres of $\sim4$ kpc (well within the central galaxy’s stellar half-mass radius), low-mass satellites can retain most of their initial orbital energy. We note that our approach is simplified in that it neglects a collisional gas component, which if present would increase the satellite mass and thus enhance dynamical friction. However, in cluster environments, ram-pressure stripping efficiently removes gas from such systems before they reach the central regions \citep[e.g.][]{Roediger2005,Tonnesen2009,Jaffe2018,Arthur2019,Kulier2023}, so the absence of a collisional gas component is unlikely to be critical.

The magnitude of this orbital decay sets an upper limit on the phase-space separation that can develop between stripped stellar and DM components. Even in the limiting case where all DM is stripped at the satellite’s highest-energy orbit and all stars are stripped only after it has reached its lowest energy, the resulting offset cannot exceed the total orbital energy decay. For satellites with mass ratios more disparate than $\sim$1:20, this decay remains small ($<10$ per cent). Consequently, minor mergers alone cannot reproduce the orbital energy offset between the ICL and DM reported by \citet{Butler2025}, in which stripped stars are shown to retain only $\sim75$ per cent of the orbital energy of the stripped DM component. Offsets of this magnitude therefore require mergers with substantially closer mass ratios, in which stronger dynamical friction and tidal torques allow more efficient loss of orbital energy by the satellite.

\subsection{Evolution of stripped stars and DM}
\label{sec:phase_space_evolution}

In this section, we analyse the dynamical properties of the stripped stellar and DM components themselves, which quantify how the debris populates the cluster potential.

\subsubsection{Phase-space distributions}

\begin{figure*}
    \centering
    \includegraphics[width=0.45\textwidth]{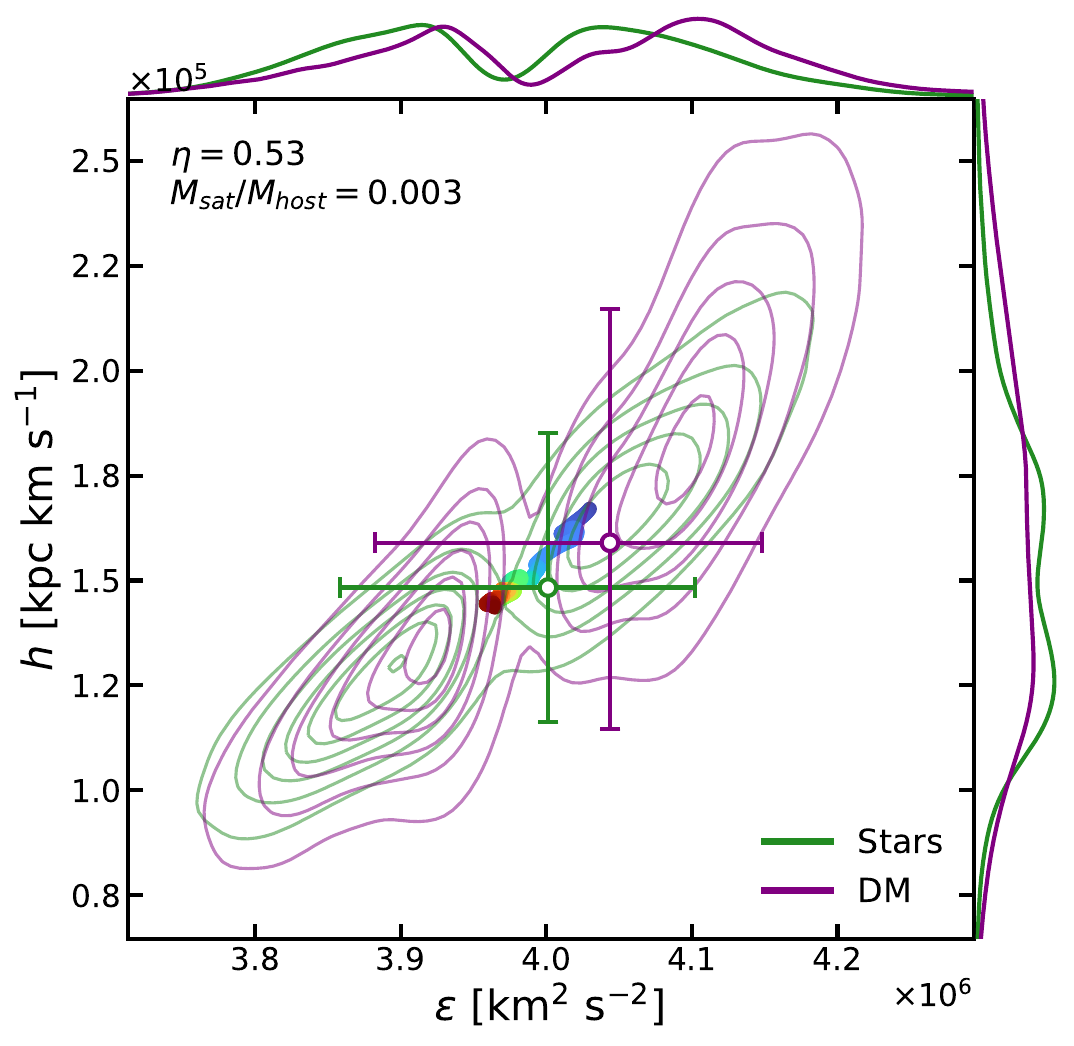}
    \includegraphics[width=0.45\textwidth]{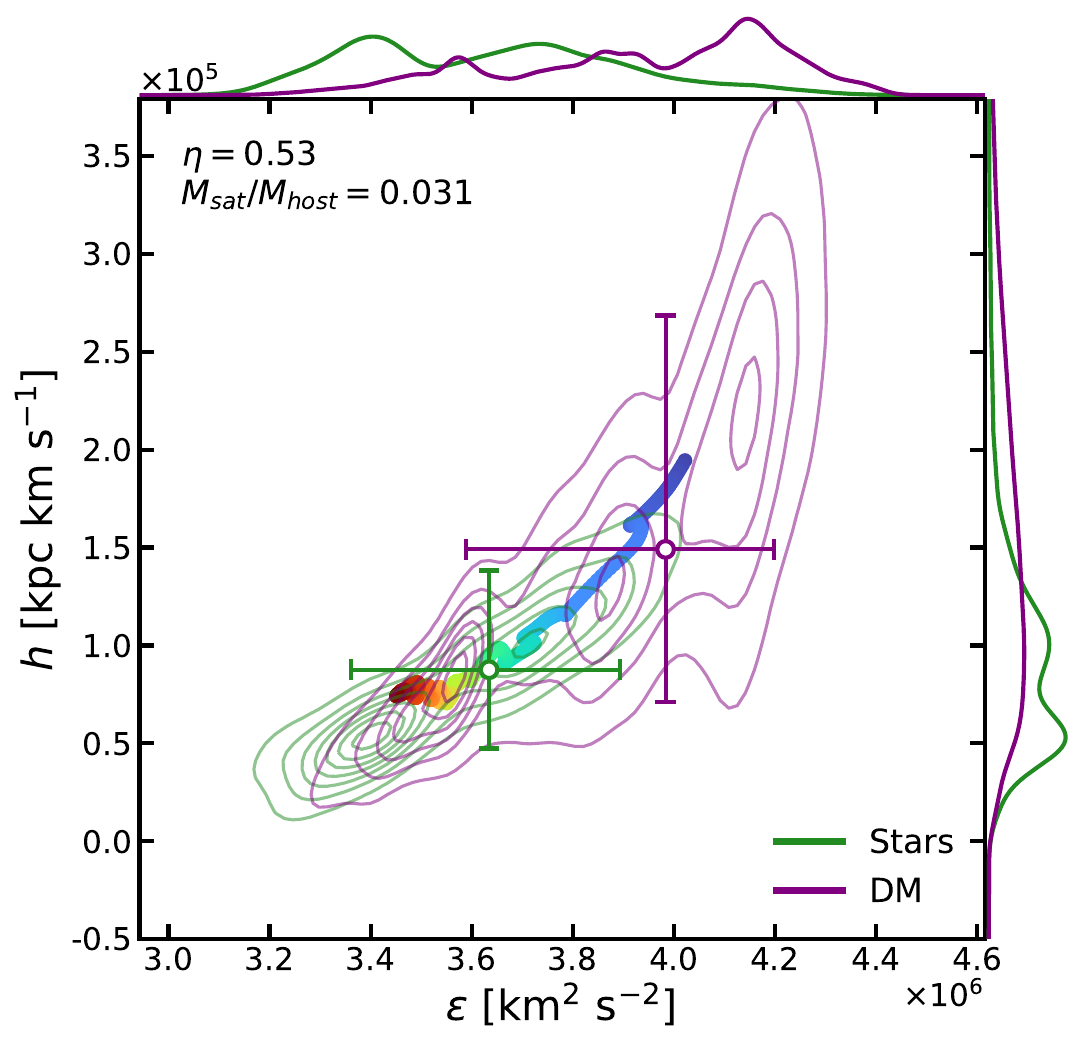}
    \includegraphics[width=0.45\textwidth]{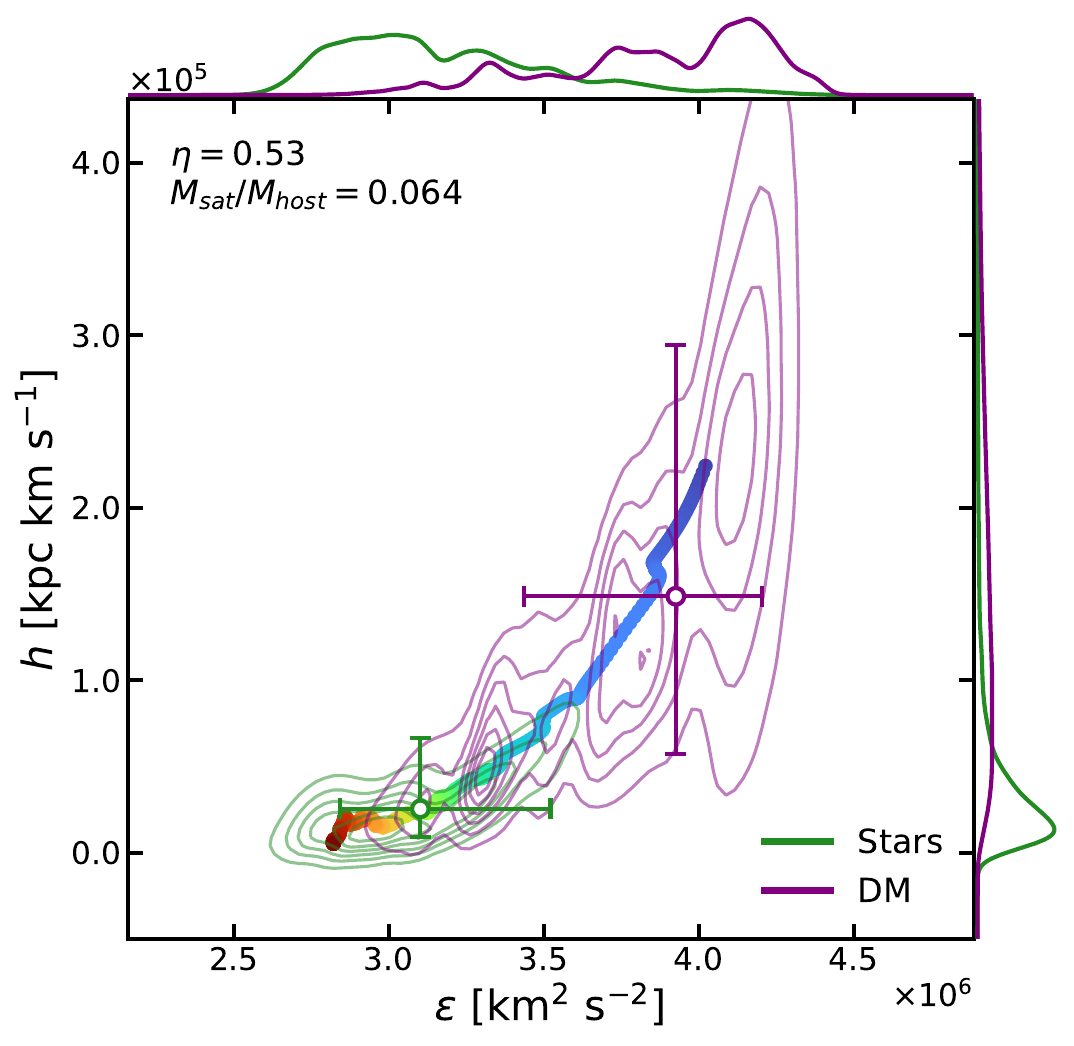}
    \includegraphics[width=0.45\textwidth]{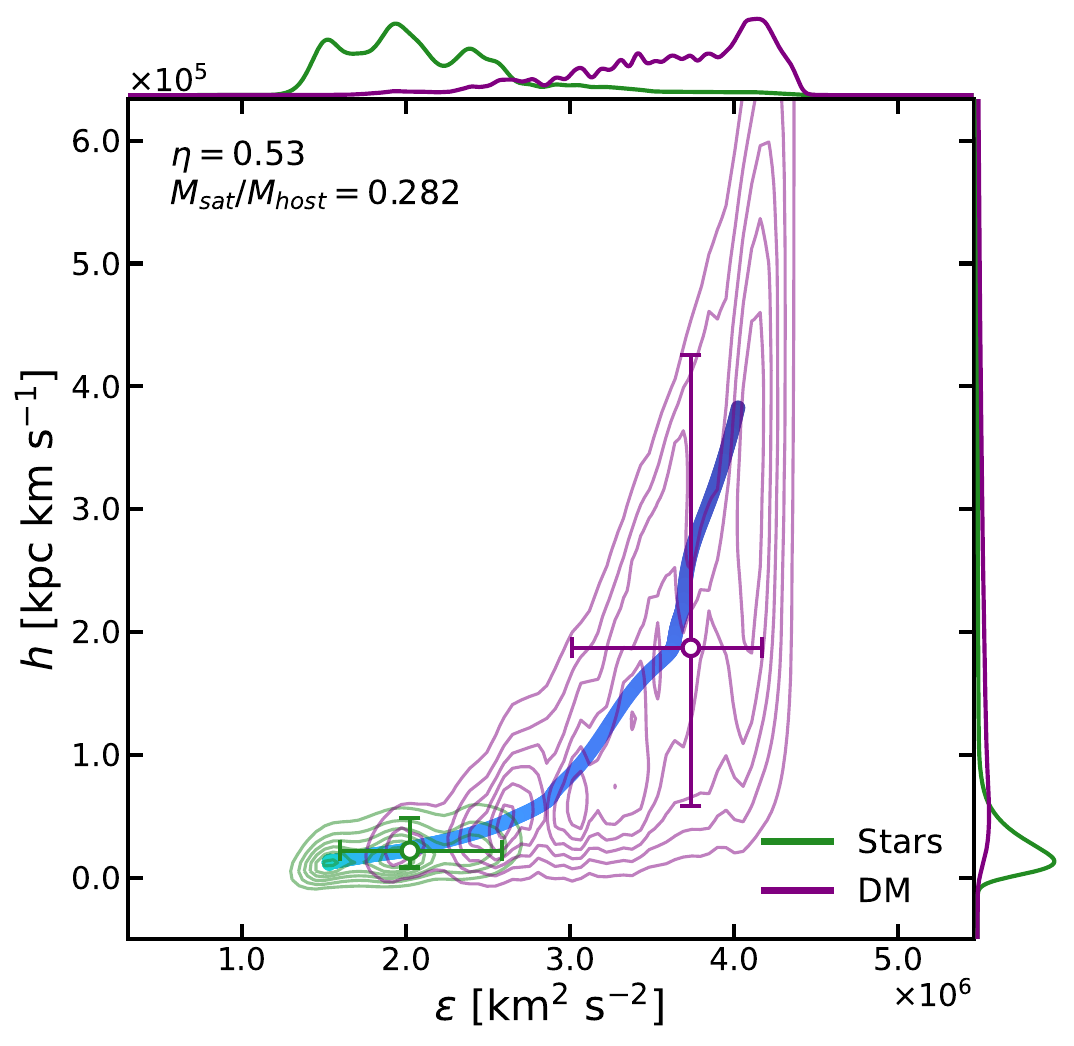}
    \caption{Contour plots showing the final distributions of specific orbital energy ($\varepsilon$) and angular momentum ($h$) for stars (green) and DM (purple) stripped from the satellite by the end of the simulation. Marginal distributions in $\varepsilon$ and $h$ are shown along the horizontal and vertical axes, respectively. Median values and 16--84th percentile dispersions are indicated by error bars for stars (green) and DM (purple). The coloured track traces the evolution of the mean $(\varepsilon,h)$ of stripped particles, from earlier (blue) to later (red) stripping times. An interactive version of these plots showing the distribution for all combinations of mass ratio and orbital circularity can be found here: \href{https://garrethmartin.github.io/interactive-profiles-ICL/index.html\#energy-am}{garrethmartin.github.io/interactive-profiles-ICL/index.html\#energy-am}.}
    \label{fig:energy_dist}
\end{figure*}

Figure~\ref{fig:energy_dist} shows the specific orbital energy--angular momentum distributions for stars and DM stripped from satellites with the mass ratio and orbital circularity indicated in the top-left of each panel. We only show four examples. An interactive version of this figure, available at the address shown in the figure caption, allows exploration of the specific orbital energy--angular momentum distributions for every combination of satellite mass ratio and orbital circularity we simulate.

Green and purple contours represent the distributions of stripped stellar and DM particles, respectively, with the marginalised distributions plotted along the top and right axes. Green and purple error bars indicate the median and 16th--84th percentile dispersion for each component. Coloured tracks show the temporal evolution of the mean energy and angular momentum of the combined DM and stars, with blue corresponding to the earliest stripped particles and red to the most recently stripped. We do not plot separate tracks for the stripped DM and stars as no significant difference is observed between the average final orbital energy or angular momentum of stars and DM provided they are stripped at the same point in time. 

Particles stripped later consistently occupy lower energy and angular momentum states than those stripped earlier, independent of the satellite mass ratio. This reflects the fact that, in the satellite's rest frame, stripped stars and DM are distributed approximately symmetrically about zero \citep[e.g.][]{Johnston1998}, acquiring no systematic offset at a fixed stripping time.
The observed offset between the host-frame distributions of stripped DM and stars is instead a consequence of their different stripping histories along the satellite's decaying orbit. Stars are stripped predominantly after the satellite has already lost orbital energy and angular momentum, while DM is stripped earlier. The magnitude of this offset scales with satellite mass ratio, since more massive satellites experience faster orbital decay (Figure \ref{fig:sat_orbit}). The declining stellar-to-halo mass ratio at high halo masses \citep{Moster2013} may also contribute, as stars in more massive satellites would be more deeply embedded and stripped later, though this effect is also modulated by the mass dependence of galaxy sizes and halo concentrations.

This effect is particularly important for high-mass satellites. As shown in Section \ref{sec:satellite_evo}, strong dynamical friction causes these satellites to lose orbital energy and angular momentum rapidly. As a result, stripped stars from massive satellites can fall to very low energies and angular momenta, often ending up close to or within the central galaxy \citep[e.g.][]{Amorisco2017}. In contrast, DM is always stripped early during the infall, retaining a broader energy and angular momentum distribution and remaining more widely distributed throughout the cluster halo. Consequently, the gap in the energies and angular momenta of stripped stars and DM increases towards more comparable satellite–to–host mass ratios.

Varying orbital circularity (which can be explored in the interactive version of this plot) does not produce any significant corresponding change in the final orbital energy distribution, although it does affect the final angular momentum distribution (partly because lower‑circularity orbits begin with lower angular momentum). However, decreasing circularity reduces the average angular momentum of both stellar and dark‑matter components in similar fashion, so this does not substantially contribute to the phase-space offset between them.

\subsubsection{Dependence on mass ratio}
\label{sec:mass_dependence}

\begin{figure}
    \centering
    \includegraphics[width=0.45\textwidth]{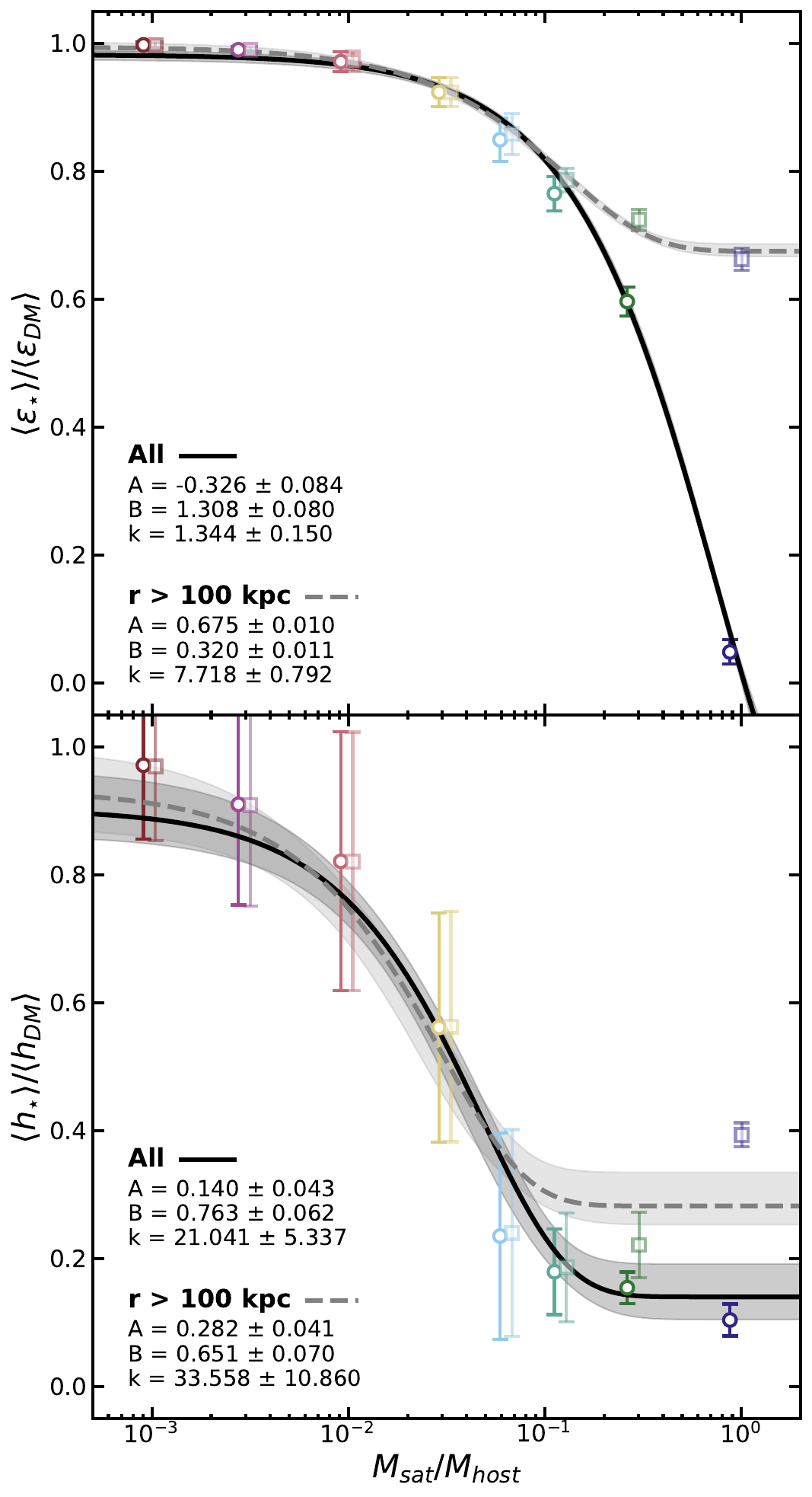}
    \caption{Top: ratio of the mean specific orbital energy of stripped stars to that of stripped DM as a function of satellite--to--host mass ratio. Bottom: equivalent ratio for specific orbital angular momentum. Error bars show the $1\sigma$ dispersion across values of orbital circularity, weighted by the circularity probability distribution. Fainter square markers and error bars indicate the result when particles within 100~kpc of the cluster central galaxy are excluded. Solid and dashed black line indicate best-fit to $y(x)=A+B\exp(-kx)$ for all stripped particles and stripped particles at $r>100$~kpc respectively with shaded grey regions showing the 16--84th percentile envelope of the posterior. Reported parameter values are the maximum a posteriori estimates, with uncertainties equal to half the 16--84th percentile credible interval.}
    \label{fig:energy_am_ratio_vs_mratio}
\end{figure}

Figure~\ref{fig:energy_am_ratio_vs_mratio} shows the ratio of the mean specific orbital energy (top) and angular momentum (bottom) of stripped stars to that of stripped DM as a function of the satellite--to--host mass ratio. Error bars denote the $1\sigma$ dispersion across orbital circularities, weighted by the circularity probability distribution. Fainter square markers and error bars indicate results obtained when excluding particles within 100~kpc of the cluster’s central galaxy, corresponding to a commonly used boundary to isolate the ICL \citep[e.g.][]{Brough2024,Butler2025}. In both panels, we fit the model
\begin{equation}
    y(x) = A + B\,\exp(-kx)
\end{equation}
using Markov Chain Monte Carlo (MCMC) sampling with uninformative priors. Data points are again weighted according to orbital circularity distribution (Figure \ref{fig:satellite_props}a), with weights applied directly to the per-point log-likelihood. Sampling is performed with the NUTS \citep{Homan2014} algorithm implemented in \textsc{PyMC} \citep{pymc2023}, and convergence is verified using standard diagnostics.

For the most unequal-mass mergers, stars and DM exhibit almost identical mean specific orbital energies and angular momenta. The two components only begin to diverge for more comparable mass ratios. At a 1:3 satellite--to--host mass ratio, stars have on average about $50\%$ of the orbital energy and only $\sim15\%$ of the angular momentum of the stripped DM, indicating that stripped stellar material becomes significantly more bound and dynamically colder than the stripped DM.

Excluding the inner 100 kpc leaves the stellar--to--DM ratios unchanged at disparate ratios but begins to diverge at mass ratios of $\sim$1:10. At closer mass ratios, the stellar--to--DM energy ratio remains higher than in the full sample, implying that the strongest energy offsets are driven by stars deposited at small radii. This reflects the fact that, in near-equal mergers, a large fraction of stripped stellar mass settles into the central galaxy, while the corresponding DM is distributed throughout the cluster halo. The angular momentum ratio shows a similar but much weaker response to the radial cut.

The evolution and dispersion in angular momentum ratios is substantially larger than for the orbital energy. This is because angular momentum is more readily modified than energy, as it responds directly to torques, whereas energy changes require coherent forces that alter the orbital speed. This difference is amplified because most stellar stripping occurs near pericentre, where tidal torques primarily redirect particle velocities, producing large reductions in angular momentum with only modest changes in orbital energy.

\subsection{Energy redistribution during satellite accretion}
\label{sec:energy_transfer}

We next examine the redistribution of orbital energy within the cluster during the interaction. Specifically, we track how orbital energy lost from the satellite's orbit is partitioned between the stripped material, the surviving satellite remnant, and the host halo.

\begin{figure}
    \centering
    \includegraphics[width=0.45\textwidth]{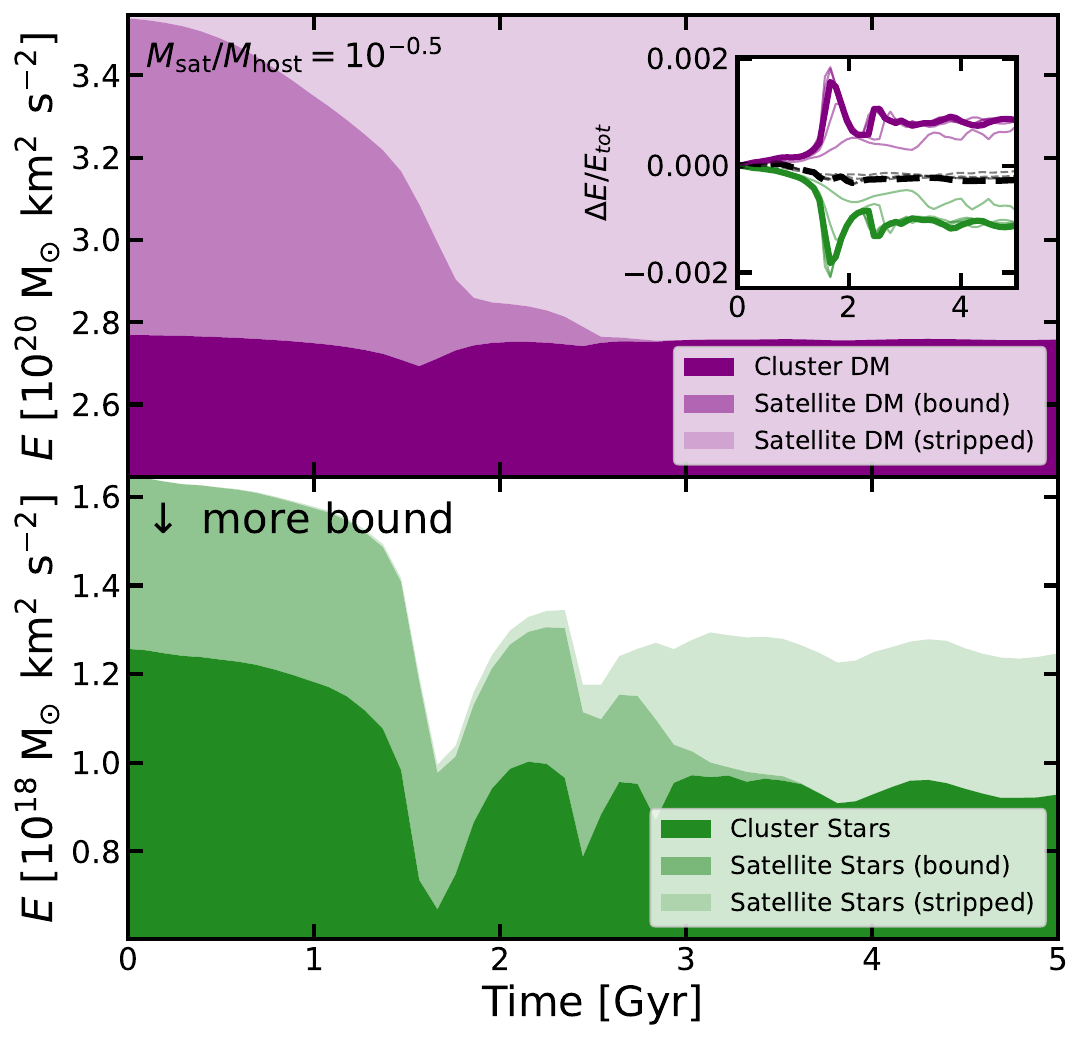}
    \caption{Example of the evolution of the total orbital energy budget $E$ for a 1:3 merger for a circularity of $\eta=0.5$. Each panel decomposes the budget into DM (purple, top sub-panels) and stars (green, bottom sub-panels), further decomposed into three categories: DM and stars associated with the cluster or central galaxy (darkest colour), DM and stars associated with the satellite halo or satellite stars, and DM and stars stripped from the satellite (lightest colour). The $y$-axis is truncated to enhance the visibility of the energy evolution. Inset: fractional change in energy relative to the combined DM+stellar budget $\Delta E / E_{\rm tot}$ (where $E_{\rm tot}$ is the total energy at the start of the simulation), shown separately for the total DM (purple) and the total stellar component (green). The black line shows the fractional change in total energy; perfect energy conservation corresponds to 0. Thin lines show the same quantities for other values of $\eta$ examined.}
    \label{fig:energy_stackplot}
\end{figure}

In Figure \ref{fig:energy_stackplot}, the exchange of orbital energy between the satellite and the cluster is shown for a 1:3 merger with orbital circularity $\eta\sim0.5$. The stacks track how the total orbital energy $E$ contained in each system evolves with time since infall, with the panels separating the DM (top) and stellar (bottom) components. Each panel further divides the energy into material associated with the central galaxy or cluster halo, material still bound to the satellite, and material stripped from the satellite and now bound to the cluster. The inset shows the fractional change in the total energy of the DM and stars, allowing the net transfer to be seen more clearly.

In the top panel, the orbital energy of the satellite DM transitions rapidly into the stripped component. The rise in the stripped contribution to the energy budget mirrors the decline in the satellite. For stripped stars (bottom panel), the behaviour is more gradual. The satellite stars lose energy early, well before the onset of significant stripping, reflecting the deepening of the potential along the inspiral. Only after most of the satellite DM has been removed do the stars begin to populate the stripped component, but this is not accompanied by any further significant decrease in their orbital energy. After $\sim5$ Gyr the stars retain less energy than at the beginning of the simulation, with the principal reduction occurring early during the initial inspiral.

In the inset plot, we show the fractional change in energy of the total stellar and total DM component of the combined satellite-cluster system.
We see that the reduction the total energy of the stars is accompanied by a corresponding increase in the total energy of the DM.
The plotted example involves a major merger, where the stellar contribution is still large enough for such a transfer to be measured. However, at lower mass ratios, while a similar reduction in stellar energy is seen, the stellar mass budget of the satellite becomes so small relative to the cluster’s DM content that the precision of the simulation’s energy conservation is insufficient to definitively attribute this lost energy to the halo.

For comparable satellite--to--host mass ratios, the orbital decay of massive satellites leads to a measurable transfer of energy into the cluster halo. However, even in these cases the net response of the cluster halo remains small ($\lesssim0.25$ per cent for both a 1:1 merger and the example 1:3 merger). This indicates that the stellar--DM phase-space offsets identified above are not the result of a strong global modification of the cluster potential. In our simulations, the offsets therefore largely reflect differences in how stars and DM populate phase-space during satellite stripping, rather than large-scale heating or restructuring of the host halo.

\section{Infalling satellite population modelling}
\label{sec:population_modeling}

\subsection{Predictive model for individual satellites}
\label{sec:model}
We now translate these single-satellite results into cluster-scale predictions and assess their sensitivity to satellite demographics. We use the simulation results to construct a model predicting, for a satellite of given satellite--to--host mass ratio and orbital circularity, the average orbital energy and angular momentum of the stripped stellar and DM components and then integrate these predictions over an ensemble of infalling satellites.
\subsubsection{Model formulation}

On the basis of the simulation outputs, we first fit a joint model that predicts six scalar quantities: the mean per-particle orbital energy and angular momentum of the stripped stars and DM (computed from the distributions shown in Figure \ref{fig:energy_dist}, with mean values presented in ratio as open circles in Figure \ref{fig:energy_am_vs_circ}), and the stripped fractions of each component,
\begin{equation}
\begin{aligned}
y = \{&\log_{10}\langle\varepsilon\rangle_{\mathrm{DM}},
        \log_{10}\langle h\rangle_{\mathrm{DM}},
        \log_{10}f_{\mathrm{strip,\,DM}},
       \log_{10}\langle\varepsilon\rangle_{\star},\\
        &\log_{10}\langle h\rangle_{\star},
        \log_{10}f_{\mathrm{strip,\,\star}}\}
\end{aligned}
\end{equation}
as a function of the satellite-to-cluster mass ratio and orbital circularity $x=\big\{\log_{10}M_\star,\ \eta\big\}$. Two independent models are fitted: one using quantities computed over all stripped particles, and a second using quantities restricted to stripped particles with final radii $r>100$~kpc, representing the boundary between the cluster's central galaxy and the ICL.

We employ a linear model of coregionalisation \citep[LMC;][]{Goulard1992, Alvarez2012} to model the joint dependence of multiple, physically related outputs on satellite-to-cluster mass ratio and orbital circularity. The LMC assumes each output is a linear combination of shared latent Gaussian processes \citep{Neal1997} that vary smoothly with the inputs. Because the latent functions are shared, outputs are correlated; the coregionalisation weights determine how strongly each output responds to each latent function, allowing the model to capture cross-output correlations while permitting each output to respond differently to the input parameters. This approach suits our problem because all predicted quantities arise from the same underlying dynamics but measure different physical outcomes.

Our implementation works with the implied covariance after integrating out the latent functions. Input arrays are augmented with an output-index dimension such that each data point is $(x,d)$, where $d$ denotes the output index. The resulting covariance between outputs $d$and $d'$ at inputs $x$ and $x'$ is then modelled as a sum of two kernels (squared exponential and Mat\'ern 5/2) acting on the physical input dimensions, each coupled to a coregionalisation matrix $B_k$:

\begin{equation}
\mathrm{Cov}\!\left(y_d(x),\,y_{d'}(x')\right)
= \sum_{ \mathclap{k\in\{\mathrm{SE},\mathrm{M52}\}} } k_{\mathrm{input},k}(x,x')\,[B_k]_{d,d'}.
\end{equation}
Each coregionalisation matrix encodes how shared latent functions contribute to different outputs, while the input kernels control how these functions vary across parameter space. Kernel amplitudes, length scales, and coregionalisation weights are given weakly informative priors.

Following identical methodology, we also employ a third model predicting the DM and stellar density of stripped stars and DM $y = \big\{ \rho_{\rm DM}, \rho_{\star} \big\}$ as a function of cluster-centric radius, satellite-to-cluster mass ratio and orbital circularity $x=\big\{ \log_{10}r, \log_{10} M_{\star}, \eta \big\}$.

For the models predicting orbital energies, angular momenta, and stripped fractions, we adopt a coregionalisation rank of two. This choice is sufficient to capture the dominant cross-output covariance present in the simulation data while avoiding unnecessary degeneracy in the coregionalisation weights. For the density profile model, which predicts stellar and DM densities tracing the same underlying radial structure, we adopt a coregionalisation rank of one, reflecting the stronger coupling between the two outputs.

\subsubsection{Model fitting and validation}
We perform Bayesian inference on the hyperparameters of the multi-output Gaussian process, including the kernel length scales and amplitudes and the coregionalisation weights that control correlations between outputs. The latent functions themselves are never explicitly fitted, as their contribution to the outputs is fully captured by the Gaussian process covariance. Inference is performed using the probabilistic programming framework \textsc{PyMC} \citep{pymc2023}. Approximate posterior estimates are first obtained via automatic differentiation variational inference (ADVI). Multiple jittered draws from these ADVI derived posterior estimates are used to initialise ten independent chains of MCMC sampling using NUTS \citep{Homan2014}. Convergence is checked with standard diagnostics: Gelman-Rubin statistic, effective sample size, trace and autocorrelation plots, and posterior predictive checks.

Posterior predictive quantities are obtained over a finely sampled grid by drawing samples of the Gaussian process hyperparameters from their posterior distribution. For each hyperparameter draw, we compute the Gaussian process predictive mean $\mu$ and covariance $\Sigma$ at the target grid conditioned on the training data, and draw a single realisation from the corresponding multivariate normal distribution $\mathcal{N}(\mu,\Sigma)$. Repeating this across 10{,}000 posterior draws produces an ensemble of plausible functions that captures hyperparameter uncertainty through the posterior samples and predictive uncertainty through the Gaussian process conditional.

The full implementation of the LMC models, including priors, likelihoods and validation is publicly available at \href{https://github.com/garrethmartin/interactive-profiles-ICL/tree/main/MCMC}{github.com/garrethmartin/interactive-profiles-ICL}.

\subsection{Convolving across infall populations}
\label{sec:populations}

To translate the single-satellite predictions to expectations for a whole cluster, the modelled per-particle energies, angular momenta and stripped fractions are convolved with plausible infalling satellite populations. Since the contribution of an individual satellite to the ICL and cluster DM halo depends both on that satellite's internal properties and on its abundance and orbital distribution, population averages can be obtained by weighting the model outputs by the stripped mass contributed by satellites and by an assumed satellite stellar mass function and orbital circularity distribution.

To quantify how the mean specific energy and angular momentum of stripped stars and DM depend on satellite demographics, we convolve the single-satellite model predictions with a grid of plausible satellite stellar mass functions $\phi(M_{\star})$ and orbital circularity distributions $p(\eta)$. For each choice of $\phi(M_{\star})$ and  $p(\eta)$, the per-particle quantities are weighted by the stripped mass contributed by satellites and integrated over mass and circularity to yield population-averaged $\langle \varepsilon \rangle$ and $\langle h \rangle$. 

We vary (1) the stellar mass function of the infalling satellites, parameterised by a Schechter function \citep{Schechter1976},
\begin{equation}
    \phi(M_{\star})\propto \left(\frac{M_{\star}}{M^{\star}}\right)^{\alpha}\exp\left(\frac{-M_{\star}}{M^{\star}}\right),
\end{equation}
with low-mass slope $\alpha$ and characteristic stellar mass $M^{\star}$, and (2) the orbital circularity distribution, which we model as a beta distribution,
\begin{equation}
    p(\eta)\propto\frac{\eta^{a-1}(1-\eta)^{b-1}}{\frac{\Gamma(a)\Gamma(b)}{\Gamma(a+b)}},
\end{equation}
with parameters $a$ and $b$ controlling the skew towards low or high circularity respectively.

\subsubsection{Population parameter choices}
In order to define a range of plausible infalling galaxy stellar mass functions, we consider a range of mass function parameters informed by the four simulations described in Appendix~\ref{sec:cosmo_sims}: \textsc{TNG100}, \textsc{Horizon-AGN}, \textsc{Hydrangea}, and \textsc{TheThreeHundred}. From each simulation we select clusters with $z=0$ halo masses in the range $M_{200}=10^{14}$--$10^{15}\,$M$_{\odot}$ as described in more detail in Brown et al., in prep. This yields 15 clusters from \textsc{TNG100} with a mean halo mass of $10^{14.2}\,$M$_{\odot}$, 8 clusters from \textsc{Horizon-AGN} with a mean halo mass of $10^{14.3}\,$M$_{\odot}$, and 16 clusters from \textsc{Hydrangea} with a mean halo mass of $10^{14.4}\,$M$_{\odot}$. Clusters in \textsc{TheThreeHundred} are systematically more massive; we retain all haloes with $M_{200}<10^{15}\,$M$_{\odot}$, resulting in a sample of 15 clusters with a mean halo mass of $10^{14.9}\,$M$_{\odot}$.

As described in Appendix \ref{app:schechter_fits}, we first measure the stellar mass distribution of all infalling satellites and then separately fit a Schechter function for each simulation. This provides an estimate of plausible $\alpha$ and $M^{\star}$ values which are broadly consistent with a range of observational measurements of the galaxy stellar mass function between $0<z<1$ \citep[e.g.][]{Baldry2012, Ilbert2013, Tomczak2014, Davidzon2017, Sedgwick2019, McLeod2021, Weaver2023}. Fiducial values are chosen near the centre of this range.

\begin{figure}
    \centering
    \includegraphics[width=0.45\textwidth]{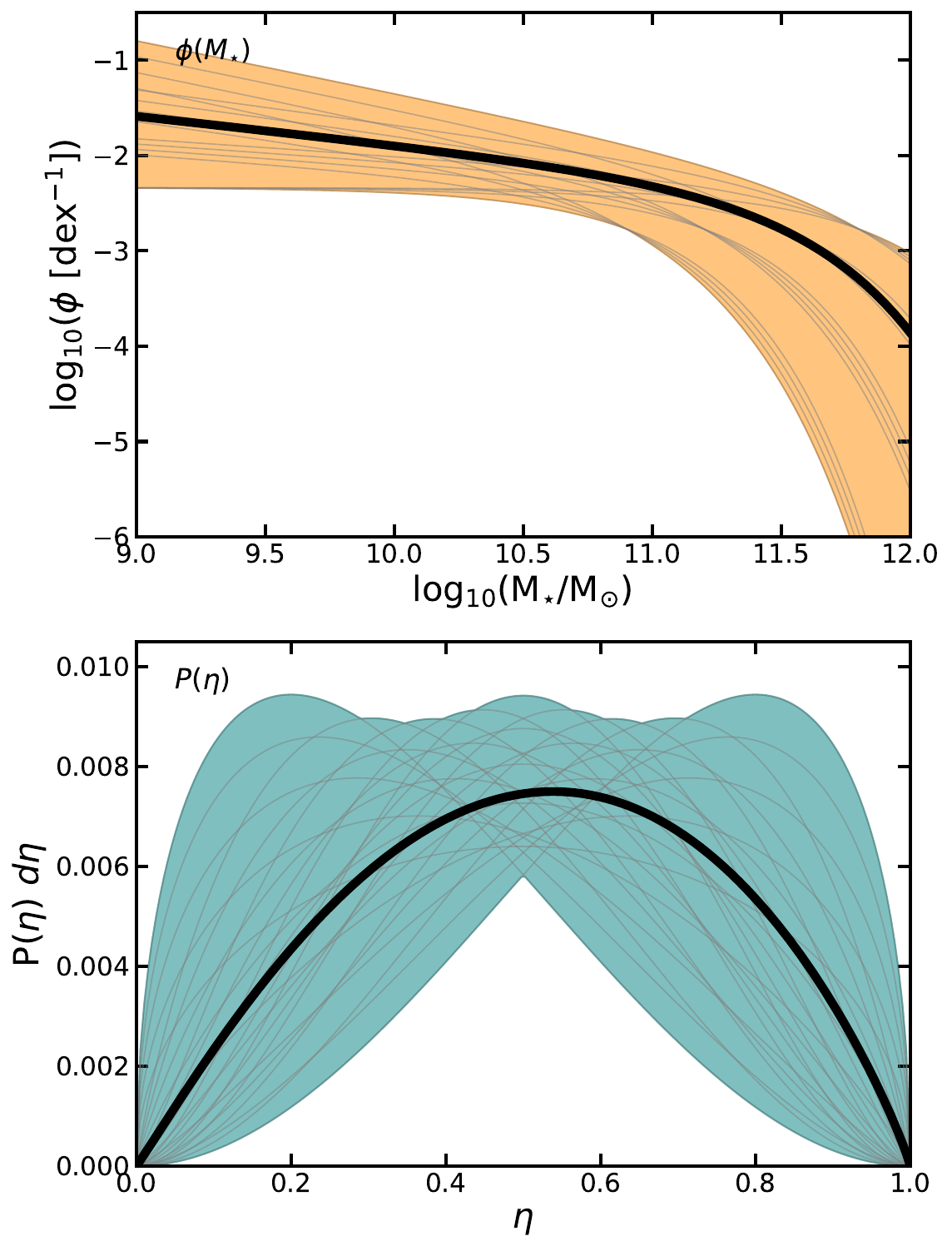}
    \caption{Parameter ranges explored in the satellite population model. Shaded regions indicate the envelopes spanned by variations in the infalling stellar mass function parameters (top, $\alpha, M^{\star}$) and the orbital circularity distribution parameters (bottom, $a,b$). The fiducial parameter choice is shown by the thick black line, while a representative subset of individual parameter realisations is indicated by thin grey lines. These ranges define the space over which population-averaged quantities are computed by marginalising over plausible satellite mass and orbital distributions.}
    \label{fig:function_params}
\end{figure}

For the initial orbital circularity distribution of infalling satellites, we fit a beta function to the results of \citet{Wetzel2011}. We adopt a broad range of values for $a$ and $b$, chosen to span reasonable variations in orbital shapes, from a moderate bias towards radial orbits to moderate bias towards circular orbits. The combined parameter ranges explored for both the stellar mass function and the orbital circularity distribution are illustrated in Figure~\ref{fig:function_params}, which shows the envelopes spanned by these variations together with the adopted fiducial model and representative individual realisations. The fiducial values and parameter ranges used are summarised in Table~\ref{tab:function_params}.

\begin{table}
\centering
\begin{tabular}{llll}
\hline
Function & Parameter & Fiducial value & Range \\ \hline
$p(\eta)$ & $a$ & 2.05 & [1.5, 3.0] \\
                        & $b$ & 1.90 & [1.5, 3.0] \\[2pt]
$\phi(M_{\star})$       & $M^{*}$ & $10^{11.5}$~M$_{\odot}$ & [$10^{10.9}$~M$_{\odot}$, $10^{11.8}$~M$_{\odot}$] \\
                        & $\alpha$ & $-1.3$ & [$-1.55$, $-1.0$] \\[2pt]
\hline
\end{tabular}
\caption{Summary of the functional parameters adopted for the infalling satellite population model, listing the fiducial values and ranges explored for each parameter.}
\label{tab:function_params}
\end{table}

\subsubsection{Population averaged quantities}

For a given stellar mass function $\phi(M_{\star})$ and orbital circularity distribution $p(\eta)$, we compute population-averaged specific energies of stripped stars and DM by summing over the contributions from all satellites, weighted simultaneously by (i) the amount of mass stripped from each satellite given by the product of $f_{\rm strip,\star}(M_\star, \eta)$ and $M_\star$ or $f_{\rm strip,DM}(M_\star, \eta)$ and  $M_{\rm h}(M_\star)$, and (ii) the product of the abundance of satellites with that stellar mass $\phi(M_\star)$ and orbital circularity $p(\eta)$. 
For the stripped stellar component the population-averaged specific energy is

\begin{equation}
\begin{aligned}
&\langle \varepsilon \rangle_{{\rm tot}, \star} = 
\\
&\frac{
\textstyle \int_{M_{\star,\rm min}}^{M_{\star,\rm max}} 
\int_{\eta_{\rm min}}^{\eta_{\rm max}} 
\varepsilon(M_\star, \eta)\, M_\star\, f_{\rm strip}(M_\star, \eta)\, \phi(M_\star)\, p(\eta)\, 
\mathrm{d}\eta\, \mathrm{d}M_\star
}{
\textstyle \int_{M_{\star,\rm min}}^{M_{\star,\rm max}} 
\int_{\eta_{\rm min}}^{\eta_{\rm max}} 
M_\star\, f_{\rm strip}(M_\star, \eta)\, \phi(M_\star)\, p(\eta)\, 
\mathrm{d}\eta\, \mathrm{d}M_\star
}.
\end{aligned}
\end{equation}

\noindent and the corresponding quantity for stripped DM is

\begin{equation}
\begin{aligned}
&\langle \varepsilon \rangle_{{\rm tot}, {\rm DM}} =
\\
&\frac{
\textstyle \int_{M_{\star,\rm min}}^{M_{\star,\rm max}} 
\int_{\eta_{\rm min}}^{\eta_{\rm max}} 
\varepsilon(M_\star, \eta)\, M_{\rm h}(M_\star)\, f_{\rm strip}(M_\star, \eta)\, \phi(M_\star)\, p(\eta)\, 
\mathrm{d}\eta\, \mathrm{d}M_\star
}{
\textstyle \int_{M_{\star,\rm min}}^{M_{\star,\rm max}} 
\int_{\eta_{\rm min}}^{\eta_{\rm max}} 
M_\star\, f_{\rm strip}(M_\star, \eta)\, \phi(M_\star)\, p(\eta)\, 
\mathrm{d}\eta\, \mathrm{d}M_\star
}.
\end{aligned}
\end{equation}

Analogous expressions are used to compute the population-averaged specific angular momentum $h(M_{\star}, \eta)$. Here $M_{h}(M_{\star})$ denotes the halo mass associated with a satellite of stellar mass $M_{\star}$ and $f_{\rm strip}$ denotes the stripped fraction of the specified component. We consider the result for particle-averaged values over all particles and restricted to $r>100$~kpc. For the latter, $f_{\rm strip}$ denotes only the stripped mass outside of 100~kpc as a fraction of the satellite's infall mass.

We further define the normalised progenitor stellar mass contribution function, $p(M_\star \mid {\rm strip},\star)$, which quantifies the amount of stellar mass entering the ICL per bin of satellite infall stellar mass, normalised by the total stripped stellar mass.

\begin{equation}
\begin{aligned}
&p(M_\star \mid {\rm stripped},\star) =
\\
&\frac{
\textstyle
\phi(M_\star)\,M_\star
\int_{\eta_{\min}}^{\eta_{\max}}
f_{{\rm strip},\star}(M_\star,\eta)\,p(\eta)\,\mathrm{d}\eta
}{
\textstyle
\int_{M_{\star,\min}}^{M_{\star,\max}}
\phi(M_\star)\,M_\star
\int_{\eta_{\min}}^{\eta_{\max}}
f_{{\rm strip},\star}(M_\star,\eta)\,p(\eta)\,\mathrm{d}\eta\,
\mathrm{d}M_\star
}.
\end{aligned}
\end{equation}

For each point in the ($\phi(M_{\star})$,$p(\eta)$) parameter space we propagate model uncertainty by drawing multiple posterior predictive realisations of our multi-output Gaussian process model. For every realisation we evaluate the integrals above. The ensemble of results yields posterior distributions for the population-averaged $\varepsilon$ and $h$ and $p(M_\star \mid {\rm stripped},\star)$, which we summarise by their median values and credible intervals. To illustrate sensitivity to the assumed infalling satellite population we vary the stellar mass-function parameters while holding $p(\eta)$ fixed at its fiducial form and then vary the circularity distribution while holding the stellar mass function fixed.

\section{Population-averaged stellar and DM phase-space offsets}
\label{sec:population_offsets}
\subsection{ICL mass budget}

\begin{figure}
    \centering
    \includegraphics[width=0.45\textwidth]{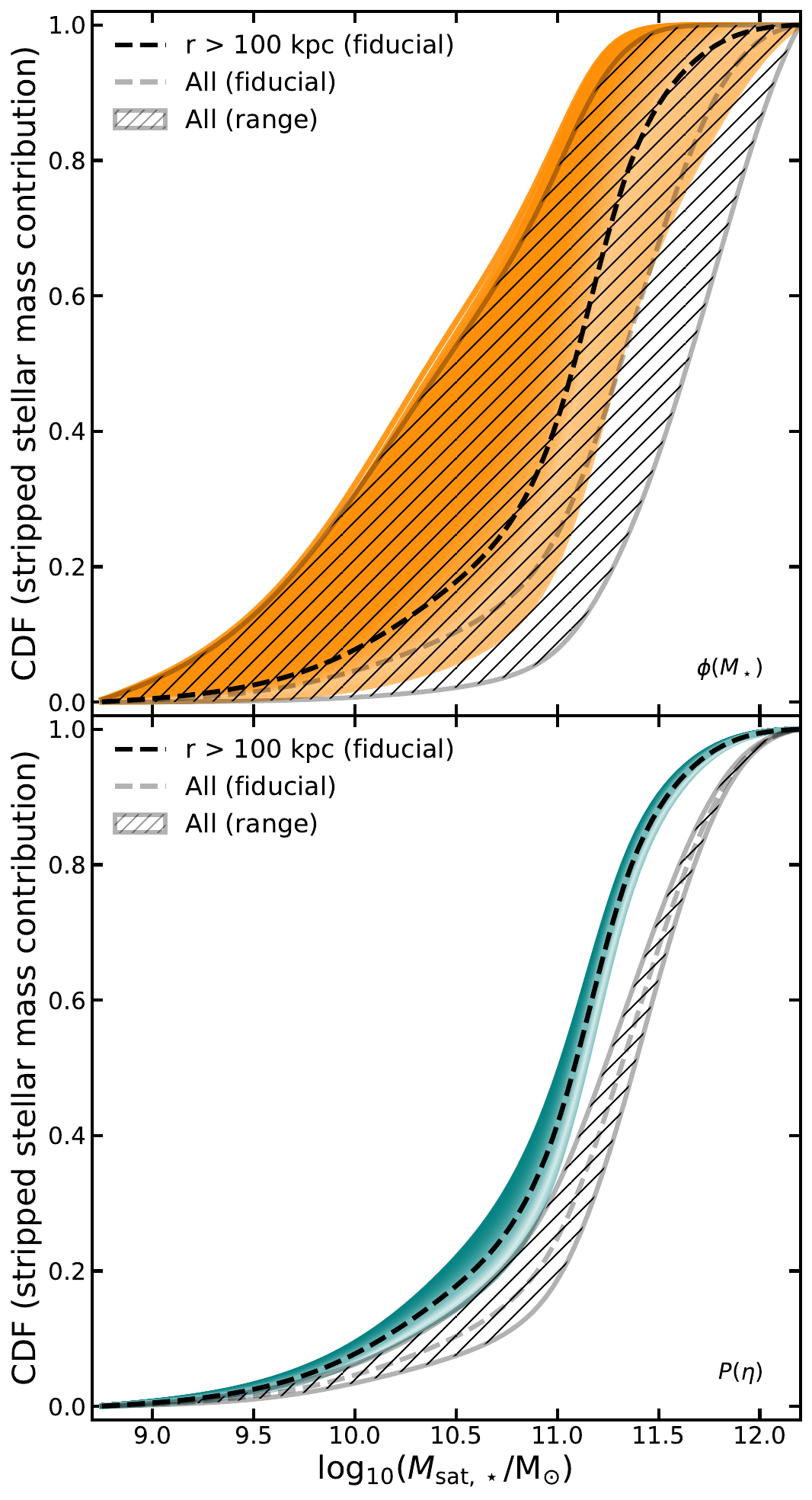}
    \caption{The ranges of cumulative distribution functions for the contribution of galaxies to the total stripped stellar mass, shown as a function of infall stellar mass over the tested parameter space (Table \ref{tab:function_params}). The shaded and hatched regions indicate the range of CDFs obtained when varying model parameters. The top panel shows variations over the stellar mass function while holding the orbital circularity distribution fixed at its fiducial value and the bottom panel shows variations over the orbital circularity distribution while holding the galaxy infall stellar mass function fixed at its fiducial value. Coloured regions correspond to stripped stars with $r>100$~kpc and hatched regions indicate the same for all stripped stars. Black and grey dashed lines show the fiducial CDF restricted to stripped stars with $r>100$~kpc and all stars, respectively. The coloured region is rendered with a graded colour to indicate the family of parameter combinations. Interactive versions of this plot can be found at \href{https://garrethmartin.github.io/interactive-profiles-ICL/index.html\#stripping}{garrethmartin.github.io/interactive-profiles-ICL/index.html\#stripping}.}
    \label{fig:mass_budget}
\end{figure}

In Figure~\ref{fig:mass_budget}, we examine how satellite stellar mass and orbital circularity distributions determine the ICL mass budget, which in turn governs how each satellite contribute to population-averaged stellar-DM phase-space offsets. We show cumulative distribution functions of the contribution of satellites to the total stripped stellar mass as a function of infall stellar mass, $p(M_\star \mid {\rm stripped},\star)$ for variations in the infalling satellite stellar mass function \textbf{(top)} and circularity distribution \textbf{(bottom)}.

Across the plausible range of satellite stellar mass functions considered, the infall stellar mass scale at which satellites contribute half of the total stripped stellar mass varies substantially, spanning approximately $M_{\star}\sim10^{10}$--$10^{11.5}\,{\rm M}_{\odot}$. This spread is driven primarily by variations in the characteristic mass $M^{\star}$ of the mass function. Increasing $M^{\star}$ shifts the typical satellites contributing to the stellar mass budget towards higher masses. In contrast, variations in the orbital circularity distribution produce only minor changes to the cumulative mass budget. Consequently, the median mass scale of the ICL contribution is insensitive to reasonable variations in $p(\eta)$ over the parameter space explored.

For the fiducial parameter choice, satellites with infall stellar masses above $M_{\star}\simeq10^{11}\,{\rm M}_{\odot}$ contribute roughly $50$ per cent of the total ICL mass ($r>100$~kpc). When considering the combined central and ICL component, the corresponding mass scale increases to $\sim10^{11.5}\,{\rm M}_{\odot}$. This difference reflects the rapid orbital decay of the most massive satellites, whose stellar mass is preferentially deposited into the central regions of the cluster, enhancing the central component relative to the diffuse ICL at large radii, as seen in Figure \ref{fig:energy_am_ratio_vs_mratio}. These higher-mass infallers therefore primarily shape the cluster-scale phase-space distributions of both the ICL and central component.

We also compute the ICL fraction from our model using the definition $f_{\rm ICL} = M_{\star}(r > 100~{\rm kpc}) / M_{\star,{\rm tot}}$, where the total stellar mass includes the central galaxy, diffuse stripped stars and surviving satellites. Our fiducial model yields $f_{\rm ICL} = 0.29$, with the value ranging from 0.23 to 0.36 when varying the satellite stellar mass function and from 0.21 to 0.38 when varying the orbital circularity distribution. Notably, while the circularity distribution has little impact on the stellar--to--DM phase-space offset (Section \ref{sec:mass_dependence}), it produces comparable variation in the ICL fraction to that from the stellar mass function. These values are in good agreement with those reported in hydrodynamic simulations using the same definition, where $f_{\rm ICL} \sim 0.15-0.35$ is found for clusters of similar mass \citep{Brough2024, Montenegro-Taborda2025}.

\subsection{Average orbital energy and angular momentum}
\label{sec:e_am_avs}

Having established how satellite mass and orbital properties shape the ICL mass budget, we now examine the population-averaged stellar--to--DM phase-space offsets obtained by convolving the single-satellite predictions over the range of plausible infalling satellite populations described in Section \ref{sec:populations}.

\begin{figure}
    \centering
    \includegraphics[width=0.45\textwidth]{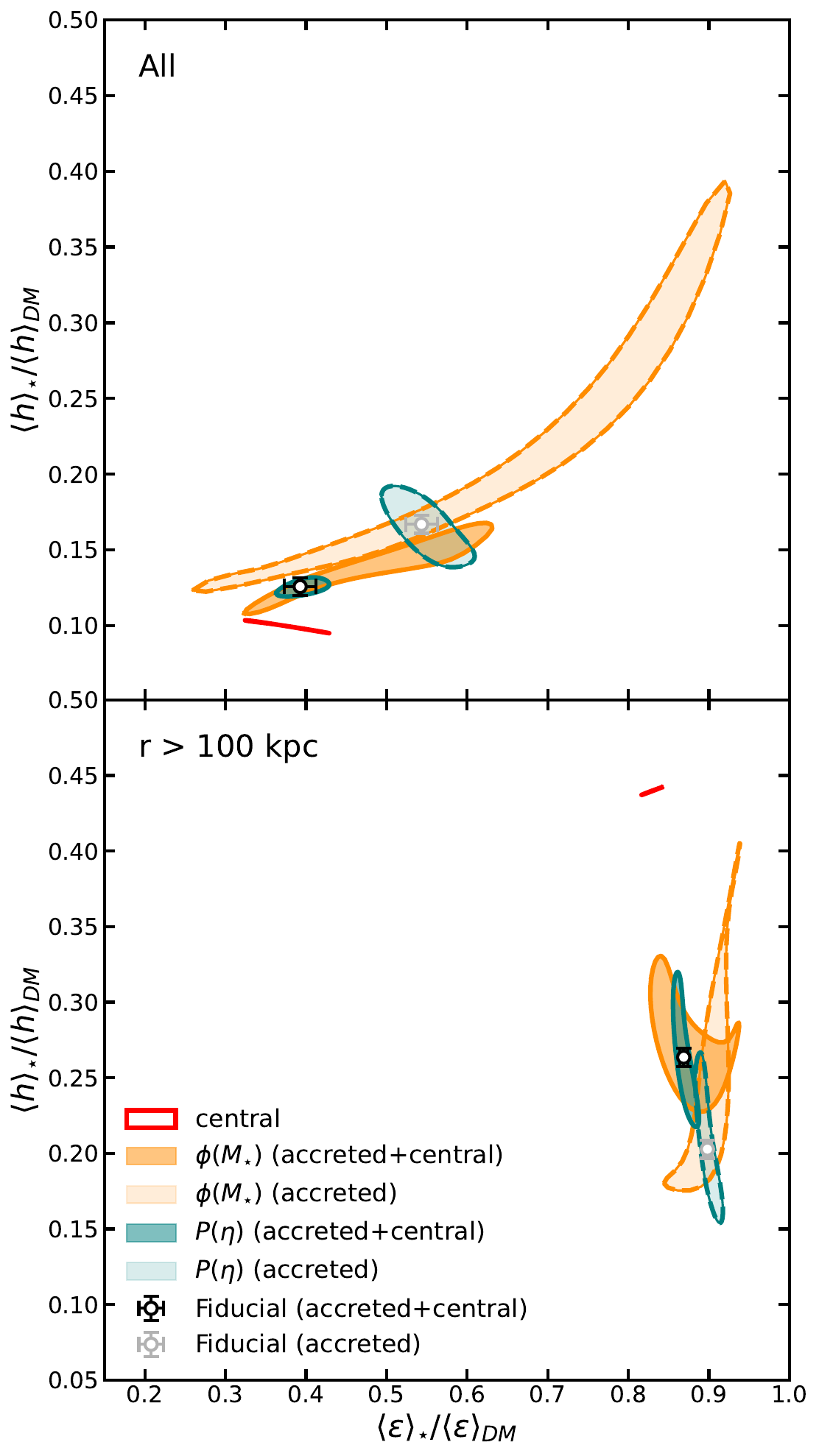}
    \caption{
    Relation between the stellar--to--DM mean energy ratio
    $\langle \varepsilon \rangle_{\star}/\langle \varepsilon \rangle_{\rm DM}$
    and the stellar--to--DM mean specific angular momentum ratio
    $\langle h \rangle_{\star}/\langle h \rangle_{\rm DM}$.
    The upper panel shows results for all particles, and the lower panel shows only particles at $r>100\,\mathrm{kpc}$.
    Filled envelopes denote the $1\sigma$ (16th--84th percentile) credible intervals for the combination of accreted matter and the cluster plus its central galaxy, over the two varied function parameters with $\phi(M_{\star})$ denoted in orange and $p(\eta)$ denoted in teal. The lighter dashed envelopes indicate the ranges for only accreted material.
    The red contour in each panel indicates the central's response to the most significant merger over the range of peak merger mass ratios. Values for our fiducial choice are shown as black points for the combined accreted and central, and grey points for the accreted-only, with error bars indicating the 16th--84th percentile credible interval. Interactive versions of this plot can be found at \href{https://garrethmartin.github.io/interactive-profiles-ICL/index.html\#average-e-am}{garrethmartin.github.io/interactive-profiles-ICL/index.html\#average-e-am}}
    \label{fig:e_am_envelopes}
\end{figure}

Figure~\ref{fig:e_am_envelopes} shows the relation between the stellar--to--DM mean specific energy ratio and the corresponding mean specific angular momentum ratio across the full range of tested infalling satellite populations. The upper panel includes all stripped particles, while the lower panel shows results restricted to particles with final radii $r>100$~kpc.

Three distinct components are shown: stars and DM stripped from accreted satellites, stars and DM associated with the pre-existing central galaxy and cluster halo, and the combination of both. For each choice of infall stellar mass function, this central response corresponds to a merger with the most massive satellite that has at least a 50 per cent probability of occurring given the total expected number of infalling satellites, capturing how the central component responds to the typical largest merger implied by each satellite population. To construct the combined accreted and central ratio, we then assume that the accreted and central components contribute equal amounts of DM mass, corresponding to a scenario in which the cluster has doubled its mass through the accretion of satellites, while allowing the stellar mass contributions of the two components to differ.

As discussed in Section \ref{sec:energy_transfer}, mergers never drive a very significant response from either the cluster halo or central galaxy in our simulations. As a result the variation due to the cluster response is small compared with the variation due to accreted material.

Across the explored parameter space, variations in the characteristic mass scale of the infalling satellites (orange contours) dominate the population-averaged offsets, while changes in orbital circularity (teal contours) play a secondary role. For material at $r>100$~kpc, the dependence on the infalling satellite population is reduced. Circularity has little influence on the energy ratio but induces a larger spread in angular momentum, while variations in the stellar mass function can shift the energy ratio in the range $\sim0.8$-$0.95$. Importantly, no plausible infalling satellite stellar mass function or orbital circularity distribution yields population-averaged ratios consistent with $\langle \varepsilon \rangle_{\star}/\langle \varepsilon \rangle_{\rm DM}=1$ and $\langle h \rangle_{\star}/\langle h \rangle_{\rm DM}=1$. Although Figure \ref{fig:e_am_envelopes} displays only the 1$\sigma$ credible regions, we have verified that this conclusion holds when extending to 3$\sigma$: even at the extremes of the explored parameter space, stripped stars remain systematically more tightly bound and lower in angular momentum than the corresponding stripped DM, and a scenario in which the ICL perfectly traces the cluster DM in phase-space is not realised.

The sensitivity of the orbital energy and angular momentum ratios to the assumed infalling satellite population underscores that robust predictions linking ICL properties to the underlying DM require well-constrained satellite stellar mass functions. Uncertainties in those inputs plausibly map onto a broad range of angular momentum and, particularly, energy offsets. Importantly, changes in the orbital energy and angular momentum offsets are driven primarily by intermediate- and high-mass satellites. Recomputing the ratios while imposing a minimum stellar mass cut shows that these offsets remain stable until a threshold of $\sim10^{10.5}$~M$_{\odot}$ over the full range of characteristic masses adopted. The energy differences are therefore driven almost entirely by massive satellites. We discuss this further in Appendix \ref{sec:energy_ratio_mmin}.

Links to an interactive version of Figure \ref{fig:e_am_envelopes} are given in the figure caption, allowing readers to examine the full parameter space and explore how changing individual stellar mass function and orbital circularity distribution parameters affects the resulting orbital energy and angular momentum ratios.

\subsection{Radial density profiles}
\label{sec:density_profiles}

In an NFW-like potential, the time-averaged radius is jointly determined by the specific energy and angular momentum of the orbit. Once $\varepsilon$ and $h$ are fixed, the range of radii explored by the orbit and the relative amount of time spent at each radius are fixed as well. The radial density profiles of the DM and ICL are therefore established by their respective phase-space distributions. Orbital energy primarily sets the apocentric scale, where the radial motion is slowest and the orbit contributes most strongly to the density, while differences in angular momentum control the orbit's inward extent. The time-averaged radius $\langle r \rangle$ is thus set by the combined effect of energy fixing the outer scale of the orbit and angular momentum determining the depth of the orbit.

Most infall orbits are far from circular (on average, $\eta \sim 0.5$: \citealt{Wetzel2011}) and clusters typically have low concentrations and hence large scale radii $r_{s}$, extending to hundreds of kpc. As a result, much of the relevant orbital evolution occurs at radii  $r\leq r_{s}$, where the potential deviates strongly from the Keplerian limit (the regime in which the enclosed mass is effectively constant and the gravitational potential approaches $\phi \propto -1/r$). In this non-Keplerian regime, changes in angular momentum at fixed energy have a significant effect on the pericentric radius: lowering $h$ allows the orbit to reach much smaller radii, thereby reducing $\langle r \rangle$.

\begin{figure}
    \centering
    \includegraphics[width=0.45\textwidth]{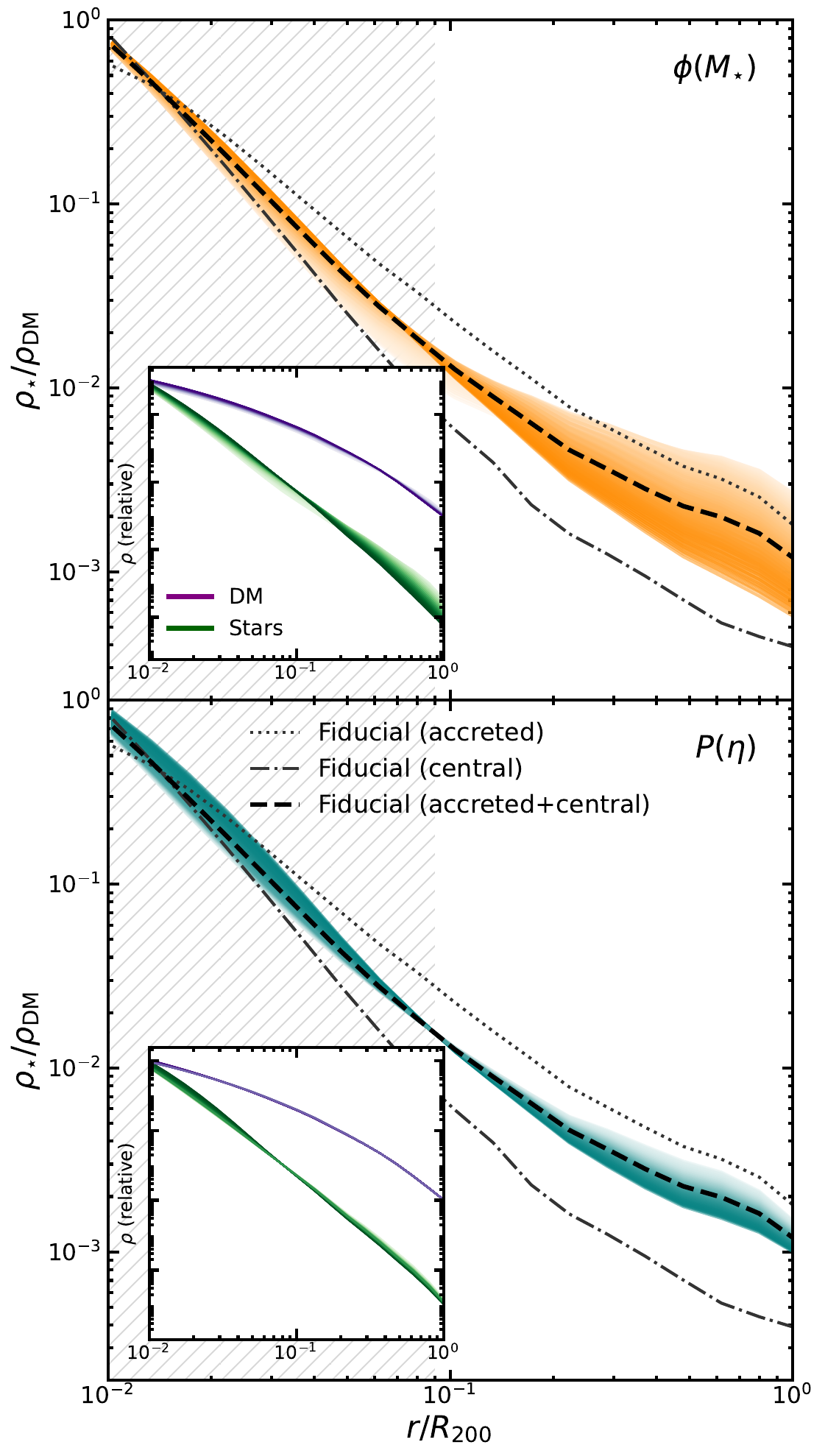}
    \caption{Radial profiles of the stellar--to--DM density ratio $\rho_{\star}/\rho_{\rm DM}$ for the full set of sampled satellite-population parameter combinations. Coloured envelopes show the range of model predictions for variations in $\phi(M_{\star})$ (top) and $p(\eta)$ (bottom). The black dotted and dot-dashed and dashed lines shows separate contributions from infalling satellites, from the cluster halo plus central galaxy, and from the combined profile respectively. The hatched region indicates radii $<100$~kpc. Insets show the underlying DM (purple) and stellar (green) radial density profiles normalised to the maximum of the DM profile to highlight relative shape differences. Interactive versions of this plot can be found at \href{https://garrethmartin.github.io/interactive-profiles-ICL/index.html\#radial}{garrethmartin.github.io/interactive-profiles-ICL/index.html\#radial}.}
    \label{fig:density_profile}
\end{figure}

In Figure \ref{fig:density_profile}, we present radial profiles of the stellar--to--DM density ratio, $\rho_{\star}/\rho_{\rm DM}$, for the set of sampled infalling satellite population parameter combinations. Each panel presents an envelope of modelled profiles rendered with a graded colour to indicate the family of parameter combinations. The black dotted and dot-dashed curves denote the fiducial contribution from infalling satellites and the cluster halo and central galaxy respectively, with the dashed curve denoting the radial profile combined in the same way as the orbital energy and angular momentum in Section \ref{sec:e_am_avs}. 

Further from the centre of the cluster, the contribution of the central galaxy declines significantly. As a result, the fiducial combined profile (dashed line) increasingly diverges from the cluster halo plus central galaxy profile (dot-dashed), and instead approaches the profile of the accreted material (dotted). Beyond $\sim100$~kpc, the stellar--to--DM density ratio is therefore primarily governed by the distribution of stripped satellite debris, with the central contribution becoming subdominant.

The stellar--to--DM density ratio shows a pronounced dependence on the properties of the infalling satellite population. In particular, variations in the satellite stellar mass function lead to substantial changes in the ratio. To quantify this dependence, we measure the logarithmic slope\footnote{Slopes measured for individual parameter combinations can be found in the interactive versions of Figure \ref{fig:density_profile}.} of $\rho_{\star}/\rho_{\rm DM}$ separately in the inner ($r < 100$~kpc) and outer ($r > 100$~kpc) regions across the explored parameter space. Varying the satellite stellar mass function produces inner slopes spanning $-2.04$ to $-1.78$ (16th--84th percentile range: $-1.88$ to $-1.81$) and outer slopes spanning $-1.28$ to $-0.54$ (16th--84th percentile: $-1.14$ to $-0.77$). The corresponding ranges when varying the orbital circularity distribution are considerably narrower: $-1.91$ to $-1.74$ for the inner slope (16th--84th percentile: $-1.85$ to $-1.76$) and $-1.07$ to $-0.97$ for the outer slope (16th--84th percentile: $-1.05$ to $-1.00$). The outer profile is thus substantially more sensitive to the characteristic mass of the infalling satellite population, while both regions show only weak dependence on the circularity distribution. In all cases, differences are driven almost exclusively by changes in the stellar radial density profile; the DM profile shown in purple in the inset panels shows very little variation across the explored parameter space.

The stellar--to--DM ratio is most sensitive to the characteristic mass of the satellite stellar mass function. Varying this parameter produces substantial shifts in the radial logarithmic slope, whereas plausible changes in the faint-end slope lead to comparatively modest effects. This behaviour is consistent with the phase-space trends identified in Section \ref{sec:phase_space_evolution}: more massive satellites undergo substantially greater losses of orbital energy and angular momentum. Increasing the characteristic mass therefore lowers the mean stellar energies and angular momenta of the stripped material, yielding a more centrally concentrated stellar distribution and hence a steeper stellar--to--DM ratio profile.

\section{Comparison with cosmological simulations}
\label{sec:cosmo_comparison}

The results presented in Section \ref{sec:population_offsets} isolate the effects of satellite mass ratio and orbital configuration on the stellar--DM phase-space offsets. However, they make several simplifying assumptions: a static, spherically symmetric cluster potential, idealised satellite structure with fixed stellar--to--halo mass relations, and satellites injected on individual orbits into an isolated cluster. Real clusters experience more complex assembly histories, with multiple satellites accreting simultaneously, ongoing major mergers that violate the adiabatic approximation, triaxial halo shapes, and correlations between satellite properties and their host environment. We therefore compare our model predictions to phase-space and radial profile measurements from cosmological simulations in order to assess whether the magnitude and scaling of their stellar--DM offsets are consistent with those expected from varying the infalling satellite population alone.

Any systematic differences between our model predictions and the cosmological simulations would therefore point to additional effects present due to the inclusion of this proper cosmological context or detailed baryonic physics. These could include: variations in halo assembly history and the timing of accretion events; baryonic back-reaction that modifies the inner density profile of the cluster; correlations between infalling satellite populations and host properties (e.g., more massive clusters accreting systematically different infalling satellite populations); the pre-processing of satellites in group environments before cluster infall; or satellite-satellite interactions within the cluster, which our model neglects \citep[e.g.][]{Knebe2006}. Differences are also expected owing to the coarser mass and force resolution of these simulations relative to our controlled simulations. At the resolutions of current large-volume cluster simulations, tidal stripping of stellar material is not perfectly converged \citep{Martin2024, Lovell2025}, which can lead to artificially enhanced stripping and potentially affect the detailed phase-space properties of the ICL.

\subsection{Cosmological simulations}

Appendix \ref{sec:cosmo_sims} describes the four independent cosmological hydrodynamical simulation suites used in this comparison (\textsc{Horizon-AGN}, \textsc{Hydrangea}, \textsc{TNG100}, and \textsc{TheThreeHundred GIZMO 7K}), their physical models, resolution, and cluster sample selection.

We measure orbital energies, angular momenta, and radial density profiles for the stellar and DM components of clusters within $R_{200}$. To isolate the diffuse intracluster component, we remove particles assigned to identified substructures following the masking procedure of \citet{Butler2025}. We do not attempt to separate the central galaxy from the diffuse ICL component, treating them as a single combined stellar population. DM particles associated with subhaloes are excluded, and star particles within four times the stellar half-mass radius $R_{1/2}$ of any subhalo are excluded. For low-mass subhaloes with $M_{\star}<10^{11}\,\mathrm{M}{\odot}$, the exclusion radius is fixed to the 84th-percentile value of each cluster’s $R_{1/2}$ distribution measured at $M_{\star}=10^{11}\,\mathrm{M}_{\odot}$ \footnote{Note that our measurements differ slightly from \citet{Butler2025}, who use the geometric mean of the projected half-mass radius in the $x$, $y$, and $z$ projections. For consistency across all simulations, we instead adopt the three-dimensional half-mass radius.}. Subhalo identification is performed using each simulation suite's native structure finder. For stellar particles, the masking procedure relies only on subhalo centroids and exclusion radii, making it robust to differences in structure finder across simulations. Given the low DM substructure mass fractions typical of clusters \citep{Giocoli2010}, the method is similarly insensitive to variations in how DM subhaloes are defined.

Specific orbital energies and angular momenta of the un-masked particles are computed in the same way as described in Section \ref{sec:setup}. We calculate the mean specific energy ratios $\langle \varepsilon \rangle_{\star}/\langle \varepsilon \rangle_{\rm DM}$ and mean specific angular momentum ratios $\langle h \rangle_{\star}/\langle h \rangle_{\rm DM}$ for the diffuse stellar and DM components of each cluster, allowing direct comparison with the population-averaged predictions from our controlled merger simulations.

\subsection{Comparison of predicted phase-space distributions}

\begin{figure}
    \centering
    \includegraphics[width=0.45\textwidth]{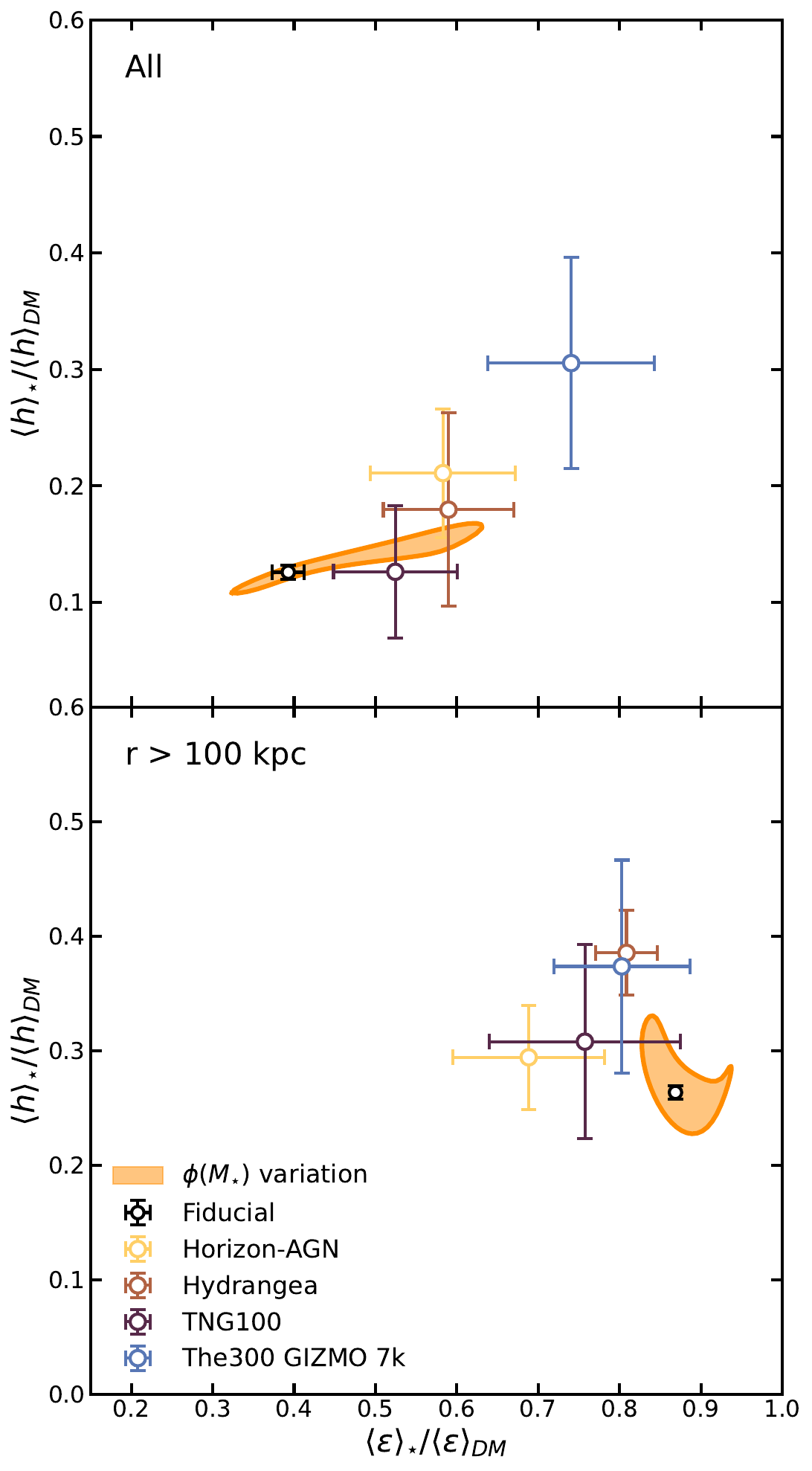}
    \caption{Comparison between the stellar--to--DM mean specific orbital energy ratio, $\langle \varepsilon \rangle_{\star}/\langle \varepsilon \rangle_{\rm DM}$, and the stellar--to--DM mean specific angular momentum ratio, $\langle h \rangle_{\star}/\langle h \rangle_{\rm DM}$, measured in the four cosmological simulations and the model envelopes shown in Figure~\ref{fig:e_am_envelopes}. The upper panel includes all particles within $R_{200}$, while the lower panel restricts the calculation to particles at radii $r>100,\mathrm{kpc}$. The filled region corresponds to the $1\sigma$ (16th--84th percentile) credible interval of the combined accreted+central model obtained by varying the satellite stellar mass function, $\phi(M_{\star})$, while holding the orbital circularity distribution, $p(\eta)$, fixed at its fiducial form. The black point and associated error bars indicate the fiducial model prediction and its credible range. Coloured points show the median ratios measured across cluster samples drawn from the \textsc{Horizon-AGN}, \textsc{Hydrangea}, \textsc{TNG100}, and \textsc{TheThreeHundred} simulation suites, as indicated in the legend. Error bars indicate the $1\sigma$ cluster--to--cluster scatter within each simulation. Stellar and DM orbital energies and angular momenta are computed following \citet{Butler2025}.}
    \label{fig:sim_comparison_energy_am}
\end{figure}

Figure \ref{fig:sim_comparison_energy_am} shows as coloured points the stellar--to--DM mean specific orbital energy ratio $\langle \varepsilon \rangle_{\star}/\langle \varepsilon \rangle_{\rm DM}$  and the mean specific angular momentum ratio $\langle h \rangle_{\star}/\langle h \rangle_{\rm DM}$ measured in cosmological simulations. The orange shaded envelopes show the same combined accreted plus central stellar and DM distributions obtained by varying the satellite stellar mass function, $\phi(M_{\star})$, as shown in Figure~\ref{fig:e_am_envelopes}.

For all particles (top panel), the simulations lie within or close to the envelope of model predictions. The energy and angular momentum ratios are slightly higher than the fiducial model predicts, but for all simulations, with the exception of \textsc{TheThreeHundred}, the $1\sigma$ scatter overlaps the range spanned by plausible variations in the infalling satellite stellar mass function. \textsc{TheThreeHundred} shows the smallest stellar--DM orbital energy and angular momentum offsets, consistent with enhanced or premature stripping of stellar material in these lower-resolution simulations. This may also be reflective of the systematically higher cluster masses of the \textsc{TheThreeHundred} sample, in which an increased fraction of the ICL mass budget is expected to originate from stars pre-processed in group haloes \citep{Mihos2017, Contini2021} and therefore already weakly bound upon cluster infall.

For particles at $r>100$~kpc (bottom panel), a larger average orbital energy offset than our model prediction is measured in all simulation suites. However, the magnitude of this offset is likely to be quite dependent on the size and mass of the central galaxy, as this influences the radius out to which the central galaxy dominates over the accreted component. We apply corrections to account for these differences in Section \ref{sec:radial_density_comparison}. However, overall, the cosmological simulations are broadly consistent with the phase-space offsets predicted by our simplified model across the explored range of infalling satellite populations. This suggests that our key simplifying assumptions capture the dominant physical processes governing stellar--DM phase-space segregation in galaxy clusters.

\subsection{Comparison of radial density profiles}

Having shown that the stellar--DM phase-space offsets measured in cosmological simulations already lie close to the range predicted by the model, we now take an additional step by conditioning the model on the infalling satellite populations realised in each simulation in order to make quantitative predictions for the radial density profiles.

\subsubsection{Conditioning the model on cosmological simulation infalling satellite populations}
\label{sec:model_calibration}

The comparison between our model and the cosmological simulations is performed by re-weighting the predicted radial profiles introduced in Section~\ref{sec:density_profiles}. These profiles already encode the response of stripped stars and DM to variations in the infalling satellite population; here we match the accreted components to the average infalling satellite stellar mass function values found for each simulation suite in Appendix \ref{app:schechter_fits} and further calibrate the normalisation of the central component to account for differences in the average star-formation efficiency and central galaxy mass measured for each simulation suite.

Specifically, for each simulation suite we construct model profiles corresponding to the fiducial orbital circularity distribution with an infall satellite stellar mass function derived from the average best-fit Schechter function values from Table \ref{tab:mf_all}. These are $\hat{\rho}_{\star,\mathrm{cen}}(r)$, $\hat{\rho}_{\star,\mathrm{acc}}(r)$, $\hat{\rho}_{\mathrm{DM},\mathrm{cen}}(r)$ and $\hat{\rho}_{\mathrm{DM},\mathrm{acc}}(r)$, which represent the radial density profiles predicted by our model for the central and accreted components of the cluster, for both the stars and DM.

The DM components $M_{\mathrm{DM},\mathrm{acc}}^{\mathrm{sim}}$ and $M_{\mathrm{DM},\mathrm{cen}}^{\mathrm{sim}}$ are combined assuming equal-mass central and accreted contributions, consistent with our previous treatment, in which we assume that the cluster has doubled its mass through the accretion of satellites between $z_{\rm form}$ and $z=0$. The stellar components are re-weighted separately: the central stellar component is normalised to the mean stellar mass of the central galaxy $M_{\star,\mathrm{cen}}^{\mathrm{sim}}$, while the accreted component is normalised to the difference between the mean total stellar mass within $R_{200}$ and $M_{\star,\mathrm{cen}}^{\mathrm{sim}}$, $M_{\star,\mathrm{acc}}^{\mathrm{sim}}$. This ensures that the relative contributions of the central and accreted stellar components reflect the stellar mass budgets realised in each simulation, while preserving the radial shapes predicted by the model for each component.

The resulting stellar--to--DM ratio profile is therefore given by
\begin{equation}
\frac{\rho_\star(r)}{\rho_{\mathrm{DM}}(r)} =
\frac{
M_{\star,\mathrm{cen}}^{\mathrm{sim}}\,
\hat{\rho}_{\star,\mathrm{cen}}(r)
+
M_{\star,\mathrm{acc}}^{\mathrm{sim}}\,
\hat{\rho}_{\star,\mathrm{acc}}(r)
}{
M_{\mathrm{DM},\mathrm{cen}}^{\mathrm{sim}}\,
\hat{\rho}_{\mathrm{DM},\mathrm{cen}}(r)
+
M_{\mathrm{DM},\mathrm{acc}}^{\mathrm{sim}}\,
\hat{\rho}_{\mathrm{DM},\mathrm{acc}}(r)
},
\end{equation}
where the DM components satisfy $M_{\mathrm{DM},\mathrm{cen}} = M_{\mathrm{DM},\mathrm{acc}}$.

A further correction is applied to account for the fraction of stellar and DM mass associated with satellites and excluded by the masking procedure of \citet{Butler2025}. For each simulation, we measure the fraction of stars and DM within $R_{200}$ that are excluded from the calculation and re-weight the corresponding central and accreted components accordingly.

\subsubsection{Radial density profiles}
\label{sec:radial_density_comparison}

\begin{figure*}
    \centering
    \includegraphics[width=0.95\textwidth]{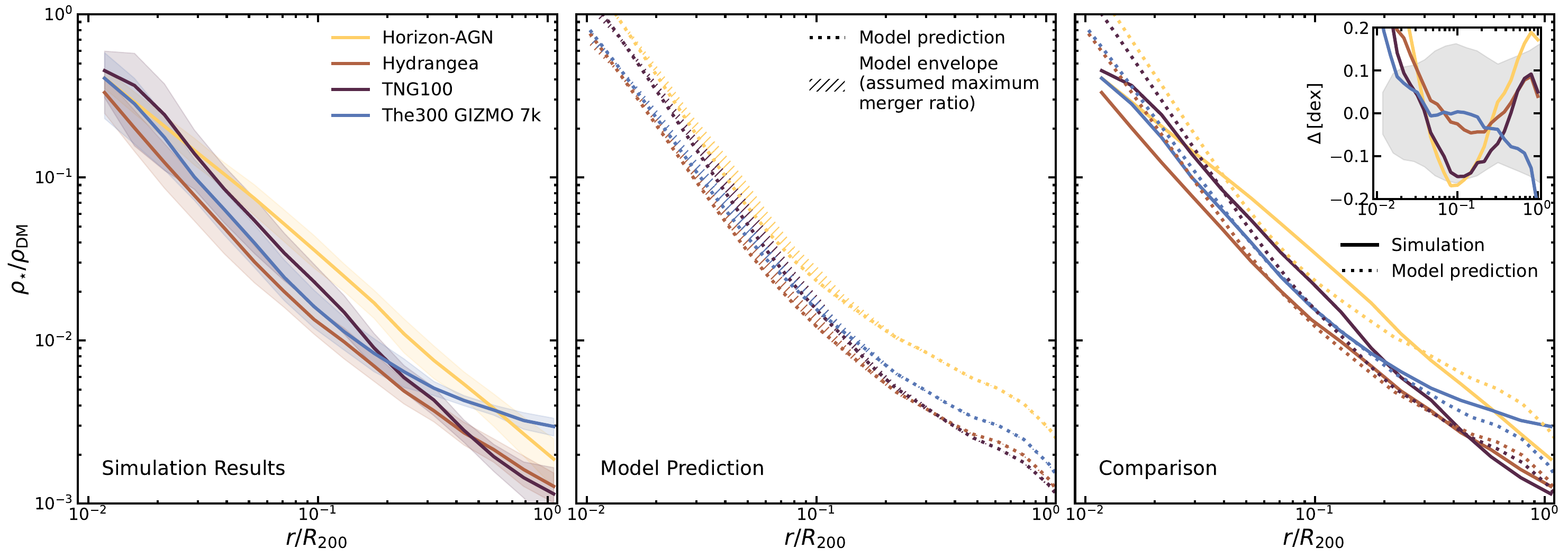}
    \caption{Radial profiles of the stellar--to--DM density ratio, $\rho_\star/\rho_{\mathrm{DM}}$, shown as a function of cluster-centric radius normalised by $R_{200}$, comparing cosmological simulations and the model predictions.
\textbf{Left panel:} Mean $\rho_\star/\rho_{\mathrm{DM}}$ profiles measured directly from the simulations (\textsc{Horizon-AGN}, \textsc{Hydrangea}, \textsc{TNG100}, and \textsc{TheThreeHundred}), with shaded regions indicating the cluster--to--cluster scatter ($\pm1\sigma$).
\textbf{Middle panel:} Model predictions for the same quantity. Hatched regions show the envelope of model central galaxy responses obtained by varying the assumed maximum satellite--to--host merger mass ratio, and dashed lines show the fiducial case corresponding to a maximum infalling satellite stellar mass of $10^{11.5}\,\mathrm{M}_\odot$.
\textbf{Right panel:} Direct comparison between simulations (solid lines) and the fiducial model predictions (dashed lines). The inset shows the logarithmic residual, $\Delta = \log_{10}\!\left[(\rho_\star/\rho_{\mathrm{DM}})_{\mathrm{model}} / (\rho_\star/\rho_{\mathrm{DM}})_{\mathrm{sim}}\right]$, as a function of radius for each simulation along with a filled grey region indicating the $1\sigma$ inter-simulation scatter measured between the four simulations.}
    \label{fig:sim_comparison_profiles}
\end{figure*}

Figure~\ref{fig:sim_comparison_profiles} compares the resulting radial profiles of the stellar--to--DM density ratio, $\rho_{\star}/\rho_{\rm DM}$, measured in the cosmological simulations, with the corresponding model predictions from Section \ref{sec:density_profiles}, after applying the normalization and masking corrections described in Section \ref{sec:model_calibration} above.

The average stellar--to--DM density ratio profiles measured for each simulation are shown in the first panel of Figure~\ref{fig:sim_comparison_profiles}, with shaded regions indicating the $1\sigma$ cluster--to--cluster scatter. The middle panel shows the corresponding model predictions from Section \ref{sec:population_offsets}, after re-weighting to match the infalling satellite populations and stellar mass budgets of each simulation suite. The hatched regions indicate the envelope of model responses obtained by varying the maximum satellite--to--host merger mass ratio, while the dashed lines show the fiducial model corresponding to a maximum infalling satellite stellar mass of $M_{\star,\max}=10^{11.5}\,\mathrm{M}_\odot$. Increasing the contribution of more massive satellites only slightly steepens the inner profile.

The right-hand panel compares the simulation profiles directly with the
fiducial model predictions and the inset shows the logarithmic residual between the each simulation and model. For all simulation suites, the model reproduces the overall radial trend of declining $\rho_\star/\rho_{\mathrm{DM}}$ with radius, though the model profiles exhibit somewhat greater curvature than seen in some simulations. Some simulations (\textsc{TheThreeHundred} and \textsc{Hydrangea}) show departures from a simple power law that are well captured by the model, while others (\textsc{Horizon-AGN} and \textsc{TNG100}) follow a more linear relation that the model does not fully reproduce.
Comparing the stellar and DM density profiles separately, we find that discrepancies in shape are driven primarily by the stellar component, reflecting its greater sensitivity to baryonic physics.

There is substantial scatter in the stellar--to--DM density ratio profiles across the different simulation suites, with \textsc{Horizon-AGN} in particular deviating significantly from the others. Individual simulation studies have shown that the intracluster stellar density profile is steeper and more centrally concentrated than the DM profile, such that the ratio $\rho_{\star}/\rho_{\rm DM}$ follows an approximate power law \citep[e.g.][]{Alonso-Asensio2020,Reina-Campos2023,Contreras-Santos2024,Butler2025,Jeon2025}, with similar trends seen observationally \citep[e.g.][]{Diego2023}. Here we find that, even when measured consistently across the four simulation suites, the normalisation and slope of the stacked stellar--to--DM density ratio differ systematically between simulations. Much of this variation likely reflects differences in the infalling satellite stellar mass function and satellite survival efficiency, both of which depend sensitively on the specific subgrid physics and numerical resolution implemented in each simulation.

Despite the simplifying assumptions of our model, including a static, spherically symmetric halo potential and idealised satellite structure, the predicted stellar--to--DM density ratio profiles reproduce the simulation results to within $<0.2$~dex over 2 orders of magnitude in radius, once differences in star-formation efficiency and central galaxy structure are accounted for. At most radii, this level of agreement is notably better than the intrinsic scatter between different simulation suites indicated as a grey filled region in the inset plot of Figure \ref{fig:sim_comparison_profiles}. This suggests that the properties of the infalling satellite population alone are sufficient to capture the dominant features of the ICL spatial distribution. Residual discrepancies may arise from several sources not accounted for in our model, including differences in inner halo structure due to baryonic back-reaction and force softening, variations in the stellar--halo mass relation, resolution-dependent satellite survival times, and the non-adiabatic evolution of cosmological haloes during mergers and rapid accretion.

Although varying the circularity distribution did not significantly affect the predicted profiles for the fiducial mass function (Figure~\ref{fig:density_profile}), some studies suggest more massive satellites may follow preferentially more radial orbits \citep{Jiang2015}, and the sensitivity to the circularity distribution may differ for mass functions with different characteristic masses. The simulations also span a range of cluster masses not explored by our fixed host halo, which may affect the typical satellite-to-host mass ratios and therefore the relative importance of dynamical friction.

\section{Discussion} 
\label{sec:discussion}

\subsection{Physical mechanisms driving the stellar--DM phase-space separation}

In this work we have sought to isolate the physical origin of the systematic phase-space and radial offsets observed between ICL and DM. We have shown that many of the observed features of the stellar--to--DM phase-space relationship arise naturally from the mass distribution and orbital evolution of infalling satellites, without invoking additional physics.

A central result is that the separation between the phase-space distributions of stripped stars and stripped DM is driven primarily by satellites with the most similar satellite--to--host mass ratios. This follows from the combination of two related effects. First, the extended DM haloes of satellites are stripped early, while the more tightly bound stellar components survive to later stages of the orbit and are deposited only after further orbital decay. Second, dynamical friction operates most efficiently on massive satellites, causing rapid loss of orbital energy and angular momentum prior to significant stellar stripping. As a result, stars are preferentially stripped from satellites that have already migrated onto more bound orbits, whereas the bulk of the DM is removed earlier at higher energies.

The most extreme case of this process corresponds to major mergers, which are expected to dominate the direct build-up of the central galaxy while also contributing a centrally concentrated component to the ICL \citep[e.g.][]{Contini2018}. For near equal mass ratios, dynamical friction and torques operate on short timescales, leading to rapid orbital decay prior to stellar stripping. Crucially, these same massive progenitors are also responsible for the largest stellar--to--DM phase-space offsets. The dynamical imprint of these events is therefore established over relatively short timescales, and is shared between the central galaxy and the most tightly bound component of the ICL.

The analysis presented in Section~\ref{sec:energy_transfer} places important context on this interpretation. Even for near equal-mass mergers, where the interaction is most violent and the orbital decay is most rapid, the net transfer of orbital energy into the cluster halo itself remains small. This indicates that the large stellar--to--DM phase-space offsets associated with major mergers do not arise from a strong global restructuring or heating of the cluster potential. Instead, they primarily reflect differences in how stars and DM are deposited in phase-space during the inspiral and stripping process, with stellar material retaining the imprint of deeper orbital decay despite minimal modification of the host halo as a whole.

\subsubsection{The role of orbital angular momentum}

During infall, the modification of both the orbital energy and angular momentum of the satellite play a role in determining where stripped stars and DM are deposited. While orbital energy primarily sets the outer scale of an orbit, angular momentum also plays a crucial role in determining the radial distribution of stripped material within the cluster. As discussed in Section~\ref{sec:density_profiles}, reducing angular momentum at fixed energy allows orbits to reach significantly smaller pericentric radii in the non-Keplerian regime characteristic of cluster potentials, thereby reducing the time-averaged radius and contributing to the radial concentration of the ICL.

This effect is particularly important because, as seen in Figure \ref{fig:energy_am_ratio_vs_mratio}, angular momentum is more readily modified than orbital energy. Because angular momentum is a vector quantity, even modest deflections of particle trajectories can produce substantial changes in $\mathbf{h} = \mathbf{r} \times \mathbf{v}$, while energy changes depend only on the component of force parallel to the satellite's velocity. The bulk of stellar mass is removed near pericentre, where tidal forces are the strongest, leading to substantial reductions in angular momentum with comparatively modest changes in orbital energy. The cluster halo's dynamical response and dynamical friction further drive coupled losses of energy and angular momentum \citep{Chandrasekhar1943, Taffoni2003}, which naturally leads to substantially larger evolution and dispersion in angular momentum ratios than in energy ratios. The combined effect of both energy and angular momentum loss is therefore important for producing the observed radial concentration of the ICL relative to the DM.

\subsection{The dominance of massive satellites and stochastic accretion}

An important implication of our results, given previous findings that the ICL is likely dominated by a relatively small number of massive progenitors \citep[e.g.][]{Murante2007,Montes2018, DeMaio2018, Montes2021,Brown2024,Contreras-Santos2024}, is that the characteristic mass scale of the infalling satellite stellar mass function emerges as the dominant parameter controlling the population-averaged offsets between the ICL and DM. Variations in the satellite mass function produce large changes in the predicted stellar--to--DM energy and angular momentum ratios (Figure \ref{fig:e_am_envelopes}), while variations in the orbital satellite infall circularity distribution have a comparatively weak effect.

This sensitivity arises because higher satellite masses lead to greater losses in both orbital energy and angular momentum prior to stellar stripping. As shown in Figure~\ref{fig:energy_am_ratio_vs_mratio}, the stellar--to--DM energy and angular momentum ratios decrease systematically with increasing satellite--to--host mass ratio. These enhanced phase-space offsets directly translate into steeper radial density profiles: stars stripped from more massive satellites are deposited at smaller radii and higher concentrations relative to the DM, producing a steeper logarithmic slope in the $\rho_\star/\rho_{\rm DM}$ profile. Consequently, infalling satellite populations with higher characteristic masses naturally produce ICL distributions that are more centrally concentrated relative to the underlying DM halo.

Importantly, the uncertainty is driven primarily by the abundance of intermediate- and high-mass satellites, which consistently dominate the ICL mass budget and the population-averaged phase-space offsets regardless of our chosen characteristic mass scale. As shown in Appendix~\ref{sec:energy_ratio_mmin} and Figure~\ref{fig:energy_ratio_vs_mmin}, the stellar--DM specific orbital energy ratios are largely insensitive to the detailed treatment of the faint end of the satellite mass function. Across the full range of characteristic masses explored here, satellites below $M_\star \sim 10^{10}$--$10^{10.5}\,\mathrm{M}_\odot$ have negligible impact on the resulting ratios.

The phase-space structure of the ICL is therefore inherently sensitive to stochastic accretion events, reflecting the dynamical imprint of a few satellites that undergo substantial orbital decay before disruption. In this sense, the ICL should be viewed both as a tracer of the cluster potential, and also as an accumulated record of a limited number of dynamically important mergers. This stochastic nature has important observational implications. Measurements of the stellar--to--DM phase-space offsets in individual clusters will be sensitive to their specific accretion histories, and in particular to whether they have recently hosted the massive satellite mergers that dominate both the ICL budget and the magnitude of the phase-space separation. Cluster dynamical state may provide some indication of recent major merger activity, though the connection between current relaxation state and integrated accretion history is not straightforward \citep[e.g.][]{Ragusa2023, Contini2024, Contreras-Santos2024, Kimmig2025, Golden-Marx2025, Canepa2025}.

\subsection{Model validation and limitations}

While our controlled simulations allow us to isolate the effects of satellite mass ratio and orbital configuration on the stellar--DM phase-space offsets, they make several simplifying assumptions that warrant discussion.

\subsubsection{Fixed satellite structural properties}

Our model adopts fixed satellite structural properties for a given stellar mass. \citet{Martin2024} shows that the efficiency of stellar stripping varies significantly depending on the extent of the satellite's stellar component relative to its halo's scale radius, while Figure~\ref{fig:satellite_props} illustrates that halo concentrations evolve systematically between $z=1$ and $z=0$, implying that satellites stars span a range of binding energies across cosmic time. A fixed prescription therefore cannot capture this evolution, nor the associated variation in stellar stripping efficiency. In more realistic models, such as the four cosmological simulations used in this study, the specific implementation of different physical processes such as feedback, ISM and star-formation physics alters the distribution and structure of stars and therefore the efficiency of stellar stripping \citep{Watkins2025,Martin2025}. These processes can also drive differential evolution of the stellar and DM components, modifying their relative distributions directly \citep[e.g.][]{Pontzen2012,Teyssier2013,Schaller2015,Jackson2025}. Consequently, variations in these physical prescriptions should modify the resulting stellar--to--DM orbital energy and angular momentum ratios.

\subsubsection{The adiabatic assumption}

Additionally, our analysis of the orbital energies and angular momenta of stars and DM neglects the growth of the cluster potential. We assume that the cluster potential evolves adiabatically and that $\langle \varepsilon \rangle_\star / \langle \varepsilon \rangle_{\rm DM}$ and $\langle h \rangle_\star / \langle h \rangle_{\rm DM}$ would therefore remain effectively constant as the cluster evolves.

Although increasingly rare for cluster-mass haloes at late times \citep{Fakhouri2010}, haloes are expected to undergo diverse evolutionary paths \citep{Onions2025} including periods of growth under non-adiabatic conditions \citep{Tormen1997,Valluri2007,Ogiya2021}, such as during major mergers, in which radial actions can deviate from invariance \citep{LyndenBell1967,Pontzen2012}. This is particularly relevant because the largest deviations in the stellar--to--DM energy and angular momentum ratios in our model are driven by the closest satellite--to--host mass ratios, the same events that most strongly violate adiabatic growth.

However, previous work by \citet{Butler2025} has shown that, in realistically evolving cluster potentials which include non-adiabatic growth, the stellar--to--DM energy ratio of tidally stripped stars and DM varies only modestly with accretion time. Therefore, while our model does not entirely account for temporal evolution of the host potential, the adiabatic assumption provides a robust baseline estimate for the expected ratios of energy and angular momentum between stars and DM. We defer a comprehensive exploration of this assumption within a realistic cosmological context to future work (Butler et al., in prep.).

\subsubsection{Validation against cosmological simulations}

To assess whether these simplifying assumptions fundamentally limit the applicability of our results, we compared our model predictions against four independent cosmological hydrodynamical simulation suites (\textsc{Horizon-AGN}, \textsc{Hydrangea}, \textsc{TNG100}, and \textsc{TheThreeHundred GIZMO 7K}; see Section~\ref{sec:cosmo_comparison}). These simulations provide a critical test of our results because they incorporate the full complexity absent from our controlled setup including non-adiabatic evolution, realistic cluster assembly histories, a smoothly accreted DM component \citep[e.g.][]{Genel2010}, and diverse physical assumptions that produce a wide range satellite properties.

Despite these complexities and the simplifying assumptions of our model, once differences in star-formation efficiency and central galaxy mass are accounted for, we are able to reproduce the radial stellar--to--DM density ratio profiles seen in a diverse set of cosmological simulations to better than the inter-simulation scatter. While recent work has demonstrated scaling relations between ICL and DM surface densities in cosmological simulations, our results suggest that the scatter in such relations, both within and between simulation suites, is driven primarily by variations in the infalling satellite stellar mass function. This level of agreement is not trivial, given the wide range of numerical methods, resolutions, and subgrid physics employed by the four simulation suites considered. It suggests that, within the cluster mass range tested ($M_{\rm h,200} < 10^{15}\,$M$_{\odot}$), the dominant factor shaping the relationship between the ICL and DM remains the infalling satellite population, with only a secondary dependence on the detailed implementation of baryonic physics, the inclusion of a fully cosmological context, or variations in host halo assembly history and dynamical state.

\subsection{Implications for using ICL as a DM tracer}

Observational constraints on the ICL--DM radial profile relationship remain limited and somewhat contradictory. \citet{Diego2023, Diego2024} found that ICL is more centrally concentrated than DM in SMACS0723, consistent with our predictions. In contrast, \citet{Zhang2019} found that ICL surface brightness profiles appear to trace the cluster radial mass distribution well in 528 DES clusters, suggesting a flatter $\rho_\star/\rho_{\rm DM}$ profile. These discrepancies may reflect differences in cluster selection, ICL measurement techniques, or intrinsic cluster--to--cluster scatter. Combining deep ICL imaging with weak lensing measurements from Euclid and LSST will soon enable systematic tests of the ICL--DM relationship across large cluster samples. Crucially, ICL offers the potential to trace mass distributions of individual clusters at high redshift \citep[e.g.][]{Bellhouse2025} and with low background source densities \citep[e.g.,][]{Hoekstra2013} where weak lensing constraints become challenging to obtain, providing a powerful independent and complementary probe of cluster mass distributions.

Our findings place clear constraints on how the ICL can be used as a tracer of cluster DM haloes in the context of such observations. Any attempt to infer the DM distribution from the spatial or kinematic structure of the ICL must account for the substantial sensitivity of the stellar--to--DM phase-space offsets to the underlying infalling satellite population. Variations in the satellite stellar mass function represent the dominant source of uncertainty, producing a wide range of possible specific orbital energy and angular momentum ratios and correspondingly large variation in the predicted radial gradients of $\rho_\star/\rho_{\rm DM}$. The comparatively weak dependence on the orbital circularity distribution indicates that constraining the infalling satellite mass function is the primary requirement for reducing uncertainty in the inferred DM density profile. Within the range of plausible infalling satellite populations, the ICL remains sufficiently uncertain to preclude a precise mapping onto the underlying DM distribution.

Importantly, the uncertainty is driven primarily by the abundance of intermediate- and high-mass satellites, rather than by the low-mass population. A relatively small number of these systems contribute a disproportionate fraction of the stripped stellar mass and dominate the population-averaged phase-space offsets. This localisation of the dominant uncertainty is encouraging, as these massive satellites are observationally accessible and are expected to leave disproportionately strong and long-lived observational signatures. Simulations show that the most massive accretion events dominate visible tidal debris and kinematic substructure, while lower-mass contributions rapidly phase-mix into a smooth background \citep{Johnston2008, Cooper2015}. These systems can generate visible tidal debris, influence the global dynamical state of the cluster \citep[e.g.][]{Kimmig2025}, and produce distinctive kinematic substructure \citep[e.g.][]{Dolag2010}.

In addition, stellar-population gradients in the ICL provide an independent estimate of the characteristic progenitor mass scale, offering a route to constrain the dominant contributors even when individual accretion events cannot be isolated \citep[e.g.][]{Morishita2017, Montes2018, DeMaio2018}. For individual clusters, improved characterisation of the intermediate- and high-mass satellite population therefore offers a promising route toward reducing uncertainty in the predicted stellar--to--DM offsets and placing more robust constraints on the radial $\rho_\star/\rho_{\rm DM}$ profile and hence on cluster-scale DM density profiles.

\section{Conclusions}
\label{sec:conclusions}

In this study we followed the tidal disruption of the stellar and DM components of model satellites evolving within a live cluster halo. Satellite--to--cluster mass ratios and orbital circularities were varied systematically, and we quantified the specific energies and angular momenta of the stripped stellar and DM material measured at the end of the simulated evolution.

These simulations were then used to construct models for the cluster's stellar--to--DM phase-space separation as a function of satellite mass ratio and orbital configuration. We integrated these models over a range of plausible satellite stellar mass functions and orbital distributions drawn from cluster simulations in order to assess how the resulting offsets emerge for realistic infalling populations and how they map onto cluster-scale observables.

Our main conclusions are as follows:

\begin{enumerate}

\item \textit{Stripped stars and DM occupy systematically different regions of phase-space.}
Across all satellite mass ratios and orbital configurations explored, stripped stars are deposited on orbits that are systematically more bound and dynamically colder than those of the stripped DM component, occupying lower specific energies and lower specific angular momenta (Figure \ref{fig:energy_dist}). These offsets arise from the differential removal of stars and DM from satellites, coupled with the subsequent orbital evolution of the satellite within the cluster potential.
\\
\item \textit{Satellite--to--host mass ratio is the primary parameter controlling the magnitude of the stellar--DM phase-space separation.}
More comparable satellite--to--host-mass ratio mergers experience substantial orbital decay prior to the deposition of the bulk of the stellar material, leading to large offsets in both specific energy and angular momentum. In contrast, low-mass satellites undergo little orbital decay and generate stripped stellar and DM debris that remain closely aligned in phase-space (Figure \ref{fig:sat_orbit}).
\\
\item \textit{Orbital circularity has only a secondary influence on the stellar-DM offsets.}
Varying the orbital circularity distribution produces modest changes in the angular momentum offsets and broadens the angular momentum distributions of the stripped material, but has little effect on the energy offsets (Figure \ref{fig:energy_am_ratio_vs_mratio}). As a result, differences in orbital distributions alone do not strongly modify the stellar--DM phase-space separation.
\\
\item \textit{Orbital energy lost by satellites is transferred to the cluster DM halo, but with only a weak global response.}
For the closest satellite--to--host mass ratios explored (1:1 and 1:3), the reduction in the orbital energy of the stellar component is accompanied by a measurable increase in the energy of the cluster DM halo, indicating that orbital energy lost during inspiral is transferred to the host system (Figure \ref{fig:energy_stackplot}). At lower mass ratios, the stellar contribution to the total energy budget becomes too small for this redistribution to be isolated above numerical uncertainties. The fact that the cluster halo response remains very small even in the most extreme mergers demonstrates that the resulting stellar--DM phase-space offsets are not driven by a strong global response of the cluster potential, at least in our simplified model.
\\
\item \textit{The stellar mass scale of the dominant ICL contributors is set by the characteristic mass of the infalling satellite stellar mass function.} When convolved with plausible infalling satellite populations, the stellar mass budget of the ICL is consistently dominated by contributions from satellites near the characteristic mass of the infalling stellar mass function. Variations in the faint-end slope and in the orbital circularity distribution produce only minor changes in the cumulative ICL mass budget, and a dominant contribution from low-mass satellites is not realised for any plausible characteristic mass (Figure \ref{fig:mass_budget}).
\\
\item \textit{Population-averaged stellar--to--DM phase-space offsets are most sensitive to the characteristic mass of the infalling satellite stellar mass function.}
When integrated over realistic infalling satellite populations, the mean stellar--to--DM energy and angular momentum ratios vary primarily in response to changes in the characteristic mass scale of the infalling satellite stellar mass function. Adjusting the low-mass slope produces comparatively little change in the population-averaged offsets, indicating that the phase-space separation is insensitive to the abundance of low-mass satellites. Variations in the orbital distribution mainly broaden the angular momentum distribution and have only a weak effect on the mean offsets, consistent with the limited sensitivity to orbital circularity seen for individual satellites (Figure \ref{fig:e_am_envelopes}).
\\
\item \textit{The corresponding radial stellar--to--DM density profiles also depend primarily on the infalling satellite stellar mass function.}
The radial dependence of the stellar--to--DM density ratio varies significantly across the allowed range of satellite stellar mass functions, with changes driven mainly by the characteristic mass scale. Shifting this mass scale alters typical energies and angular momenta of stripped stars (with a much weaker influence on the DM) producing corresponding changes in the radial concentration and logarithmic slope of the ICL profile, with higher characteristic masses resulting in steeper slopes. In contrast, variations in the orbital distribution induce only weak modifications to these radial trends (Figure \ref{fig:density_profile}). As a result, the mapping between ICL density profiles and the underlying DM distribution is governed primarily by the mass scale of the infalling satellite population rather than by its orbital structure.
\\
\item \textit{Stellar--to--DM energy and angular momentum offsets in cosmological simulations are broadly captured by the model.}
Across all four cosmological simulations considered, the stellar--to--DM specific angular momentum ratios are comparable to those predicted by the model, indicating that the magnitude of the angular momentum offset is well reproduced when a full range of plausible infalling satellite populations are considered (Figure \ref{fig:sim_comparison_energy_am}). The corresponding stellar--to--DM energy ratios in the simulations fall within or close to the envelope of model predictions, though offset on average from the fiducial model.
\\
\item \textit{Systematic differences exist in stacked stellar--to--DM density profiles across simulation suites.}  
Even when measured consistently using the same methodology, the normalisation and slope of the stacked stellar--to--DM density ratio differ systematically between the four cosmological simulation suites. This extends previous studies showing that intracluster stellar profiles are typically steeper and more centrally concentrated than the DM, and demonstrates that differences in the cosmic environments and physical models employed by each simulation imprint measurable variations on the relative radial distribution of stars and DM (Figure \ref{fig:sim_comparison_profiles}, left panel).
\\
\item \textit{The model reproduces radial stellar--to--DM density profiles within the inter-simulation scatter.}  
Despite these systematic differences between simulation suites, once adjustments are made for variations in star-formation efficiency and the mass and size of the central galaxy, and the infalling satellite stellar mass function is matched, our model reproduces the radial dependence of the stellar--to--DM density ratio to within the scatter between simulation suites (Figure \ref{fig:sim_comparison_profiles}, right panel). This indicates that the dominant radial offsets between stars and DM in clusters are largely set by the infalling satellite stellar mass function.
\end{enumerate}

Our results establish that the phase-space and spatial offsets between intracluster stars and DM arise naturally from the demographics of the infalling satellite population. The characteristic mass scale of infalling satellites emerges as the key parameter controlling both the population-averaged energy and angular momentum offsets and the radial profile of the ICL. When the infalling satellite stellar mass function is matched, the resulting stellar--to--DM phase-space properties and radial density profiles agree with those measured in cosmological simulations to within the inter-simulation scatter. These results indicate that ICL encodes a robust record of satellite accretion, and that its spatial and kinematic structure can be interpreted in terms of the underlying DM halo once the properties of the infalling satellite population are properly accounted for.

\section*{Acknowledgements}

G.~M, F.~R.~P, and N.~A.~H acknowledge support from the UK STFC under grant ST/X000982/1. H.~J.~B  acknowledges support from the UK STFC under grant ST/Y509437/1. N.~A.~H and J.~B gratefully acknowledge support from the Leverhulme Trust through a Research Leadership Award. Y.~M.~B acknowledges support from UK Research and Innovation through a Future Leaders Fellowship (grant agreement MR/X035166/1) and financial support from the Swiss National Science Foundation (SNSF) under project ``Galaxy evolution in the cosmic web''  (200021\_213076). W.~C gratefully thanks Comunidad de Madrid for the Atracci\'{o}n de Talento fellowship no. 2020-T1/TIC19882 and Agencia Estatal de Investigación (AEI) for the Consolidación Investigadora Grant CNS2024-154838. He further acknowledges the Project PID2024-156100NB-C21 financed by MICIU/AEI /10.13039/501100011033/FEDER, EU and ERC: HORIZON-TMA-MSCA-SE for supporting the LACEGAL-III (Latin American Chinese European Galaxy Formation Network) project with grant number 101086388 and the science research grants from the China Manned Space Project. AK is supported by project PID2024-156100NB-C21 financed by MICIU /AEI/10.13039/501100011033 / FEDER, UE and further thanks The Mary Onettes for Islands. G.~M is thankful to Mathias Urbano, Callum Bellhouse, Tutku Kolcu and Jesse Golden-Marx for helpful comments.

The \textsc{Horizon-AGN} simulation was performed using HPC resources of CINES under allocations 2013047012, 2014047012, and 2015047012 made by GENCI. The \textsc{IllustrisTNG} simulations, of which \textsc{TNG100} is part, were undertaken with compute time awarded by the Gauss Centre for Supercomputing (GCS) under GCS Large-Scale Projects GCS-ILLU and GCS-DWAR on the GCS share of the supercomputer Hazel Hen at the High Performance Computing Center Stuttgart (HLRS), as well as on the machines of the Max Planck Computing and Data Facility (MPCDF) in Garching, Germany. The \textsc{Hydrangea} simulations were in part performed on the German federal maximum performance computer ‘HazelHen’ at the maximum performance computing centre Stuttgart (HLRS), under project GCS-HYDA / ID 44067 financed through the large-scale project ‘\textit{Hydrangea}’ of the Gauss Center for Supercomputing. Further simulations were performed at the Max Planck Computing and Data Facility (MPCDF) in Garching, Germany. Part of this work has been made possible by \textit{TheThreeHundred} collaboration, which benefits from financial support of the European Union’s Horizon 2020 Research and Innovation programme under the Marie Skłodowska-Curie grant agreement number 734374, i.e. the LACEGAL project. \textsc{TheThreeHundred} simulations used in this paper have been performed in the MareNostrum Supercomputer at the Barcelona Supercomputing Center, thanks to CPU time granted by the Red Española de Supercomputación. The high-resolution (7K) simulations from \textsc{TheThreeHundred} were performed on multiple Supercomputers: MareNostrum, Finisterrae3, and Cibeles through The Red Española de Supercomputación grants, DIaL3 -- DiRAC Data Intensive service at the University of Leicester through the RAC15 grant, and the Niagara supercomputer at the SciNet HPC Consortium.

This work made use of the following software: \textsc{GADGET-4} \citep[][]{Springel2021}, \textsc{GalIC} \citep{Yurin2014}, \textsc{PyMC} \citep{pymc2023}, \textsc{Matplotlib} \citep{Hunter2007}, \textsc{NumPy} \citep{Harris2020} and \textsc{SciPy} \citep{Virtanen2020}.

This work has made use of the Infinity cluster, hosted by the Institut d’Astrophysique de Paris. We warmly thank S. Rouberol for running it smoothly.

\section*{Data Availability}

Simulation data analysed in this paper can be obtained from the following sources: Data from \textsc{TheThreeHundred} galaxy clusters sample are available on request following the guidelines of \textit{TheThreeHundred} collaboration, at https://www.the300-project.org. \textsc{TNG100} data is publicly available and can be accessed at \url{https://tng-project.org/}. The \textsc{Hydrangea} simulations are publicly available at \url{https://ftp.strw.leidenuniv.nl/bahe/Hydrangea}. The raw data products of the \textsc{Horizon-AGN} simulation are available upon reasonable request through the collaboration’s website: \url{https:// www.horizon-simulation.org/}. Other data can be shared on request to the corresponding author.
 


\bibliographystyle{mnras}
\bibliography{paper_mnras} 

@ARTICLE{Pillepich2018,
       author = {{Pillepich}, Annalisa and {Nelson}, Dylan and {Hernquist}, Lars and {Springel}, Volker and {Pakmor}, R{\"u}diger and {Torrey}, Paul and {Weinberger}, Rainer and {Genel}, Shy and {Naiman}, Jill P. and {Marinacci}, Federico and {Vogelsberger}, Mark},
        title = "{First results from the IllustrisTNG simulations: the stellar mass content of groups and clusters of galaxies}",
      journal = {\mnras},
         year = 2018,
        month = mar,
       volume = {475},
       number = {1},
        pages = {648-675},
          doi = {10.1093/mnras/stx3112},
archivePrefix = {arXiv},
       eprint = {1707.03406},
 primaryClass = {astro-ph.GA},
       adsurl = {https://ui.adsabs.harvard.edu/abs/2018MNRAS.475..648P}
}

@ARTICLE{Dubois2014,
   author = {{Dubois}, Y. and {Pichon}, C. and {Welker}, C. and {Le Borgne}, D. and 
	{Devriendt}, J. and {Laigle}, C. and {Codis}, S. and {Pogosyan}, D. and 
	{Arnouts}, S. and {Benabed}, K. and {Bertin}, E. and {Blaizot}, J. and 
	{Bouchet}, F. and {Cardoso}, J.-F. and {Colombi}, S. and {de Lapparent}, V. and 
	{Desjacques}, V. and {Gavazzi}, R. and {Kassin}, S. and {Kimm}, T. and 
	{McCracken}, H. and {Milliard}, B. and {Peirani}, S. and {Prunet}, S. and 
	{Rouberol}, S. and {Silk}, J. and {Slyz}, A. and {Sousbie}, T. and 
	{Teyssier}, R. and {@ARTICLE{2016MNRAS.457.4340K,
       author = {{Klypin}, Anatoly and {Yepes}, Gustavo and {Gottl{\"o}ber}, Stefan and {Prada}, Francisco and {He{\ss}}, Steffen},
        title = "{MultiDark simulations: the story of dark matter halo concentrations and density profiles}",
      journal = {\mnras},
     keywords = {methods: numerical, galaxies: haloes, dark matter, Astrophysics - Cosmology and Nongalactic Astrophysics},
         year = 2016,
        month = apr,
       volume = {457},
       number = {4},
        pages = {4340-4359},
          doi = {10.1093/mnras/stw248},
archivePrefix = {arXiv},
       eprint = {1411.4001},
 primaryClass = {astro-ph.CO},
       adsurl = {https://ui.adsabs.harvard.edu/abs/2016MNRAS.457.4340K},
      adsnote = {Provided by the SAO/NASA Astrophysics Data System}
}

Tresse}, L. and {Treyer}, M. and {Vibert}, D. and 
	{Volonteri}, M.},
    title = "{Dancing in the dark: galactic properties trace spin swings along the cosmic web}",
  journal = {\mnras},
archivePrefix = "arXiv",
   eprint = {1402.1165},
     year = 2014,
    month = oct,
   volume = 444,
    pages = {1453-1468},
      doi = {10.1093/mnras/stu1227},
   adsurl = {http://adsabs.harvard.edu/abs/2014MNRAS.444.1453D}
}

@ARTICLE{Schaye2015,
   author = {{Schaye}, J. and {Crain}, R.~A. and {Bower}, R.~G. and {Furlong}, M. and 
	{Schaller}, M. and {Theuns}, T. and {Dalla Vecchia}, C. and 
	{Frenk}, C.~S. and {McCarthy}, I.~G. and {Helly}, J.~C. and 
	{Jenkins}, A. and {Rosas-Guevara}, Y.~M. and {White}, S.~D.~M. and 
	{Baes}, M. and {Booth}, C.~M. and {Camps}, P. and {Navarro}, J.~F. and 
	{Qu}, Y. and {Rahmati}, A. and {Sawala}, T. and {Thomas}, P.~A. and 
	{Trayford}, J.},
    title = "{The EAGLE project: simulating the evolution and assembly of galaxies and their environments}",
  journal = {\mnras},
archivePrefix = "arXiv",
   eprint = {1407.7040},
     year = 2015,
    month = jan,
   volume = 446,
    pages = {521-554},
      doi = {10.1093/mnras/stu2058},
   adsurl = {http://adsabs.harvard.edu/abs/2015MNRAS.446..521S}
}

@ARTICLE{Teyssier2002,
   author = {{Teyssier}, R.},
    title = "{Cosmological hydrodynamics with adaptive mesh refinement. A new high resolution code called RAMSES}",
  journal = {\aap},
   eprint = {astro-ph/0111367},
     year = 2002,
    month = apr,
   volume = 385,
    pages = {337-364},
      doi = {10.1051/0004-6361:20011817},
   adsurl = {http://adsabs.harvard.edu/abs/2002A%26A...385..337T}
}

@ARTICLE{Komatsu2011,
   author = {{Komatsu}, E. and {Smith}, K.~M. and {Dunkley}, J. and {Bennett}, C.~L. and 
	{Gold}, B. and {Hinshaw}, G. and {Jarosik}, N. and {Larson}, D. and 
	{Nolta}, M.~R. and {Page}, L. and {Spergel}, D.~N. and {Halpern}, M. and 
	{Hill}, R.~S. and {Kogut}, A. and {Limon}, M. and {Meyer}, S.~S. and 
	{Odegard}, N. and {Tucker}, G.~S. and {Weiland}, J.~L. and {Wollack}, E. and 
	{Wright}, E.~L.},
    title = "{Seven-year Wilkinson Microwave Anisotropy Probe (WMAP) Observations: Cosmological Interpretation}",
  journal = {\apjs},
archivePrefix = "arXiv",
   eprint = {1001.4538},
 primaryClass = "astro-ph.CO",
     year = 2011,
    month = feb,
   volume = 192,
      eid = {18},
    pages = {18},
      doi = {10.1088/0067-0049/192/2/18},
   adsurl = {http://adsabs.harvard.edu/abs/2011ApJS..192...18K}
}

@ARTICLE{Dubois2016,
   author = {{Dubois}, Y. and {Peirani}, S. and {Pichon}, C. and {Devriendt}, J. and 
	{Gavazzi}, R. and {Welker}, C. and {Volonteri}, M.},
    title = "{The HORIZON-AGN simulation: morphological diversity of galaxies promoted by AGN feedback}",
  journal = {\mnras},
archivePrefix = "arXiv",
   eprint = {1606.03086},
     year = 2016,
    month = dec,
   volume = 463,
    pages = {3948-3964},
      doi = {10.1093/mnras/stw2265},
   adsurl = {http://adsabs.harvard.edu/abs/2016MNRAS.463.3948D}
}

@ARTICLE{Kaviraj2017,
   author = {{Kaviraj}, S. and {Laigle}, C. and {Kimm}, T. and {Devriendt}, J.~E.~G. and 
	{Dubois}, Y. and {Pichon}, C. and {Slyz}, A. and {Chisari}, E. and 
	{Peirani}, S.},
    title = "{The Horizon-AGN simulation: evolution of galaxy properties over cosmic time}",
  journal = {\mnras},
archivePrefix = "arXiv",
   eprint = {1605.09379},
     year = 2017,
    month = jun,
   volume = 467,
    pages = {4739-4752},
      doi = {10.1093/mnras/stx126},
   adsurl = {http://adsabs.harvard.edu/abs/2017MNRAS.467.4739K}
}

@ARTICLE{Laureijs2011,
   author = {{Laureijs}, R. and {Amiaux}, J. and {Arduini}, S. and {Augu{\`e}res}, J.~-. and 
	{Brinchmann}, J. and {Cole}, R. and {Cropper}, M. and {Dabin}, C. and 
	{Duvet}, L. and {Ealet}, A. and et al.},
    title = "{Euclid Definition Study Report}",
  journal = {ArXiv e-prints},
archivePrefix = "arXiv",
   eprint = {1110.3193},
 primaryClass = "astro-ph.CO",
     year = 2011,
    month = oct,
   adsurl = {http://adsabs.harvard.edu/abs/2011arXiv1110.3193L}
}

@ARTICLE{Byrd1990,
       author = {{Byrd}, Gene and {Valtonen}, Mauri},
        title = "{Tidal Generation of Active Spirals and S0 Galaxies by Rich Clusters}",
      journal = {\apj},
         year = 1990,
        month = Feb,
       volume = {350},
        pages = {89},
          doi = {10.1086/168362},
       adsurl = {https://ui.adsabs.harvard.edu/#abs/1990ApJ...350...89B}
}

@ARTICLE{Gnedin2004,
       author = {{Gnedin}, Oleg Y. and {Kravtsov}, Andrey V. and {Klypin}, Anatoly A. and
        {Nagai}, Daisuke},
        title = "{Response of Dark Matter Halos to Condensation of Baryons: Cosmological
        Simulations and Improved Adiabatic Contraction Model}",
      journal = {\apj},
         year = 2004,
        month = Nov,
       volume = {616},
        pages = {16-26},
          doi = {10.1086/424914},
archivePrefix = {arXiv},
       eprint = {astro-ph/0406247},
       adsurl = {https://ui.adsabs.harvard.edu/#abs/2004ApJ...616...16G}
}

@ARTICLE{Pontzen2012,
       author = {{Pontzen}, Andrew and {Governato}, Fabio},
        title = "{How supernova feedback turns dark matter cusps into cores}",
      journal = {\mnras},
         year = 2012,
        month = Apr,
       volume = {421},
        pages = {3464-3471},
          doi = {10.1111/j.1365-2966.2012.20571.x},
archivePrefix = {arXiv},
       eprint = {1106.0499},
       adsurl = {https://ui.adsabs.harvard.edu/#abs/2012MNRAS.421.3464P}
}

@ARTICLE{Teyssier2013,
       author = {{Teyssier}, Romain and {Pontzen}, Andrew and {Dubois}, Yohan and {Read},
        Justin I.},
        title = "{Cusp-core transformations in dwarf galaxies: observational predictions}",
      journal = {\mnras},
         year = 2013,
        month = Mar,
       volume = {429},
        pages = {3068-3078},
          doi = {10.1093/mnras/sts563},
archivePrefix = {arXiv},
       eprint = {1206.4895},
       adsurl = {https://ui.adsabs.harvard.edu/#abs/2013MNRAS.429.3068T}
}

@ARTICLE{Schechter1976,
       author = {{Schechter}, P.},
        title = "{An analytic expression for the luminosity function for galaxies.}",
      journal = {\apj},
         year = 1976,
        month = Jan,
       volume = {203},
        pages = {297-306},
          doi = {10.1086/154079},
       adsurl = {https://ui.adsabs.harvard.edu/#abs/1976ApJ...203..297S}
}

@ARTICLE{Dave2019,
       author = {{Dav{\'e}}, Romeel and {Angl{\'e}s-Alc{\'a}zar}, Daniel and
         {Narayanan}, Desika and {Li}, Qi and {Rafieferantsoa}, Mika H. and
         {Appleby}, Sarah},
        title = "{SIMBA: Cosmological simulations with black hole growth and feedback}",
      journal = {\mnras},
         year = 2019,
        month = jun,
       volume = {486},
       number = {2},
        pages = {2827-2849},
          doi = {10.1093/mnras/stz937},
archivePrefix = {arXiv},
       eprint = {1901.10203},
 primaryClass = {astro-ph.GA},
       adsurl = {https://ui.adsabs.harvard.edu/abs/2019MNRAS.486.2827D}
}

@ARTICLE{Nelson2019,
       author = {{Nelson}, Dylan and {Pillepich}, Annalisa and {Springel}, Volker and
         {Pakmor}, R{\"u}diger and {Weinberger}, Rainer and {Genel}, Shy and
         {Torrey}, Paul and {Vogelsberger}, Mark and {Marinacci}, Federico and
         {Hernquist}, Lars},
        title = "{First results from the TNG50 simulation: galactic outflows driven by supernovae and black hole feedback}",
      journal = {\mnras},
         year = 2019,
        month = dec,
       volume = {490},
       number = {3},
        pages = {3234-3261},
          doi = {10.1093/mnras/stz2306},
archivePrefix = {arXiv},
       eprint = {1902.05554},
 primaryClass = {astro-ph.GA},
       adsurl = {https://ui.adsabs.harvard.edu/abs/2019MNRAS.490.3234N}
}

@ARTICLE{Moster2013,
       author = {{Moster}, Benjamin P. and {Naab}, Thorsten and {White}, Simon D.~M.},
        title = "{Galactic star formation and accretion histories from matching galaxies to dark matter haloes}",
      journal = {\mnras},
         year = 2013,
        month = feb,
       volume = {428},
       number = {4},
        pages = {3121-3138},
          doi = {10.1093/mnras/sts261},
archivePrefix = {arXiv},
       eprint = {1205.5807},
 primaryClass = {astro-ph.CO},
       adsurl = {https://ui.adsabs.harvard.edu/abs/2013MNRAS.428.3121M}
}

@ARTICLE{Johnston2008,
       author = {{Johnston}, Kathryn V. and {Bullock}, James S. and {Sharma}, Sanjib and {Font}, Andreea and {Robertson}, Brant E. and {Leitner}, Samuel N.},
        title = "{Tracing Galaxy Formation with Stellar Halos. II. Relating Substructure in Phase and Abundance Space to Accretion Histories}",
      journal = {\apj},
         year = 2008,
        month = dec,
       volume = {689},
       number = {2},
        pages = {936-957},
          doi = {10.1086/592228},
archivePrefix = {arXiv},
       eprint = {0807.3911},
 primaryClass = {astro-ph},
       adsurl = {https://ui.adsabs.harvard.edu/abs/2008ApJ...689..936J}
}

@ARTICLE{Ivezic2019,
       author = {{Ivezi{\'c}}, {\v{Z}}eljko and {Kahn}, Steven M. and {Tyson}, J. Anthony and {Abel}, Bob and {Acosta}, Emily and {Allsman}, Robyn and {Alonso}, David and {AlSayyad}, Yusra and {Anderson}, Scott F. and {Andrew}, John and {Angel}, James Roger P. and {Angeli}, George Z. and {Ansari}, Reza and {Antilogus}, Pierre and {Araujo}, Constanza and {Armstrong}, Robert and {Arndt}, Kirk T. and {Astier}, Pierre and {Aubourg}, {\'E}ric and {Auza}, Nicole and {Axelrod}, Tim S. and {Bard}, Deborah J. and {Barr}, Jeff D. and {Barrau}, Aurelian and {Bartlett}, James G. and {Bauer}, Amanda E. and {Bauman}, Brian J. and {Baumont}, Sylvain and {Bechtol}, Ellen and {Bechtol}, Keith and {Becker}, Andrew C. and {Becla}, Jacek and {Beldica}, Cristina and {Bellavia}, Steve and {Bianco}, Federica B. and {Biswas}, Rahul and {Blanc}, Guillaume and {Blazek}, Jonathan and {Blandford}, Roger D. and {Bloom}, Josh S. and {Bogart}, Joanne and {Bond}, Tim W. and {Booth}, Michael T. and {Borgland}, Anders W. and {Borne}, Kirk and {Bosch}, James F. and {Boutigny}, Dominique and {Brackett}, Craig A. and {Bradshaw}, Andrew and {Brandt}, William Nielsen and {Brown}, Michael E. and {Bullock}, James S. and {Burchat}, Patricia and {Burke}, David L. and {Cagnoli}, Gianpietro and {Calabrese}, Daniel and {Callahan}, Shawn and {Callen}, Alice L. and {Carlin}, Jeffrey L. and {Carlson}, Erin L. and {Chandrasekharan}, Srinivasan and {Charles-Emerson}, Glenaver and {Chesley}, Steve and {Cheu}, Elliott C. and {Chiang}, Hsin-Fang and {Chiang}, James and {Chirino}, Carol and {Chow}, Derek and {Ciardi}, David R. and {Claver}, Charles F. and {Cohen-Tanugi}, Johann and {Cockrum}, Joseph J. and {Coles}, Rebecca and {Connolly}, Andrew J. and {Cook}, Kem H. and {Cooray}, Asantha and {Covey}, Kevin R. and {Cribbs}, Chris and {Cui}, Wei and {Cutri}, Roc and {Daly}, Philip N. and {Daniel}, Scott F. and {Daruich}, Felipe and {Daubard}, Guillaume and {Daues}, Greg and {Dawson}, William and {Delgado}, Francisco and {Dellapenna}, Alfred and {de Peyster}, Robert and {de Val-Borro}, Miguel and {Digel}, Seth W. and {Doherty}, Peter and {Dubois}, Richard and {Dubois-Felsmann}, Gregory P. and {Durech}, Josef and {Economou}, Frossie and {Eifler}, Tim and {Eracleous}, Michael and {Emmons}, Benjamin L. and {Fausti Neto}, Angelo and {Ferguson}, Henry and {Figueroa}, Enrique and {Fisher-Levine}, Merlin and {Focke}, Warren and {Foss}, Michael D. and {Frank}, James and {Freemon}, Michael D. and {Gangler}, Emmanuel and {Gawiser}, Eric and {Geary}, John C. and {Gee}, Perry and {Geha}, Marla and {Gessner}, Charles J.~B. and {Gibson}, Robert R. and {Gilmore}, D. Kirk and {Glanzman}, Thomas and {Glick}, William and {Goldina}, Tatiana and {Goldstein}, Daniel A. and {Goodenow}, Iain and {Graham}, Melissa L. and {Gressler}, William J. and {Gris}, Philippe and {Guy}, Leanne P. and {Guyonnet}, Augustin and {Haller}, Gunther and {Harris}, Ron and {Hascall}, Patrick A. and {Haupt}, Justine and {Hernandez}, Fabio and {Herrmann}, Sven and {Hileman}, Edward and {Hoblitt}, Joshua and {Hodgson}, John A. and {Hogan}, Craig and {Howard}, James D. and {Huang}, Dajun and {Huffer}, Michael E. and {Ingraham}, Patrick and {Innes}, Walter R. and {Jacoby}, Suzanne H. and {Jain}, Bhuvnesh and {Jammes}, Fabrice and {Jee}, M. James and {Jenness}, Tim and {Jernigan}, Garrett and {Jevremovi{\'c}}, Darko and {Johns}, Kenneth and {Johnson}, Anthony S. and {Johnson}, Margaret W.~G. and {Jones}, R. Lynne and {Juramy-Gilles}, Claire and {Juri{\'c}}, Mario and {Kalirai}, Jason S. and {Kallivayalil}, Nitya J. and {Kalmbach}, Bryce and {Kantor}, Jeffrey P. and {Karst}, Pierre and {Kasliwal}, Mansi M. and {Kelly}, Heather and {Kessler}, Richard and {Kinnison}, Veronica and {Kirkby}, David and {Knox}, Lloyd and {Kotov}, Ivan V. and {Krabbendam}, Victor L. and {Krughoff}, K. Simon and {Kub{\'a}nek}, Petr and {Kuczewski}, John and {Kulkarni}, Shri and {Ku}, John and {Kurita}, Nadine R. and {Lage}, Craig S. and {Lambert}, Ron and {Lange}, Travis and {Langton}, J. Brian and {Le Guillou}, Laurent and {Levine}, Deborah and {Liang}, Ming and {Lim}, Kian-Tat and {Lintott}, Chris J. and {Long}, Kevin E. and {Lopez}, Margaux and {Lotz}, Paul J. and {Lupton}, Robert H. and {Lust}, Nate B. and {MacArthur}, Lauren A. and {Mahabal}, Ashish and {Mandelbaum}, Rachel and {Markiewicz}, Thomas W. and {Marsh}, Darren S. and {Marshall}, Philip J. and {Marshall}, Stuart and {May}, Morgan and {McKercher}, Robert and {McQueen}, Michelle and {Meyers}, Joshua and {Migliore}, Myriam and {Miller}, Michelle and {Mills}, David J. and {Miraval}, Connor and {Moeyens}, Joachim and {Moolekamp}, Fred E. and {Monet}, David G. and {Moniez}, Marc and {Monkewitz}, Serge and {Montgomery}, Christopher and {Morrison}, Christopher B. and {Mueller}, Fritz and {Muller}, Gary P. and {Mu{\~n}oz Arancibia}, Freddy and {Neill}, Douglas R. and {Newbry}, Scott P. and {Nief}, Jean-Yves and {Nomerotski}, Andrei and {Nordby}, Martin and {O'Connor}, Paul and {Oliver}, John and {Olivier}, Scot S. and {Olsen}, Knut and {O'Mullane}, William and {Ortiz}, Sandra and {Osier}, Shawn and {Owen}, Russell E. and {Pain}, Reynald and {Palecek}, Paul E. and {Parejko}, John K. and {Parsons}, James B. and {Pease}, Nathan M. and {Peterson}, J. Matt and {Peterson}, John R. and {Petravick}, Donald L. and {Libby Petrick}, M.~E. and {Petry}, Cathy E. and {Pierfederici}, Francesco and {Pietrowicz}, Stephen and {Pike}, Rob and {Pinto}, Philip A. and {Plante}, Raymond and {Plate}, Stephen and {Plutchak}, Joel P. and {Price}, Paul A. and {Prouza}, Michael and {Radeka}, Veljko and {Rajagopal}, Jayadev and {Rasmussen}, Andrew P. and {Regnault}, Nicolas and {Reil}, Kevin A. and {Reiss}, David J. and {Reuter}, Michael A. and {Ridgway}, Stephen T. and {Riot}, Vincent J. and {Ritz}, Steve and {Robinson}, Sean and {Roby}, William and {Roodman}, Aaron and {Rosing}, Wayne and {Roucelle}, Cecille and {Rumore}, Matthew R. and {Russo}, Stefano and {Saha}, Abhijit and {Sassolas}, Benoit and {Schalk}, Terry L. and {Schellart}, Pim and {Schindler}, Rafe H. and {Schmidt}, Samuel and {Schneider}, Donald P. and {Schneider}, Michael D. and {Schoening}, William and {Schumacher}, German and {Schwamb}, Megan E. and {Sebag}, Jacques and {Selvy}, Brian and {Sembroski}, Glenn H. and {Seppala}, Lynn G. and {Serio}, Andrew and {Serrano}, Eduardo and {Shaw}, Richard A. and {Shipsey}, Ian and {Sick}, Jonathan and {Silvestri}, Nicole and {Slater}, Colin T. and {Smith}, J. Allyn and {Smith}, R. Chris and {Sobhani}, Shahram and {Soldahl}, Christine and {Storrie-Lombardi}, Lisa and {Stover}, Edward and {Strauss}, Michael A. and {Street}, Rachel A. and {Stubbs}, Christopher W. and {Sullivan}, Ian S. and {Sweeney}, Donald and {Swinbank}, John D. and {Szalay}, Alexander and {Takacs}, Peter and {Tether}, Stephen A. and {Thaler}, Jon J. and {Thayer}, John Gregg and {Thomas}, Sandrine and {Thornton}, Adam J. and {Thukral}, Vaikunth and {Tice}, Jeffrey and {Trilling}, David E. and {Turri}, Max and {Van Berg}, Richard and {Vanden Berk}, Daniel and {Vetter}, Kurt and {Virieux}, Francoise and {Vucina}, Tomislav and {Wahl}, William and {Walkowicz}, Lucianne and {Walsh}, Brian and {Walter}, Christopher W. and {Wang}, Daniel L. and {Wang}, Shin-Yawn and {Warner}, Michael and {Wiecha}, Oliver and {Willman}, Beth and {Winters}, Scott E. and {Wittman}, David and {Wolff}, Sidney C. and {Wood-Vasey}, W. Michael and {Wu}, Xiuqin and {Xin}, Bo and {Yoachim}, Peter and {Zhan}, Hu},
        title = "{LSST: From Science Drivers to Reference Design and Anticipated Data Products}",
      journal = {\apj},
         year = 2019,
        month = mar,
       volume = {873},
       number = {2},
          eid = {111},
        pages = {111},
          doi = {10.3847/1538-4357/ab042c},
archivePrefix = {arXiv},
       eprint = {0805.2366},
 primaryClass = {astro-ph},
       adsurl = {https://ui.adsabs.harvard.edu/abs/2019ApJ...873..111I}
}

@ARTICLE{Hendel2015,
       author = {{Hendel}, David and {Johnston}, Kathryn V.},
        title = "{Tidal debris morphology and the orbits of satellite galaxies}",
      journal = {\mnras},
         year = 2015,
        month = dec,
       volume = {454},
       number = {3},
        pages = {2472-2485},
          doi = {10.1093/mnras/stv2035},
archivePrefix = {arXiv},
       eprint = {1509.06369},
 primaryClass = {astro-ph.GA},
       adsurl = {https://ui.adsabs.harvard.edu/abs/2015MNRAS.454.2472H}
}

@ARTICLE{Bullock2005,
       author = {{Bullock}, James S. and {Johnston}, Kathryn V.},
        title = "{Tracing Galaxy Formation with Stellar Halos. I. Methods}",
      journal = {\apj},
         year = 2005,
        month = dec,
       volume = {635},
       number = {2},
        pages = {931-949},
          doi = {10.1086/497422},
archivePrefix = {arXiv},
       eprint = {astro-ph/0506467},
 primaryClass = {astro-ph},
       adsurl = {https://ui.adsabs.harvard.edu/abs/2005ApJ...635..931B}
}

@ARTICLE{Mihos2017,
       author = {{Mihos}, J. Christopher and {Harding}, Paul and {Feldmeier}, John J. and {Rudick}, Craig and {Janowiecki}, Steven and {Morrison}, Heather and {Slater}, Colin and {Watkins}, Aaron},
        title = "{The Burrell Schmidt Deep Virgo Survey: Tidal Debris, Galaxy Halos, and Diffuse Intracluster Light in the Virgo Cluster}",
      journal = {\apj},
         year = 2017,
        month = jan,
       volume = {834},
       number = {1},
          eid = {16},
        pages = {16},
          doi = {10.3847/1538-4357/834/1/16},
archivePrefix = {arXiv},
       eprint = {1611.04435},
 primaryClass = {astro-ph.GA},
       adsurl = {https://ui.adsabs.harvard.edu/abs/2017ApJ...834...16M}
}

@ARTICLE{Martin2022,
       author = {{Martin}, G. and {Bazkiaei}, A.~E. and {Spavone}, M. and {Iodice}, E. and {Mihos}, J.~C. and {Montes}, M. and {Benavides}, J.~A. and {Brough}, S. and {Carlin}, J.~L. and {Collins}, C.~A. and {Duc}, P.~A. and {G{\'o}mez}, F.~A. and {Galaz}, G. and {Hern{\'a}ndez-Toledo}, H.~M. and {Jackson}, R.~A. and {Kaviraj}, S. and {Knapen}, J.~H. and {Mart{\'\i}nez-Lombilla}, C. and {McGee}, S. and {O'Ryan}, D. and {Prole}, D.~J. and {Rich}, R.~M. and {Rom{\'a}n}, J. and {Shah}, E.~A. and {Starkenburg}, T.~K. and {Watkins}, A.~E. and {Zaritsky}, D. and {Pichon}, C. and {Armus}, L. and {Bianconi}, M. and {Buitrago}, F. and {Bus{\'a}}, I. and {Davis}, F. and {Demarco}, R. and {Desmons}, A. and {Garc{\'\i}a}, P. and {Graham}, A.~W. and {Holwerda}, B. and {Hon}, D.~S. -H. and {Khalid}, A. and {Klehammer}, J. and {Klutse}, D.~Y. and {Lazar}, I. and {Nair}, P. and {Noakes-Kettel}, E.~A. and {Rutkowski}, M. and {Saha}, K. and {Sahu}, N. and {Sola}, E. and {V{\'a}zquez-Mata}, J.~A. and {Vera-Casanova}, A. and {Yoon}, I.},
        title = "{Preparing for low surface brightness science with the Vera C. Rubin Observatory: Characterization of tidal features from mock images}",
      journal = {\mnras},
         year = 2022,
        month = jun,
       volume = {513},
       number = {1},
        pages = {1459-1487},
          doi = {10.1093/mnras/stac1003},
archivePrefix = {arXiv},
       eprint = {2203.07675},
 primaryClass = {astro-ph.GA},
       adsurl = {https://ui.adsabs.harvard.edu/abs/2022MNRAS.513.1459M}
}

@ARTICLE{Montes2021,
       author = {{Montes}, Mireia and {Brough}, Sarah and {Owers}, Matt S. and {Santucci}, Giulia},
        title = "{The Buildup of the Intracluster Light of A85 as Seen by Subaru's Hyper Suprime-Cam}",
      journal = {\apj},
         year = 2021,
        month = mar,
       volume = {910},
       number = {1},
          eid = {45},
        pages = {45},
          doi = {10.3847/1538-4357/abddb6},
archivePrefix = {arXiv},
       eprint = {2101.08290},
 primaryClass = {astro-ph.GA},
       adsurl = {https://ui.adsabs.harvard.edu/abs/2021ApJ...910...45M}
}

@ARTICLE{Springel2010,
       author = {{Springel}, Volker},
        title = "{E pur si muove: Galilean-invariant cosmological hydrodynamical simulations on a moving mesh}",
      journal = {\mnras},
         year = 2010,
        month = jan,
       volume = {401},
       number = {2},
        pages = {791-851},
          doi = {10.1111/j.1365-2966.2009.15715.x},
archivePrefix = {arXiv},
       eprint = {0901.4107},
 primaryClass = {astro-ph.CO},
       adsurl = {https://ui.adsabs.harvard.edu/abs/2010MNRAS.401..791S}
}

@ARTICLE{Planck2016,
       author = {{Planck Collaboration XIII}},
        title = "{Planck 2015 results. XIII. Cosmological parameters}",
      journal = {\aap},
         year = 2016,
        month = sep,
       volume = {594},
          eid = {A13},
        pages = {A13},
          doi = {10.1051/0004-6361/201525830},
archivePrefix = {arXiv},
       eprint = {1502.01589},
 primaryClass = {astro-ph.CO},
       adsurl = {https://ui.adsabs.harvard.edu/abs/2016A&A...594A..13P}
}

@ARTICLE{Rudick2011,
       author = {{Rudick}, Craig S. and {Mihos}, J. Christopher and {McBride}, Cameron K.},
        title = "{The Quantity of Intracluster Light: Comparing Theoretical and Observational Measurement Techniques using Simulated Clusters}",
      journal = {\apj},
         year = 2011,
        month = may,
       volume = {732},
       number = {1},
          eid = {48},
        pages = {48},
          doi = {10.1088/0004-637X/732/1/48},
archivePrefix = {arXiv},
       eprint = {1103.1215},
 primaryClass = {astro-ph.CO},
       adsurl = {https://ui.adsabs.harvard.edu/abs/2011ApJ...732...48R}
}

@ARTICLE{Contini2021,
       author = {{Contini}, Emanuele},
        title = "{On the Origin and Evolution of the Intra-Cluster Light: A Brief Review of the Most Recent Developments}",
      journal = {Galaxies},
         year = 2021,
        month = aug,
       volume = {9},
       number = {3},
          eid = {60},
        pages = {60},
          doi = {10.3390/galaxies9030060},
archivePrefix = {arXiv},
       eprint = {2107.04180},
 primaryClass = {astro-ph.GA},
       adsurl = {https://ui.adsabs.harvard.edu/abs/2021Galax...9...60C}
}

@ARTICLE{Brough2024,
       author = {{Brough}, Sarah and {Ahad}, Syeda Lammim and {Bah{\'e}}, Yannick M. and {Ellien}, Ama{\"e}l and {Gonzalez}, Anthony H. and {Jim{\'e}nez-Teja}, Yolanda and {Kimmig}, Lucas C. and {Martin}, Garreth and {Mart{\'\i}nez-Lombilla}, Cristina and {Montes}, Mireia and {Pillepich}, Annalisa and {Ragusa}, Rossella and {Remus}, Rhea-Silvia and {Collins}, Chris A. and {Knapen}, Johan H. and {Mihos}, J. Christopher},
        title = "{Preparing for low surface brightness science with the Vera C. Rubin Observatory: a comparison of observable and simulated intracluster light fractions}",
      journal = {\mnras},
         year = 2024,
        month = feb,
       volume = {528},
       number = {1},
        pages = {771-795},
          doi = {10.1093/mnras/stad3810},
archivePrefix = {arXiv},
       eprint = {2311.18016},
 primaryClass = {astro-ph.CO},
       adsurl = {https://ui.adsabs.harvard.edu/abs/2024MNRAS.528..771B}
}

@ARTICLE{Montes2018,
       author = {{Montes}, Mireia and {Trujillo}, Ignacio},
        title = "{Intracluster light at the Frontier - II. The Frontier Fields Clusters}",
      journal = {\mnras},
         year = 2018,
        month = feb,
       volume = {474},
       number = {1},
        pages = {917-932},
          doi = {10.1093/mnras/stx2847},
archivePrefix = {arXiv},
       eprint = {1710.03240},
 primaryClass = {astro-ph.CO},
       adsurl = {https://ui.adsabs.harvard.edu/abs/2018MNRAS.474..917M}
}

@ARTICLE{Zhang2019,
       author = {{Zhang}, Y. and {Yanny}, B. and {Palmese}, A. and {Gruen}, D. and {To}, C. and {Rykoff}, E.~S. and {Leung}, Y. and {Collins}, C. and {Hilton}, M. and {Abbott}, T.~M.~C. and {Annis}, J. and {Avila}, S. and {Bertin}, E. and {Brooks}, D. and {Burke}, D.~L. and {Carnero Rosell}, A. and {Carrasco Kind}, M. and {Carretero}, J. and {Cunha}, C.~E. and {D'Andrea}, C.~B. and {da Costa}, L.~N. and {De Vicente}, J. and {Desai}, S. and {Diehl}, H.~T. and {Dietrich}, J.~P. and {Doel}, P. and {Drlica-Wagner}, A. and {Eifler}, T.~F. and {Evrard}, A.~E. and {Flaugher}, B. and {Fosalba}, P. and {Frieman}, J. and {Garc{\'\i}a-Bellido}, J. and {Gaztanaga}, E. and {Gerdes}, D.~W. and {Gruendl}, R.~A. and {Gschwend}, J. and {Gutierrez}, G. and {Hartley}, W.~G. and {Hollowood}, D.~L. and {Honscheid}, K. and {Hoyle}, B. and {James}, D.~J. and {Jeltema}, T. and {Kuehn}, K. and {Kuropatkin}, N. and {Li}, T.~S. and {Lima}, M. and {Maia}, M.~A.~G. and {March}, M. and {Marshall}, J.~L. and {Melchior}, P. and {Menanteau}, F. and {Miller}, C.~J. and {Miquel}, R. and {Mohr}, J.~J. and {Ogando}, R.~L.~C. and {Plazas}, A.~A. and {Romer}, A.~K. and {Sanchez}, E. and {Scarpine}, V. and {Schubnell}, M. and {Serrano}, S. and {Sevilla-Noarbe}, I. and {Smith}, M. and {Soares-Santos}, M. and {Sobreira}, F. and {Suchyta}, E. and {Swanson}, M.~E.~C. and {Tarle}, G. and {Thomas}, D. and {Wester}, W. and {DES Collaboration}},
        title = "{Dark Energy Survey Year 1 Results: Detection of Intracluster Light at Redshift {\ensuremath{\sim}} 0.25}",
      journal = {\apj},
         year = 2019,
        month = apr,
       volume = {874},
       number = {2},
          eid = {165},
        pages = {165},
          doi = {10.3847/1538-4357/ab0dfd},
archivePrefix = {arXiv},
       eprint = {1812.04004},
 primaryClass = {astro-ph.CO},
       adsurl = {https://ui.adsabs.harvard.edu/abs/2019ApJ...874..165Z}
}

@ARTICLE{Gonzalez2007,
       author = {{Gonzalez}, Anthony H. and {Zaritsky}, Dennis and {Zabludoff}, Ann I.},
        title = "{A Census of Baryons in Galaxy Clusters and Groups}",
      journal = {\apj},
         year = 2007,
        month = sep,
       volume = {666},
       number = {1},
        pages = {147-155},
          doi = {10.1086/519729},
archivePrefix = {arXiv},
       eprint = {0705.1726},
 primaryClass = {astro-ph},
       adsurl = {https://ui.adsabs.harvard.edu/abs/2007ApJ...666..147G}
}

@ARTICLE{Rudick2006,
       author = {{Rudick}, Craig S. and {Mihos}, J. Christopher and {McBride}, Cameron},
        title = "{The Formation and Evolution of Intracluster Light}",
      journal = {\apj},
         year = 2006,
        month = sep,
       volume = {648},
       number = {2},
        pages = {936-946},
          doi = {10.1086/506176},
archivePrefix = {arXiv},
       eprint = {astro-ph/0605603},
 primaryClass = {astro-ph},
       adsurl = {https://ui.adsabs.harvard.edu/abs/2006ApJ...648..936R}
}

@ARTICLE{Contini2014,
       author = {{Contini}, E. and {De Lucia}, G. and {Villalobos}, {\'A}. and {Borgani}, S.},
        title = "{On the formation and physical properties of the intracluster light in hierarchical galaxy formation models}",
      journal = {\mnras},
         year = 2014,
        month = feb,
       volume = {437},
       number = {4},
        pages = {3787-3802},
          doi = {10.1093/mnras/stt2174},
archivePrefix = {arXiv},
       eprint = {1311.2076},
 primaryClass = {astro-ph.CO},
       adsurl = {https://ui.adsabs.harvard.edu/abs/2014MNRAS.437.3787C}
}

@ARTICLE{Murante2007,
       author = {{Murante}, Giuseppe and {Giovalli}, Martina and {Gerhard}, Ortwin and {Arnaboldi}, Magda and {Borgani}, Stefano and {Dolag}, Klaus},
        title = "{The importance of mergers for the origin of intracluster stars in cosmological simulations of galaxy clusters}",
      journal = {\mnras},
         year = 2007,
        month = may,
       volume = {377},
       number = {1},
        pages = {2-16},
          doi = {10.1111/j.1365-2966.2007.11568.x},
archivePrefix = {arXiv},
       eprint = {astro-ph/0701925},
 primaryClass = {astro-ph},
       adsurl = {https://ui.adsabs.harvard.edu/abs/2007MNRAS.377....2M}
}

@ARTICLE{Smith2016,
       author = {{Smith}, Rory and {Choi}, Hoseung and {Lee}, Jaehyun and {Rhee}, Jinsu and {Sanchez-Janssen}, Ruben and {Yi}, Sukyoung K.},
        title = "{The Preferential Tidal Stripping of Dark Matter versus Stars in Galaxies}",
      journal = {\apj},
         year = 2016,
        month = dec,
       volume = {833},
       number = {1},
          eid = {109},
        pages = {109},
          doi = {10.3847/1538-4357/833/1/109},
archivePrefix = {arXiv},
       eprint = {1610.04264},
 primaryClass = {astro-ph.GA},
       adsurl = {https://ui.adsabs.harvard.edu/abs/2016ApJ...833..109S}
}

@ARTICLE{vandenBosch2018,
       author = {{van den Bosch}, Frank C. and {Ogiya}, Go},
        title = "{Dark matter substructure in numerical simulations: a tale of discreteness noise, runaway instabilities, and artificial disruption}",
      journal = {\mnras},
         year = 2018,
        month = apr,
       volume = {475},
       number = {3},
        pages = {4066-4087},
          doi = {10.1093/mnras/sty084},
archivePrefix = {arXiv},
       eprint = {1801.05427},
 primaryClass = {astro-ph.GA},
       adsurl = {https://ui.adsabs.harvard.edu/abs/2018MNRAS.475.4066V}
}

@ARTICLE{Chun2023,
       author = {{Chun}, Kyungwon and {Shin}, Jihye and {Smith}, Rory and {Ko}, Jongwan and {Yoo}, Jaewon},
        title = "{The Formation of the Brightest Cluster Galaxy and Intracluster Light in Cosmological N-body Simulations with the Galaxy Replacement Technique}",
      journal = {\apj},
         year = 2023,
        month = feb,
       volume = {943},
       number = {2},
          eid = {148},
        pages = {148},
          doi = {10.3847/1538-4357/aca890},
archivePrefix = {arXiv},
       eprint = {2212.02510},
 primaryClass = {astro-ph.GA},
       adsurl = {https://ui.adsabs.harvard.edu/abs/2023ApJ...943..148C}
}

@ARTICLE{Arnaboldi2004,
       author = {{Arnaboldi}, Magda and {Gerhard}, Ortwin and {Aguerri}, J. Alfonso L. and {Freeman}, Kenneth C. and {Napolitano}, Nicola R. and {Okamura}, Sadanori and {Yasuda}, Naoki},
        title = "{The Line-of-Sight Velocity Distributions of Intracluster Planetary Nebulae in the Virgo Cluster Core}",
      journal = {\apjl},
         year = 2004,
        month = oct,
       volume = {614},
       number = {1},
        pages = {L33-L36},
          doi = {10.1086/425417},
archivePrefix = {arXiv},
       eprint = {astro-ph/0502421},
 primaryClass = {astro-ph},
       adsurl = {https://ui.adsabs.harvard.edu/abs/2004ApJ...614L..33A}
}

@ARTICLE{Yoo2024,
       author = {{Yoo}, Jaewon and {Park}, Changbom and {Sabiu}, Cristiano G. and {Singh}, Ankit and {Ko}, Jongwan and {Lee}, Jaehyun and {Pichon}, Christophe and {Jee}, M. James and {Gibson}, Brad K. and {Snaith}, Owain and {Kim}, Juhan and {Shin}, Jihye and {Kim}, Yonghwi and {Kim}, Hyowon},
        title = "{Spatial Distribution of Intracluster Light versus Dark Matter in Horizon Run 5}",
      journal = {\apj},
         year = 2024,
        month = apr,
       volume = {965},
       number = {2},
          eid = {145},
        pages = {145},
          doi = {10.3847/1538-4357/ad2df8},
archivePrefix = {arXiv},
       eprint = {2402.17958},
 primaryClass = {astro-ph.CO},
       adsurl = {https://ui.adsabs.harvard.edu/abs/2024ApJ...965..145Y}
}

@ARTICLE{Ogiya2021,
       author = {{Ogiya}, Go and {Taylor}, James E. and {Hudson}, Michael J.},
        title = "{Evolution of subhalo orbits in a smoothly growing host halo potential}",
      journal = {\mnras},
         year = 2021,
        month = may,
       volume = {503},
       number = {1},
        pages = {1233-1247},
          doi = {10.1093/mnras/stab361},
archivePrefix = {arXiv},
       eprint = {2102.02786},
 primaryClass = {astro-ph.GA},
       adsurl = {https://ui.adsabs.harvard.edu/abs/2021MNRAS.503.1233O}
}

@ARTICLE{Hernquist1990,
       author = {{Hernquist}, Lars},
        title = "{An Analytical Model for Spherical Galaxies and Bulges}",
      journal = {\apj},
         year = 1990,
        month = jun,
       volume = {356},
        pages = {359},
          doi = {10.1086/168845},
       adsurl = {https://ui.adsabs.harvard.edu/abs/1990ApJ...356..359H}
}

@ARTICLE{Yurin2014,
       author = {{Yurin}, Denis and {Springel}, Volker},
        title = "{An iterative method for the construction of N-body galaxy models in collisionless equilibrium}",
      journal = {\mnras},
         year = 2014,
        month = oct,
       volume = {444},
       number = {1},
        pages = {62-79},
          doi = {10.1093/mnras/stu1421},
archivePrefix = {arXiv},
       eprint = {1402.1623},
 primaryClass = {astro-ph.CO},
       adsurl = {https://ui.adsabs.harvard.edu/abs/2014MNRAS.444...62Y}
}

@ARTICLE{Bullock2001,
       author = {{Bullock}, J.~S. and {Dekel}, A. and {Kolatt}, T.~S. and {Kravtsov}, A.~V. and {Klypin}, A.~A. and {Porciani}, C. and {Primack}, J.~R.},
        title = "{A Universal Angular Momentum Profile for Galactic Halos}",
      journal = {\apj},
         year = 2001,
        month = jul,
       volume = {555},
       number = {1},
        pages = {240-257},
          doi = {10.1086/321477},
archivePrefix = {arXiv},
       eprint = {astro-ph/0011001},
 primaryClass = {astro-ph},
       adsurl = {https://ui.adsabs.harvard.edu/abs/2001ApJ...555..240B}
}

@ARTICLE{McBride2009,
       author = {{McBride}, James and {Fakhouri}, Onsi and {Ma}, Chung-Pei},
        title = "{Mass accretion rates and histories of dark matter haloes}",
      journal = {\mnras},
         year = 2009,
        month = oct,
       volume = {398},
       number = {4},
        pages = {1858-1868},
          doi = {10.1111/j.1365-2966.2009.15329.x},
archivePrefix = {arXiv},
       eprint = {0902.3659},
 primaryClass = {astro-ph.CO},
       adsurl = {https://ui.adsabs.harvard.edu/abs/2009MNRAS.398.1858M}
}

@ARTICLE{DeMaio2018,
       author = {{DeMaio}, Tahlia and {Gonzalez}, Anthony H. and {Zabludoff}, Ann and {Zaritsky}, Dennis and {Connor}, Thomas and {Donahue}, Megan and {Mulchaey}, John S.},
        title = "{Lost but not forgotten: intracluster light in galaxy groups and clusters}",
      journal = {\mnras},
         year = 2018,
        month = mar,
       volume = {474},
       number = {3},
        pages = {3009-3031},
          doi = {10.1093/mnras/stx2946},
archivePrefix = {arXiv},
       eprint = {1710.11313},
 primaryClass = {astro-ph.GA},
       adsurl = {https://ui.adsabs.harvard.edu/abs/2018MNRAS.474.3009D}
}

@ARTICLE{Wetzel2011,
       author = {{Wetzel}, Andrew R.},
        title = "{On the orbits of infalling satellite haloes}",
      journal = {\mnras},
         year = 2011,
        month = mar,
       volume = {412},
       number = {1},
        pages = {49-58},
          doi = {10.1111/j.1365-2966.2010.17877.x},
archivePrefix = {arXiv},
       eprint = {1001.4792},
 primaryClass = {astro-ph.CO},
       adsurl = {https://ui.adsabs.harvard.edu/abs/2011MNRAS.412...49W}
}

@ARTICLE{Sedgwick2019,
       author = {{Sedgwick}, Thomas M. and {Baldry}, Ivan K. and {James}, Philip A. and {Kelvin}, Lee S.},
        title = "{The galaxy stellar mass function and low surface brightness galaxies from core-collapse supernovae}",
      journal = {\mnras},
         year = 2019,
        month = apr,
       volume = {484},
       number = {4},
        pages = {5278-5295},
          doi = {10.1093/mnras/stz186},
archivePrefix = {arXiv},
       eprint = {1901.05020},
 primaryClass = {astro-ph.GA},
       adsurl = {https://ui.adsabs.harvard.edu/abs/2019MNRAS.484.5278S}
}

@ARTICLE{Bahe2017,
       author = {{Bah{\'e}}, Yannick M. and {Barnes}, David J. and {Dalla Vecchia}, Claudio and {Kay}, Scott T. and {White}, Simon D.~M. and {McCarthy}, Ian G. and {Schaye}, Joop and {Bower}, Richard G. and {Crain}, Robert A. and {Theuns}, Tom and {Jenkins}, Adrian and {McGee}, Sean L. and {Schaller}, Matthieu and {Thomas}, Peter A. and {Trayford}, James W.},
        title = "{The Hydrangea simulations: galaxy formation in and around massive clusters}",
      journal = {\mnras},
         year = 2017,
        month = oct,
       volume = {470},
       number = {4},
        pages = {4186-4208},
          doi = {10.1093/mnras/stx1403},
archivePrefix = {arXiv},
       eprint = {1703.10610},
 primaryClass = {astro-ph.GA},
       adsurl = {https://ui.adsabs.harvard.edu/abs/2017MNRAS.470.4186B}
}

@ARTICLE{Knebe2006,
       author = {{Knebe}, Alexander and {Power}, Chris and {Gill}, Stuart P.~D. and {Gibson}, Brad K.},
        title = "{The importance of interactions for mass loss from satellite galaxies in cold dark matter haloes}",
      journal = {\mnras},
         year = 2006,
        month = may,
       volume = {368},
       number = {2},
        pages = {741-750},
          doi = {10.1111/j.1365-2966.2006.10161.x},
archivePrefix = {arXiv},
       eprint = {astro-ph/0507380},
 primaryClass = {astro-ph},
       adsurl = {https://ui.adsabs.harvard.edu/abs/2006MNRAS.368..741K}
}

@ARTICLE{Cooper2015,
       author = {{Cooper}, A.~P. and {Gao}, L. and {Guo}, Q. and {Frenk}, C.~S. and {Jenkins}, A. and {Springel}, V. and {White}, S.~D.~M.},
        title = "{Surface photometry of brightest cluster galaxies and intracluster stars in {\ensuremath{\Lambda}}CDM}",
      journal = {\mnras},
         year = 2015,
        month = aug,
       volume = {451},
       number = {3},
        pages = {2703-2722},
          doi = {10.1093/mnras/stv1042},
archivePrefix = {arXiv},
       eprint = {1407.5627},
 primaryClass = {astro-ph.GA},
       adsurl = {https://ui.adsabs.harvard.edu/abs/2015MNRAS.451.2703C}
}

@ARTICLE{Kulier2023,
       author = {{Kulier}, Andrea and {Poggianti}, Bianca and {Tonnesen}, Stephanie and {Smith}, Rory and {Ignesti}, Alessandro and {Akerman}, Nina and {Marasco}, Antonino and {Vulcani}, Benedetta and {Moretti}, Alessia and {Wolter}, Anna},
        title = "{Ram Pressure Stripping in the EAGLE Simulation}",
      journal = {\apj},
         year = 2023,
        month = sep,
       volume = {954},
       number = {2},
          eid = {177},
        pages = {177},
          doi = {10.3847/1538-4357/aceda3},
archivePrefix = {arXiv},
       eprint = {2305.03758},
 primaryClass = {astro-ph.GA},
       adsurl = {https://ui.adsabs.harvard.edu/abs/2023ApJ...954..177K}
}

@ARTICLE{LyndenBell1967,
       author = {{Lynden-Bell}, D.},
        title = "{Statistical mechanics of violent relaxation in stellar systems}",
      journal = {\mnras},
         year = 1967,
        month = jan,
       volume = {136},
        pages = {101},
          doi = {10.1093/mnras/136.1.101},
       adsurl = {https://ui.adsabs.harvard.edu/abs/1967MNRAS.136..101L}
}

@ARTICLE{Cui2022,
       author = {{Cui}, Weiguang and {Dave}, Romeel and {Knebe}, Alexander and {Rasia}, Elena and {Gray}, Meghan and {Pearce}, Frazer and {Power}, Chris and {Yepes}, Gustavo and {Anbajagane}, Dhayaa and {Ceverino}, Daniel and {Contreras-Santos}, Ana and {de Andres}, Daniel and {De Petris}, Marco and {Ettori}, Stefano and {Haggar}, Roan and {Li}, Qingyang and {Wang}, Yang and {Yang}, Xiaohu and {Borgani}, Stefano and {Dolag}, Klaus and {Zu}, Ying and {Kuchner}, Ulrike and {Ca{\~n}as}, Rodrigo and {Ferragamo}, Antonio and {Gianfagna}, Giulia},
        title = "{THE THREE HUNDRED project: The GIZMO-SIMBA run}",
      journal = {\mnras},
         year = 2022,
        month = jul,
       volume = {514},
       number = {1},
        pages = {977-996},
          doi = {10.1093/mnras/stac1402},
archivePrefix = {arXiv},
       eprint = {2202.14038},
 primaryClass = {astro-ph.GA},
       adsurl = {https://ui.adsabs.harvard.edu/abs/2022MNRAS.514..977C}
}

@ARTICLE{Brown2024,
       author = {{Brown}, Harley J. and {Martin}, Garreth and {Pearce}, Frazer R. and {Hatch}, Nina A. and {Bah{\'e}}, Yannick M. and {Dubois}, Yohan},
        title = "{Assembly of the intracluster light in the HORIZON-AGN simulation}",
      journal = {\mnras},
         year = 2024,
        month = oct,
       volume = {534},
       number = {1},
        pages = {431-443},
          doi = {10.1093/mnras/stae2084},
archivePrefix = {arXiv},
       eprint = {2409.10607},
 primaryClass = {astro-ph.GA},
       adsurl = {https://ui.adsabs.harvard.edu/abs/2024MNRAS.534..431B}
}

@ARTICLE{Cui2018,
       author = {{Cui}, Weiguang and {Knebe}, Alexander and {Yepes}, Gustavo and {Pearce}, Frazer and {Power}, Chris and {Dave}, Romeel and {Arth}, Alexander and {Borgani}, Stefano and {Dolag}, Klaus and {Elahi}, Pascal and {Mostoghiu}, Robert and {Murante}, Giuseppe and {Rasia}, Elena and {Stoppacher}, Doris and {Vega-Ferrero}, Jesus and {Wang}, Yang and {Yang}, Xiaohu and {Benson}, Andrew and {Cora}, Sof{\'\i}a A. and {Croton}, Darren J. and {Sinha}, Manodeep and {Stevens}, Adam R.~H. and {Vega-Mart{\'\i}nez}, Cristian A. and {Arthur}, Jake and {Baldi}, Anna S. and {Ca{\~n}as}, Rodrigo and {Cialone}, Giammarco and {Cunnama}, Daniel and {De Petris}, Marco and {Durando}, Giacomo and {Ettori}, Stefano and {Gottl{\"o}ber}, Stefan and {Nuza}, Sebasti{\'a}n E. and {Old}, Lyndsay J. and {Pilipenko}, Sergey and {Sorce}, Jenny G. and {Welker}, Charlotte},
        title = "{The Three Hundred project: a large catalogue of theoretically modelled galaxy clusters for cosmological and astrophysical applications}",
      journal = {\mnras},
         year = 2018,
        month = nov,
       volume = {480},
       number = {3},
        pages = {2898-2915},
          doi = {10.1093/mnras/sty2111},
archivePrefix = {arXiv},
       eprint = {1809.04622},
 primaryClass = {astro-ph.GA},
       adsurl = {https://ui.adsabs.harvard.edu/abs/2018MNRAS.480.2898C}
}

@ARTICLE{Cui2014,
       author = {{Cui}, Weiguang and {Murante}, G. and {Monaco}, P. and {Borgani}, S. and {Granato}, G.~L. and {Killedar}, M. and {De Lucia}, G. and {Presotto}, V. and {Dolag}, K.},
        title = "{Characterizing diffused stellar light in simulated galaxy clusters}",
      journal = {\mnras},
         year = 2014,
        month = jan,
       volume = {437},
       number = {1},
        pages = {816-830},
          doi = {10.1093/mnras/stt1940},
archivePrefix = {arXiv},
       eprint = {1310.1396},
 primaryClass = {astro-ph.CO},
       adsurl = {https://ui.adsabs.harvard.edu/abs/2014MNRAS.437..816C}
}

@ARTICLE{Arthur2019,
       author = {{Arthur}, Jake and {Pearce}, Frazer R. and {Gray}, Meghan E. and {Knebe}, Alexander and {Cui}, Weiguang and {Elahi}, Pascal J. and {Power}, Chris and {Yepes}, Gustavo and {Arth}, Alexander and {De Petris}, Marco and {Dolag}, Klaus and {Garratt-Smithson}, Lilian and {Old}, Lyndsay J. and {Rasia}, Elena and {Stevens}, Adam R.~H.},
        title = "{THETHREEHUNDRED Project: ram pressure and gas content of haloes and subhaloes in the phase-space plane}",
      journal = {Monthly Notices of the Royal Astronomical Society},
         year = 2019,
        month = apr,
       volume = {484},
       number = {3},
        pages = {3968-3983},
          doi = {10.1093/mnras/stz212},
archivePrefix = {arXiv},
       eprint = {1901.05969},
 primaryClass = {astro-ph.GA},
       adsurl = {https://ui.adsabs.harvard.edu/abs/2019MNRAS.484.3968A}
}

@ARTICLE{Haggar2021,
       author = {{Haggar}, Roan and {Pearce}, Frazer R. and {Gray}, Meghan E. and {Knebe}, Alexander and {Yepes}, Gustavo},
        title = "{The Three Hundred Project: Substructure in hydrodynamical and dark matter simulations of galaxy groups around clusters}",
      journal = {Monthly Notices of the Royal Astronomical Society},
         year = 2021,
        month = mar,
       volume = {502},
       number = {1},
        pages = {1191-1204},
          doi = {10.1093/mnras/stab064},
archivePrefix = {arXiv},
       eprint = {2101.03178},
 primaryClass = {astro-ph.GA},
       adsurl = {https://ui.adsabs.harvard.edu/abs/2021MNRAS.502.1191H}
}

@ARTICLE{Ragusa2023,
       author = {{Ragusa}, R. and {Iodice}, E. and {Spavone}, M. and {Montes}, M. and {Forbes}, D.~A. and {Brough}, S. and {Mirabile}, M. and {Cantiello}, M. and {Paolillo}, M. and {Schipani}, P.},
        title = "{Does the virial mass drive the intra-cluster light?. Relationship between the ICL and M$_{vir}$ from VEGAS}",
      journal = {\aap},
         year = 2023,
        month = feb,
       volume = {670},
          eid = {L20},
        pages = {L20},
          doi = {10.1051/0004-6361/202245530},
archivePrefix = {arXiv},
       eprint = {2212.06164},
 primaryClass = {astro-ph.GA},
       adsurl = {https://ui.adsabs.harvard.edu/abs/2023A&A...670L..20R}
}

@ARTICLE{Contini2018,
       author = {{Contini}, E. and {Yi}, S.~K. and {Kang}, X.},
        title = "{The different growth pathways of brightest cluster galaxies and intracluster light}",
      journal = {\mnras},
         year = 2018,
        month = sep,
       volume = {479},
       number = {1},
        pages = {932-944},
          doi = {10.1093/mnras/sty1518},
archivePrefix = {arXiv},
       eprint = {1806.01480},
 primaryClass = {astro-ph.GA},
       adsurl = {https://ui.adsabs.harvard.edu/abs/2018MNRAS.479..932C}
}

@ARTICLE{Morishita2017,
       author = {{Morishita}, Takahiro and {Abramson}, Louis E. and {Treu}, Tommaso and {Schmidt}, Kasper B. and {Vulcani}, Benedetta and {Wang}, Xin},
        title = "{Characterizing Intracluster Light in the Hubble Frontier Fields}",
      journal = {\apj},
         year = 2017,
        month = sep,
       volume = {846},
       number = {2},
          eid = {139},
        pages = {139},
          doi = {10.3847/1538-4357/aa8403},
archivePrefix = {arXiv},
       eprint = {1610.08503},
 primaryClass = {astro-ph.GA},
       adsurl = {https://ui.adsabs.harvard.edu/abs/2017ApJ...846..139M}
}

@ARTICLE{Contini2019,
       author = {{Contini}, E. and {Yi}, S.~K. and {Kang}, X.},
        title = "{Theoretical Predictions of Colors and Metallicity of the Intracluster Light}",
      journal = {\apj},
         year = 2019,
        month = jan,
       volume = {871},
       number = {1},
          eid = {24},
        pages = {24},
          doi = {10.3847/1538-4357/aaf41f},
archivePrefix = {arXiv},
       eprint = {1811.03253},
 primaryClass = {astro-ph.GA},
       adsurl = {https://ui.adsabs.harvard.edu/abs/2019ApJ...871...24C}
}

@ARTICLE{Contini2024,
       author = {{Contini}, Emanuele and {Rhee}, Jinsu and {Han}, San and {Jeon}, Seyoung and {Yi}, Sukyoung K.},
        title = "{The Connection between the Intracluster Light and its Host Halo: Formation Time and Contribution from Different Channels}",
      journal = {\aj},
         year = 2024,
        month = jan,
       volume = {167},
       number = {1},
          eid = {7},
        pages = {7},
          doi = {10.3847/1538-3881/ad0894},
archivePrefix = {arXiv},
       eprint = {2310.20135},
 primaryClass = {astro-ph.GA},
       adsurl = {https://ui.adsabs.harvard.edu/abs/2024AJ....167....7C}
}

@ARTICLE{Mowla2019,
       author = {{Mowla}, Lamiya A. and {van Dokkum}, Pieter and {Brammer}, Gabriel B. and {Momcheva}, Ivelina and {van der Wel}, Arjen and {Whitaker}, Katherine and {Nelson}, Erica and {Bezanson}, Rachel and {Muzzin}, Adam and {Franx}, Marijn and {MacKenty}, John and {Leja}, Joel and {Kriek}, Mariska and {Marchesini}, Danilo},
        title = "{COSMOS-DASH: The Evolution of the Galaxy Size-Mass Relation since z {\ensuremath{\sim}} 3 from New Wide-field WFC3 Imaging Combined with CANDELS/3D-HST}",
      journal = {\apj},
         year = 2019,
        month = jul,
       volume = {880},
       number = {1},
          eid = {57},
        pages = {57},
          doi = {10.3847/1538-4357/ab290a},
archivePrefix = {arXiv},
       eprint = {1808.04379},
 primaryClass = {astro-ph.GA},
       adsurl = {https://ui.adsabs.harvard.edu/abs/2019ApJ...880...57M}
}

@ARTICLE{Prada2012,
       author = {{Prada}, Francisco and {Klypin}, Anatoly A. and {Cuesta}, Antonio J. and {Betancort-Rijo}, Juan E. and {Primack}, Joel},
        title = "{Halo concentrations in the standard {\ensuremath{\Lambda}} cold dark matter cosmology}",
      journal = {\mnras},
         year = 2012,
        month = jul,
       volume = {423},
       number = {4},
        pages = {3018-3030},
          doi = {10.1111/j.1365-2966.2012.21007.x},
archivePrefix = {arXiv},
       eprint = {1104.5130},
 primaryClass = {astro-ph.CO},
       adsurl = {https://ui.adsabs.harvard.edu/abs/2012MNRAS.423.3018P}
}

@ARTICLE{Harker2006,
       author = {{Harker}, Geraint and {Cole}, Shaun and {Helly}, John and {Frenk}, Carlos and {Jenkins}, Adrian},
        title = "{A marked correlation function analysis of halo formation times in the Millennium Simulation}",
      journal = {\mnras},
         year = 2006,
        month = apr,
       volume = {367},
       number = {3},
        pages = {1039-1049},
          doi = {10.1111/j.1365-2966.2006.10022.x},
archivePrefix = {arXiv},
       eprint = {astro-ph/0510488},
 primaryClass = {astro-ph},
       adsurl = {https://ui.adsabs.harvard.edu/abs/2006MNRAS.367.1039H}
}

@ARTICLE{Martin2024,
       author = {{Martin}, G. and {Pearce}, F.~R. and {Hatch}, N.~A. and {Contreras-Santos}, A. and {Knebe}, A. and {Cui}, W.},
        title = "{Stellar stripping efficiencies of satellites in numerical simulations: the effect of resolution, satellite properties, and numerical disruption}",
      journal = {\mnras},
         year = 2024,
        month = dec,
       volume = {535},
       number = {3},
        pages = {2375-2393},
          doi = {10.1093/mnras/stae2488},
archivePrefix = {arXiv},
       eprint = {2410.19292},
 primaryClass = {astro-ph.GA},
       adsurl = {https://ui.adsabs.harvard.edu/abs/2024MNRAS.535.2375M}
}

@BOOK{Binney2008,
       author = {{Binney}, James and {Tremaine}, Scott},
        title = "{Galactic Dynamics: Second Edition}",
         year = 2008,
       adsurl = {https://ui.adsabs.harvard.edu/abs/2008gady.book.....B}
}

@ARTICLE{Blumenthal1986,
       author = {{Blumenthal}, G.~R. and {Faber}, S.~M. and {Flores}, R. and {Primack}, J.~R.},
        title = "{Contraction of Dark Matter Galactic Halos Due to Baryonic Infall}",
      journal = {\apj},
         year = 1986,
        month = feb,
       volume = {301},
        pages = {27},
          doi = {10.1086/163867},
       adsurl = {https://ui.adsabs.harvard.edu/abs/1986ApJ...301...27B}
}

@ARTICLE{Jackson2025,
       author = {{Jackson}, R.~A. and {Navarro}, J.~F. and {Santos-Santos}, I.~M.~E. and {Kaviraj}, S. and {Yi}, S.~K. and {Peirani}, S. and {Dubois}, Y. and {Martin}, G. and {Devriendt}, J.~E.~G. and {Slyz}, A. and {Pichon}, C. and {Volonteri}, M. and {Kimm}, T. and {Kraljic}, K.},
        title = "{The diversity of rotation curves of galaxies in the NEWHORIZON cosmological simulation}",
      journal = {\mnras},
         year = 2025,
        month = jun,
       volume = {539},
       number = {4},
        pages = {3797-3807},
          doi = {10.1093/mnras/staf667},
       adsurl = {https://ui.adsabs.harvard.edu/abs/2025MNRAS.539.3797J}
}

@ARTICLE{Butler2025,
       author = {{Butler}, J. and {Martin}, G. and {Hatch}, N.~A. and {Pearce}, F. and {Brough}, S. and {Dubois}, Y.},
        title = "{Intracluster light is a biased tracer of the dark matter distribution in clusters}",
      journal = {\mnras},
         year = 2025,
        month = may,
       volume = {539},
       number = {3},
        pages = {2279-2291},
          doi = {10.1093/mnras/staf615},
archivePrefix = {arXiv},
       eprint = {2504.03518},
 primaryClass = {astro-ph.GA},
       adsurl = {https://ui.adsabs.harvard.edu/abs/2025MNRAS.539.2279B}
}

@ARTICLE{Springel2021,
       author = {{Springel}, Volker and {Pakmor}, R{\"u}diger and {Zier}, Oliver and {Reinecke}, Martin},
        title = "{Simulating cosmic structure formation with the GADGET-4 code}",
      journal = {\mnras},
         year = 2021,
        month = sep,
       volume = {506},
       number = {2},
        pages = {2871-2949},
          doi = {10.1093/mnras/stab1855},
archivePrefix = {arXiv},
       eprint = {2010.03567},
 primaryClass = {astro-ph.IM},
       adsurl = {https://ui.adsabs.harvard.edu/abs/2021MNRAS.506.2871S}
}

@ARTICLE{Gomez2025,
       author = {{G{\'o}mez}, Jonathan S. and {Hough}, Tomas and {Jim{\'e}nez Mu{\~n}oz}, Alejandro and {Yepes}, Gustavo and {Cui}, Weiguang and {Cora}, Sof{\'\i}a A.},
        title = "{The Three Hundred Project: A fast semi-analytic model emulator of hydrodynamical galaxy cluster simulations}",
      journal = {\aap},
         year = 2025,
        month = may,
       volume = {697},
          eid = {A171},
        pages = {A171},
          doi = {10.1051/0004-6361/202554064},
archivePrefix = {arXiv},
       eprint = {2504.03519},
 primaryClass = {astro-ph.GA},
       adsurl = {https://ui.adsabs.harvard.edu/abs/2025A&A...697A.171G}
}

@article{pymc2023,
  title = {{PyMC}: A Modern and Comprehensive Probabilistic Programming Framework in {P}ython},
  author = {Oriol Abril-Pla and Virgile Andreani and Colin Carroll and Larry Dong and Christopher J. Fonnesbeck and Maxim Kochurov and Ravin Kumar and Junpeng Lao and Christian C. Luhmann and Osvaldo A. Martin and Michael Osthege and Ricardo Vieira and Thomas Wiecki and Robert Zinkov },
  journal = {{PeerJ} Computer Science},
  volume = {9},
  number = {e1516},
  doi = {10.7717/peerj-cs.1516},
  year = {2023}
}

@Article{Hunter2007,
  Author    = {Hunter, J. D.},
  Title     = {Matplotlib: A 2D graphics environment},
  Journal   = {Computing in Science \& Engineering},
  Volume    = {9},
  Number    = {3},
  Pages     = {90--95},
  publisher = {IEEE COMPUTER SOC},
  doi       = {10.1109/MCSE.2007.55},
  year      = 2007
}

@Article{         Harris2020,
 title         = {Array programming with {NumPy}},
 author        = {Charles R. Harris and K. Jarrod Millman and St{\'{e}}fan J.
                 van der Walt and Ralf Gommers and Pauli Virtanen and David
                 Cournapeau and Eric Wieser and Julian Taylor and Sebastian
                 Berg and Nathaniel J. Smith and Robert Kern and Matti Picus
                 and Stephan Hoyer and Marten H. van Kerkwijk and Matthew
                 Brett and Allan Haldane and Jaime Fern{\'{a}}ndez del
                 R{\'{i}}o and Mark Wiebe and Pearu Peterson and Pierre
                 G{\'{e}}rard-Marchant and Kevin Sheppard and Tyler Reddy and
                 Warren Weckesser and Hameer Abbasi and Christoph Gohlke and
                 Travis E. Oliphant},
 year          = {2020},
 month         = sep,
 journal       = {Nature},
 volume        = {585},
 number        = {7825},
 pages         = {357--362},
 doi           = {10.1038/s41586-020-2649-2},
 publisher     = {Springer Science and Business Media {LLC}},
 url           = {https://doi.org/10.1038/s41586-020-2649-2}
}

@ARTICLE{Virtanen2020,
  author  = {Virtanen, Pauli and Gommers, Ralf and Oliphant, Travis E. and
            Haberland, Matt and Reddy, Tyler and Cournapeau, David and
            Burovski, Evgeni and Peterson, Pearu and Weckesser, Warren and
            Bright, Jonathan and {van der Walt}, St{\'e}fan J. and
            Brett, Matthew and Wilson, Joshua and Millman, K. Jarrod and
            Mayorov, Nikolay and Nelson, Andrew R. J. and Jones, Eric and
            Kern, Robert and Larson, Eric and Carey, C J and
            Polat, {\.I}lhan and Feng, Yu and Moore, Eric W. and
            {VanderPlas}, Jake and Laxalde, Denis and Perktold, Josef and
            Cimrman, Robert and Henriksen, Ian and Quintero, E. A. and
            Harris, Charles R. and Archibald, Anne M. and
            Ribeiro, Ant{\^o}nio H. and Pedregosa, Fabian and
            {van Mulbregt}, Paul and {SciPy 1.0 Contributors}},
  title   = {{{SciPy} 1.0: Fundamental Algorithms for Scientific
            Computing in Python}},
  journal = {Nature Methods},
  year    = {2020},
  volume  = {17},
  pages   = {261--272},
  adsurl  = {https://rdcu.be/b08Wh},
  doi     = {10.1038/s41592-019-0686-2},
}

@ARTICLE{Allen2011,
       author = {{Allen}, Steven W. and {Evrard}, August E. and {Mantz}, Adam B.},
        title = "{Cosmological Parameters from Observations of Galaxy Clusters}",
      journal = {\araa},
         year = 2011,
        month = sep,
       volume = {49},
       number = {1},
        pages = {409-470},
          doi = {10.1146/annurev-astro-081710-102514},
archivePrefix = {arXiv},
       eprint = {1103.4829},
 primaryClass = {astro-ph.CO},
       adsurl = {https://ui.adsabs.harvard.edu/abs/2011ARA&A..49..409A}
}

@ARTICLE{Kravtsov2012,
       author = {{Kravtsov}, Andrey V. and {Borgani}, Stefano},
        title = "{Formation of Galaxy Clusters}",
      journal = {\araa},
         year = 2012,
        month = sep,
       volume = {50},
        pages = {353-409},
          doi = {10.1146/annurev-astro-081811-125502},
archivePrefix = {arXiv},
       eprint = {1205.5556},
 primaryClass = {astro-ph.CO},
       adsurl = {https://ui.adsabs.harvard.edu/abs/2012ARA&A..50..353K}
}

@ARTICLE{Hoekstra2013,
       author = {{Hoekstra}, Henk and {Bartelmann}, Matthias and {Dahle}, H{\r{a}}kon and {Israel}, Holger and {Limousin}, Marceau and {Meneghetti}, Massimo},
        title = "{Masses of Galaxy Clusters from Gravitational Lensing}",
      journal = {\ssr},
         year = 2013,
        month = aug,
       volume = {177},
       number = {1-4},
        pages = {75-118},
          doi = {10.1007/s11214-013-9978-5},
archivePrefix = {arXiv},
       eprint = {1303.3274},
 primaryClass = {astro-ph.CO},
       adsurl = {https://ui.adsabs.harvard.edu/abs/2013SSRv..177...75H}
}

@ARTICLE{Montes2019b,
       author = {{Montes}, Mireia and {Trujillo}, Ignacio},
        title = "{Intracluster light: a luminous tracer for dark matter in clusters of galaxies}",
      journal = {\mnras},
         year = 2019,
        month = jan,
       volume = {482},
       number = {2},
        pages = {2838-2851},
          doi = {10.1093/mnras/sty2858},
archivePrefix = {arXiv},
       eprint = {1807.11488},
 primaryClass = {astro-ph.GA},
       adsurl = {https://ui.adsabs.harvard.edu/abs/2019MNRAS.482.2838M}
}

@ARTICLE{Borgani2001,
       author = {{Borgani}, Stefano and {Guzzo}, Luigi},
        title = "{X-ray clusters of galaxies as tracers of structure in the Universe}",
      journal = {\nat},
         year = 2001,
        month = jan,
       volume = {409},
       number = {6816},
        pages = {39-45},
          doi = {10.1038/409039A010.1038/35051000},
archivePrefix = {arXiv},
       eprint = {astro-ph/0012439},
 primaryClass = {astro-ph},
       adsurl = {https://ui.adsabs.harvard.edu/abs/2001Natur.409...39B}
}

@ARTICLE{Gifford2013,
       author = {{Gifford}, Daniel and {Miller}, Christopher and {Kern}, Nicholas},
        title = "{A Systematic Analysis of Caustic Methods for Galaxy Cluster Masses}",
      journal = {\apj},
         year = 2013,
        month = aug,
       volume = {773},
       number = {2},
          eid = {116},
        pages = {116},
          doi = {10.1088/0004-637X/773/2/116},
archivePrefix = {arXiv},
       eprint = {1307.0017},
 primaryClass = {astro-ph.CO},
       adsurl = {https://ui.adsabs.harvard.edu/abs/2013ApJ...773..116G}
}

@ARTICLE{Gonzalez2013,
       author = {{Gonzalez}, Anthony H. and {Sivanandam}, Suresh and {Zabludoff}, Ann I. and {Zaritsky}, Dennis},
        title = "{Galaxy Cluster Baryon Fractions Revisited}",
      journal = {\apj},
         year = 2013,
        month = nov,
       volume = {778},
       number = {1},
          eid = {14},
        pages = {14},
          doi = {10.1088/0004-637X/778/1/14},
archivePrefix = {arXiv},
       eprint = {1309.3565},
 primaryClass = {astro-ph.CO},
       adsurl = {https://ui.adsabs.harvard.edu/abs/2013ApJ...778...14G}
}

@article{Scaramella2022, title={Euclid preparation: I. The Euclid Wide Survey}, volume={662}, ISSN={0004-6361, 1432-0746}, DOI={10.1051/0004-6361/202141938}, note={arXiv:2108.01201 [astro-ph]}, journal={\aap}, author={Scaramella, R. and Amiaux, J. and Mellier, Y. and Burigana, C. and Carvalho, C. S. and Cuillandre, J.-C. and Silva, A. Da and Derosa, A. and Dinis, J. and Maiorano, E. and Maris, M. and Tereno, I. and Laureijs, R. and Boenke, T. and Buenadicha, G. and Dupac, X. and Venancio, L. M. Gaspar and Gómez-Álvarez, P. and Hoar, J. and Alvarez, J. Lorenzo and Racca, G. D. and Saavedra-Criado, G. and Schwartz, J. and Vavrek, R. and Schirmer, M. and Aussel, H. and Azzollini, R. and Cardone, V. F. and Cropper, M. and Ealet, A. and Garilli, B. and Gillard, W. and Granett, B. R. and Guzzo, L. and Hoekstra, H. and Jahnke, K. and Kitching, T. and Meneghetti, M. and Miller, L. and Nakajima, R. and Niemi, S. M. and Pasian, F. and Percival, W. J. and Sauvage, M. and Scodeggio, M. and Wachter, S. and Zacchei, A. and Aghanim, N. and Amara, A. and Auphan, T. and Auricchio, N. and Awan, S. and Balestra, A. and Bender, R. and Bodendorf, C. and Bonino, D. and Branchini, E. and Brau-Nogue, S. and Brescia, M. and Candini, G. P. and Capobianco, V. and Carbone, C. and Carlberg, R. G. and Carretero, J. and Casas, R. and Castander, F. J. and Castellano, M. and Cavuoti, S. and Cimatti, A. and Cledassou, R. and Congedo, G. and Conselice, C. J. and Conversi, L. and Copin, Y. and Corcione, L. and Costille, A. and Courbin, F. and Degaudenzi, H. and Douspis, M. and Dubath, F. and Duncan, C. A. J. and Dusini, S. and Farrens, S. and Ferriol, S. and Fosalba, P. and Fourmanoit, N. and Frailis, M. and Franceschi, E. and Franzetti, P. and Fumana, M. and Gillis, B. and Giocoli, C. and Grazian, A. and Grupp, F. and Haugan, S. V. H. and Holmes, W. and Hormuth, F. and Hudelot, P. and Kermiche, S. and Kiessling, A. and Kilbinger, M. and Kohley, R. and Kubik, B. and Kümmel, M. and Kunz, M. and Kurki-Suonio, H. and Ligori, S. and Lilje, P. B. and Lloro, I. and Mansutti, O. and Marggraf, O. and Markovic, K. and Marulli, F. and Massey, R. and Maurogordato, S. and Melchior, M. and Merlin, E. and Meylan, G. and Mohr, J. J. and Moresco, M. and Morin, B. and Moscardini, L. and Munari, E. and Nichol, R. C. and Padilla, C. and Paltani, S. and Peacock, J. and Pedersen, K. and Pettorino, V. and Pires, S. and Poncet, M. and Popa, L. and Pozzetti, L. and Raison, F. and Rebolo, R. and Rhodes, J. and Rix, H.-W. and Roncarelli, M. and Rossetti, E. and Saglia, R. and Schneider, P. and Schrabback, T. and Secroun, A. and Seidel, G. and Serrano, S. and Sirignano, C. and Sirri, G. and Skottfelt, J. and Stanco, L. and Starck, J. L. and Tallada-Crespí, P. and Tavagnacco, D. and Taylor, A. N. and Teplitz, H. I. and Toledo-Moreo, R. and Torradeflot, F. and Trifoglio, M. and Valentijn, E. A. and Valenziano, L. and Kleijn, G. A. Verdoes and Wang, Y. and Welikala, N. and Weller, J. and Wetzstein, M. and Zamorani, G. and Zoubian, J. and Andreon, S. and Baldi, M. and Bardelli, S. and Boucaud, A. and Camera, S. and Fabbian, G. and Farinelli, R. and Graciá-Carpio, J. and Maino, D. and Medinaceli, E. and Mei, S. and Neissner, C. and Polenta, G. and Renzi, A. and Romelli, E. and Rosset, C. and Sureau, F. and Tenti, M. and Vassallo, T. and Zucca, E. and Baccigalupi, C. and Balaguera-Antolínez, A. and Battaglia, P. and Biviano, A. and Borgani, S. and Bozzo, E. and Cabanac, R. and Cappi, A. and Casas, S. and Castignani, G. and Colodro-Conde, C. and Coupon, J. and Courtois, H. M. and Cuby, J. and Torre, S. de la and Desai, S. and Ferdinando, D. Di and Dole, H. and Fabricius, M. and Farina, M. and Ferreira, P. G. and Finelli, F. and Flose-Reimberg, P. and Fotopoulou, S. and Galeotta, S. and Ganga, K. and Gozaliasl, G. and Hook, I. M. and Keihanen, E. and Kirkpatrick, C. C. and Liebing, P. and Lindholm, V. and Mainetti, G. and Martinelli, M. and Martinet, N. and Maturi, M. and McCracken, H. J. and Metcalf, R. B. and Morgante, G. and Nightingale, J. and Nucita, A. and Patrizii, L. and Potter, D. and Riccio, G. and Sánchez, A. G. and Sapone, D. and Schewtschenko, J. A. and Schultheis, M. and Scottez, V. and Teyssier, R. and Tutusaus, I. and Valiviita, J. and Viel, M. and Vriend, W. and Whittaker, L.}, year={2022}, month=jun, pages={A112}, language={en} }

@ARTICLE{Bellhouse2025,
       author = {{Euclid Collaboration, Bellhouse}, C. and {Euclid Collaboration} and  {Golden-Marx}, J.~B. and {Bamford}, S.~P. and {Hatch}, N.~A. and {Kluge}, M. and {Ellien}, A. and {Ahad}, S.~L. and {Dimauro}, P. and {Durret}, F. and {Gonzalez}, A.~H. and {Jimenez-Teja}, Y. and {Montes}, M. and {Sereno}, M. and {Slezak}, E. and {Bolzonella}, M. and {Castignani}, G. and {Cucciati}, O. and {De Lucia}, G. and {Ghaffari}, Z. and {Moscardini}, L. and {Pello}, R. and {Pozzetti}, L. and {Saifollahi}, T. and {Borlaff}, A.~S. and {Aghanim}, N. and {Altieri}, B. and {Amara}, A. and {Andreon}, S. and {Baccigalupi}, C. and {Baldi}, M. and {Bardelli}, S. and {Basset}, A. and {Battaglia}, P. and {Bender}, R. and {Bonino}, D. and {Branchini}, E. and {Brescia}, M. and {Caillat}, A. and {Camera}, S. and {Capobianco}, V. and {Carbone}, C. and {Cardone}, V.~F. and {Carretero}, J. and {Casas}, S. and {Castellano}, M. and {Cavuoti}, S. and {Cimatti}, A. and {Colodro-Conde}, C. and {Congedo}, G. and {Conselice}, C.~J. and {Conversi}, L. and {Copin}, Y. and {Courbin}, F. and {Courtois}, H.~M. and {Cuillandre}, J.-C. and {Da Silva}, A. and {Degaudenzi}, H. and {Di Giorgio}, A.~M. and {Dinis}, J. and {Dubath}, F. and {Duncan}, C.~A.~J. and {Dupac}, X. and {Dusini}, S. and {Farina}, M. and {Farrens}, S. and {Faustini}, F. and {Ferriol}, S. and {Fotopoulou}, S. and {Frailis}, M. and {Franceschi}, E. and {Fumana}, M. and {Galeotta}, S. and {George}, K. and {Gillis}, B. and {Giocoli}, C. and {G{\'o}mez-Alvarez}, P. and {Grazian}, A. and {Grupp}, F. and {Haugan}, S.~V.~H. and {Hoekstra}, H. and {Holliman}, M.~S. and {Holmes}, W. and {Hook}, I. and {Hormuth}, F. and {Hornstrup}, A. and {Hudelot}, P. and {Jahnke}, K. and {Jhabvala}, M. and {Keih{\"a}nen}, E. and {Kermiche}, S. and {Kiessling}, A. and {Kilbinger}, M. and {Kubik}, B. and {K{\"u}mmel}, M. and {Kunz}, M. and {Kurki-Suonio}, H. and {Liebing}, P. and {Ligori}, S. and {Lilje}, P.~B. and {Lindholm}, V. and {Lloro}, I. and {Mainetti}, G. and {Maino}, D. and {Maiorano}, E. and {Mansutti}, O. and {Marggraf}, O. and {Markovic}, K. and {Martinelli}, M. and {Martinet}, N. and {Marulli}, F. and {Massey}, R. and {Maurogordato}, S. and {Medinaceli}, E. and {Mei}, S. and {Melchior}, M. and {Meneghetti}, M. and {Merlin}, E. and {Meylan}, G. and {Moresco}, M. and {Nakajima}, R. and {Neissner}, C. and {Niemi}, S.-M. and {Padilla}, C. and {Paltani}, S. and {Pasian}, F. and {Pedersen}, K. and {Pettorino}, V. and {Pires}, S. and {Polenta}, G. and {Poncet}, M. and {Popa}, L.~A. and {Raison}, F. and {Renzi}, A. and {Rhodes}, J. and {Riccio}, G. and {Romelli}, E. and {Roncarelli}, M. and {Rossetti}, E. and {Saglia}, R. and {Sakr}, Z. and {Sapone}, D. and {Sartoris}, B. and {Schneider}, P. and {Schrabback}, T. and {Seidel}, G. and {Serrano}, S. and {Sirignano}, C. and {Sirri}, G. and {Stanco}, L. and {Steinwagner}, J. and {Tallada-Cresp{\'\i}}, P. and {Tereno}, I. and {Toledo-Moreo}, R. and {Torradeflot}, F. and {Tsyganov}, A. and {Tutusaus}, I. and {Valenziano}, L. and {Vassallo}, T. and {Verdoes Kleijn}, G. and {Veropalumbo}, A. and {Wang}, Y. and {Weller}, J. and {Zamorani}, G. and {Zucca}, E. and {Biviano}, A. and {Bozzo}, E. and {Burigana}, C. and {Calabrese}, M. and {Di Ferdinando}, D. and {Escartin Vigo}, J.~A. and {Farinelli}, R. and {Finelli}, F. and {Gabarra}, L. and {Gracia-Carpio}, J. and {Matthew}, S. and {Mauri}, N. and {Mora}, A. and {P{\"o}ntinen}, M. and {Scottez}, V. and {Simon}, P. and {Tenti}, M. and {Viel}, M. and {Wiesmann}, M. and {Akrami}, Y. and {Andika}, I.~T. and {Anselmi}, S. and {Archidiacono}, M. and {Atrio-Barandela}, F. and {Ballardini}, M. and {Bethermin}, M. and {Blanchard}, A. and {Blot}, L. and {B{\"o}hringer}, H. and {Borgani}, S. and {Brown}, M.~L. and {Bruton}, S. and {Cabanac}, R. and {Calabro}, A. and {Ca{\~n}as-Herrera}, G.},
        title = "{Euclid preparation: LXX. Forecasting detection limits for intracluster light in the Euclid Wide Survey}",
      journal = {\aap},
         year = 2025,
        month = jun,
       volume = {698},
          eid = {A14},
        pages = {A14},
          doi = {10.1051/0004-6361/202553887},
archivePrefix = {arXiv},
       eprint = {2503.17455},
 primaryClass = {astro-ph.GA},
       adsurl = {https://ui.adsabs.harvard.edu/abs/2025A&A...698A..14E}
}

@ARTICLE{Englert2025,
       author = {{Englert}, Anthony M. and {Dell'Antonio}, Ian and {Montes}, Mireia},
        title = "{The Intracluster Light of Abell 3667: Unveiling an Optical Bridge in LSST Precursor Data}",
      journal = {\apjl},
         year = 2025,
        month = aug,
       volume = {989},
       number = {1},
          eid = {L2},
        pages = {L2},
          doi = {10.3847/2041-8213/ade8f1},
archivePrefix = {arXiv},
       eprint = {2505.23551},
 primaryClass = {astro-ph.CO},
       adsurl = {https://ui.adsabs.harvard.edu/abs/2025ApJ...989L...2E}
}

@ARTICLE{Alonso-Asensio2020,
       author = {{Alonso Asensio}, Isaac and {Dalla Vecchia}, Claudio and {Bah{\'e}}, Yannick M. and {Barnes}, David J. and {Kay}, Scott T.},
        title = "{The intracluster light as a tracer of the total matter density distribution: a view from simulations}",
      journal = {\mnras},
         year = 2020,
        month = may,
       volume = {494},
       number = {2},
        pages = {1859-1864},
          doi = {10.1093/mnras/staa861},
archivePrefix = {arXiv},
       eprint = {2003.04662},
 primaryClass = {astro-ph.GA},
       adsurl = {https://ui.adsabs.harvard.edu/abs/2020MNRAS.494.1859A}
}

@ARTICLE{Contreras-Santos2024,
       author = {{Contreras-Santos}, A. and {Knebe}, A. and {Cui}, W. and {Alonso Asensio}, I. and {Dalla Vecchia}, C. and {Ca{\~n}as}, R. and {Haggar}, R. and {Mostoghiu Paun}, R.~A. and {Pearce}, F.~R. and {Rasia}, E.},
        title = "{Characterising the intra-cluster light in The Three Hundred simulations}",
      journal = {\aap},
         year = 2024,
        month = mar,
       volume = {683},
          eid = {A59},
        pages = {A59},
          doi = {10.1051/0004-6361/202348474},
archivePrefix = {arXiv},
       eprint = {2401.08283},
 primaryClass = {astro-ph.CO},
       adsurl = {https://ui.adsabs.harvard.edu/abs/2024A&A...683A..59C}
}

@ARTICLE{Contreras-Santos2025,
       author = {{Contreras-Santos}, A. and {Knebe}, A. and {Cui}, W. and {Alonso Asensio}, I. and {Dalla Vecchia}, C. and {Haggar}, R. and {Mostoghiu Paun}, R.~A. and {Pearce}, F.~R. and {Rasia}, E. and {Martin}, G. and {Nuza}, S.~E. and {Yepes}, G.},
        title = "{The origin of the intracluster light in The Three Hundred simulations}",
      journal = {\aap},
         year = 2025,
        month = nov,
       volume = {703},
          eid = {A85},
        pages = {A85},
          doi = {10.1051/0004-6361/202554248},
archivePrefix = {arXiv},
       eprint = {2509.17831},
 primaryClass = {astro-ph.CO},
       adsurl = {https://ui.adsabs.harvard.edu/abs/2025A&A...703A..85C}
}

@ARTICLE{Diego2023,
       author = {{Diego}, J.~M. and {Pascale}, M. and {Frye}, B. and {Zitrin}, A. and {Broadhurst}, T. and {Mahler}, G. and {Caminha}, G.~B. and {Jauzac}, M. and {Lee}, M.~G. and {Bae}, J.~H. and {Jang}, I.~S. and {Montes}, M.},
        title = "{Exploring the correlation between dark matter, intracluster light, and globular cluster distribution in SMACS0723}",
      journal = {\aap},
         year = 2023,
        month = nov,
       volume = {679},
          eid = {A159},
        pages = {A159},
          doi = {10.1051/0004-6361/202345868},
archivePrefix = {arXiv},
       eprint = {2301.03629},
 primaryClass = {astro-ph.CO}
}

@ARTICLE{Diego2024,
       author = {{Diego}, Jose M. and {Adams}, Nathan J. and {Willner}, Steven P. and {Harvey}, Tom and {Broadhurst}, Tom and {Cohen}, Seth H. and {Jansen}, Rolf A. and {Summers}, Jake and {Windhorst}, Rogier A. and {D'Silva}, Jordan C.~J. and {Koekemoer}, Anton M. and {Coe}, Dan and {Conselice}, Christopher J. and {Driver}, Simon P. and {Frye}, Brenda and {Grogin}, Norman A. and {Marshall}, Madeline A. and {Nonino}, Mario and {Ortiz}, Rafael and {Pirzkal}, Nor and {Robotham}, Aaron and {Ryan}, Russell E. and {Willmer}, Christopher N.~A. and {Yan}, Haojing and {Sun}, Fengwu and {Hainline}, Kevin and {Berkheimer}, Jessica and {Polletta}, Maria del Carmen and {Zitrin}, Adi},
        title = "{JWST's PEARLS: 119 multiply imaged galaxies behind MACS0416, lensing properties of caustic crossing galaxies, and the relation between halo mass and number of globular clusters at z = 0.4}",
      journal = {\aap},
         year = 2024,
        month = oct,
       volume = {690},
          eid = {A114},
        pages = {A114},
          doi = {10.1051/0004-6361/202349119},
archivePrefix = {arXiv},
       eprint = {2312.11603},
 primaryClass = {astro-ph.CO}
}

@ARTICLE{Chen2022,
       author = {{Chen}, Xiaokai and {Zu}, Ying and {Shao}, Zhiwei and {Shan}, Huanyuan},
        title = "{The sphere of influence of the bright central galaxies in the diffuse light of SDSS clusters}",
      journal = {\mnras},
         year = 2022,
        month = aug,
       volume = {514},
       number = {2},
        pages = {2692},
          doi = {10.1093/mnras/stac1456},
archivePrefix = {arXiv},
       eprint = {2112.03934},
 primaryClass = {astro-ph.GA}
}

@ARTICLE{Kluge2025,
       author = {{Kluge}, M. and {Hatch}, N.~A. and {Montes}, M. and {Golden-Marx}, J.~B. and {Gonzalez}, A.~H. and {Cuillandre}, J.-C. and {Bolzonella}, M. and {Lan{\c{c}}on}, A. and {Laureijs}, R. and {Saifollahi}, T. and {Schirmer}, M. and {Stone}, C. and {Boselli}, A. and {Cantiello}, M. and {Sorce}, J.~G. and {Marleau}, F.~R. and {Duc}, P.-A. and {Sola}, E. and {Urbano}, M. and {Ahad}, S.~L. and {Bah{\'e}}, Y.~M. and {Bamford}, S.~P. and {Bellhouse}, C. and {Buitrago}, F. and {Dimauro}, P. and {Durret}, F. and {Ellien}, A. and {Jimenez-Teja}, Y. and {Slezak}, E. and {Aghanim}, N. and {Altieri}, B. and {Andreon}, S. and {Auricchio}, N. and {Baldi}, M. and {Balestra}, A. and {Bardelli}, S. and {Bender}, R. and {Bonino}, D. and {Branchini}, E. and {Brescia}, M. and {Brinchmann}, J. and {Camera}, S. and {Candini}, G.~P. and {Capobianco}, V. and {Carbone}, C. and {Carretero}, J. and {Casas}, S. and {Castellano}, M. and {Cavuoti}, S. and {Cimatti}, A. and {Congedo}, G. and {Conselice}, C.~J. and {Conversi}, L. and {Copin}, Y. and {Courbin}, F. and {Courtois}, H.~M. and {Cropper}, M. and {Da Silva}, A. and {Degaudenzi}, H. and {Dinis}, J. and {Duncan}, C.~A.~J. and {Dupac}, X. and {Dusini}, S. and {Farina}, M. and {Farrens}, S. and {Ferriol}, S. and {Fosalba}, P. and {Frailis}, M. and {Franceschi}, E. and {Fumana}, M. and {Galeotta}, S. and {Garilli}, B. and {Gillard}, W. and {Gillis}, B. and {Giocoli}, C. and {G{\'o}mez-Alvarez}, P. and {Granett}, B.~R. and {Grazian}, A. and {Grupp}, F. and {Guzzo}, L. and {Haugan}, S.~V.~H. and {Hoar}, J. and {Hoekstra}, H. and {Holmes}, W. and {Hook}, I. and {Hormuth}, F. and {Hornstrup}, A. and {Hudelot}, P. and {Jahnke}, K. and {Keih{\"a}nen}, E. and {Kermiche}, S. and {Kiessling}, A. and {Kitching}, T. and {Kohley}, R. and {Kubik}, B. and {K{\"u}mmel}, M. and {Kunz}, M. and {Kurki-Suonio}, H. and {Lahav}, O. and {Ligori}, S. and {Lilje}, P.~B. and {Lindholm}, V. and {Lloro}, I. and {Maiorano}, E. and {Mansutti}, O. and {Marggraf}, O. and {Markovic}, K. and {Martinet}, N. and {Marulli}, F. and {Massey}, R. and {Maurogordato}, S. and {McCracken}, H.~J. and {Medinaceli}, E. and {Mei}, S. and {Melchior}, M. and {Mellier}, Y. and {Meneghetti}, M. and {Merlin}, E. and {Meylan}, G. and {Moresco}, M. and {Moscardini}, L. and {Munari}, E. and {Nichol}, R.~C. and {Niemi}, S.-M. and {Nightingale}, J.~W. and {Padilla}, C. and {Paltani}, S. and {Pasian}, F. and {Pedersen}, K. and {Percival}, W.~J. and {Pettorino}, V. and {Pires}, S. and {Polenta}, G. and {Poncet}, M. and {Popa}, L.~A. and {Pozzetti}, L. and {Racca}, G.~D. and {Raison}, F. and {Rebolo}, R. and {Renzi}, A. and {Rhodes}, J. and {Riccio}, G. and {Rix}, H.-W. and {Romelli}, E. and {Roncarelli}, M. and {Rossetti}, E. and {Saglia}, R. and {Sapone}, D. and {Sartoris}, B. and {Sauvage}, M. and {Scaramella}, R. and {Schneider}, P. and {Schrabback}, T. and {Secroun}, A. and {Seidel}, G. and {Seiffert}, M. and {Serrano}, S. and {Sirignano}, C. and {Sirri}, G. and {Skottfelt}, J. and {Stanco}, L. and {Tallada-Cresp{\'\i}}, P. and {Taylor}, A.~N. and {Teplitz}, H.~I. and {Tereno}, I. and {Toledo-Moreo}, R. and {Torradeflot}, F. and {Tutusaus}, I. and {Valentijn}, E.~A. and {Valenziano}, L. and {Vassallo}, T. and {Verdoes Kleijn}, G. and {Veropalumbo}, A. and {Wang}, Y. and {Weller}, J. and {Williams}, O.~R. and {Zamorani}, G. and {Zucca}, E. and {Biviano}, A. and {Burigana}, C. and {De Lucia}, G. and {George}, K. and {Scottez}, V. and {Simon}, P. and {Mora}, A. and {Mart{\'\i}n-Fleitas}, J. and {Ruppin}, F. and {Scott}, D.},
        title = "{Euclid: Early Release Observations {\textendash} The intracluster light and intracluster globular clusters of the Perseus cluster}",
      journal = {\aap},
         year = 2025,
        month = may,
       volume = {697},
          eid = {A13},
        pages = {A13},
          doi = {10.1051/0004-6361/202450772},
archivePrefix = {arXiv},
       eprint = {2405.13503},
 primaryClass = {astro-ph.GA},
       adsurl = {https://ui.adsabs.harvard.edu/abs/2025A&A...697A..13K}
}

@ARTICLE{West1995,
       author = {{West}, Michael J. and {Cote}, Patrick and {Jones}, Christine and {Forman}, William and {Marzke}, Rondald O.},
        title = "{Intracluster Globular Clusters}",
      journal = {\apjl},
         year = 1995,
        month = nov,
       volume = {453},
        pages = {L77},
          doi = {10.1086/309748},
archivePrefix = {arXiv},
       eprint = {astro-ph/9508141},
 primaryClass = {astro-ph},
       adsurl = {https://ui.adsabs.harvard.edu/abs/1995ApJ...453L..77W}
}

@ARTICLE{Chandrasekhar1943,
       author = {{Chandrasekhar}, S.},
        title = "{Dynamical Friction. I. General Considerations: the Coefficient of Dynamical Friction.}",
      journal = {\apj},
         year = 1943,
        month = mar,
       volume = {97},
        pages = {255},
          doi = {10.1086/144517}
}

@ARTICLE{Jeon2025,
       author = {{Jeon}, Seyoung and {Contini}, Emanuele and {Han}, San and {Rhee}, Jinsu and {Martin}, Garreth and {Kim}, Juhan and {Lee}, Jaehyun and {Kimm}, Taysun and {Pichon}, Christophe and {Byun}, Gyeong-Hwan and {Dubois}, Yohan and {Cadiou}, Corentin and {Jang}, J.~K. and {Yi}, Sukyoung K.},
        title = "{On the Origin of Intracluster Light Based on the High-resolution Simulation, NEWCLUSTER}",
      journal = {\apj},
         year = 2026,
        month = feb,
       volume = {998},
       number = {1},
          eid = {30},
        pages = {30},
          doi = {10.3847/1538-4357/ae2eaa},
archivePrefix = {arXiv},
       eprint = {2512.06098},
 primaryClass = {astro-ph.GA},
       adsurl = {https://ui.adsabs.harvard.edu/abs/2026ApJ...998...30J}
}

@ARTICLE{Han2025,
       author = {{Han}, S. and {Yi}, S.~K. and {Dubois}, Y. and {Rhee}, J. and {Jeon}, S. and {Jang}, J.~K. and {Byun}, G.-H. and {Cadiou}, C. and {Kim}, J. and {Kimm}, T. and {Pichon}, C.},
        title = "{Introducing NewCluster: First half of the history of a high-resolution cluster simulation}",
      journal = {\aap},
         year = 2026,
        month = jan,
       volume = {705},
          eid = {A169},
        pages = {A169},
          doi = {10.1051/0004-6361/202556291},
archivePrefix = {arXiv},
       eprint = {2507.06301},
 primaryClass = {astro-ph.GA},
       adsurl = {https://ui.adsabs.harvard.edu/abs/2026A&A...705A.169H}
}

@ARTICLE{Chiang2025,
       author = {{Chiang}, Barry T. and {van den Bosch}, Frank C. and {Schive}, Hsi-Yu},
        title = "{Universal numerical convergence criteria for subhalo tidal evolution}",
      journal = {The Open Journal of Astrophysics},
         year = 2026,
        month = jan,
       volume = {9},
        pages = {55367},
          doi = {10.33232/001c.155367},
archivePrefix = {arXiv},
       eprint = {2510.26901},
 primaryClass = {astro-ph.CO},
       adsurl = {https://ui.adsabs.harvard.edu/abs/2026OJAp....955367C}
}

@ARTICLE{Lovell2025,
       author = {{Lovell}, Mark R. and {Pillepich}, Annalisa and {Engler}, Christoph and {Nelson}, Dylan and {Ramesh}, Rahul and {Springel}, Volker and {Hernquist}, Lars},
        title = "{Numerical effects on the stripping of dark matter and stars in IllustrisTNG galaxy groups and clusters}",
      journal = {\mnras},
         year = 2025,
        month = dec,
       volume = {544},
       number = {4},
        pages = {4367-4389},
          doi = {10.1093/mnras/staf2012},
archivePrefix = {arXiv},
       eprint = {2509.07078},
 primaryClass = {astro-ph.GA},
       adsurl = {https://ui.adsabs.harvard.edu/abs/2025MNRAS.544.4367L}
}

@ARTICLE{Taffoni2003,
       author = {{Taffoni}, Giuliano and {Mayer}, Lucio and {Colpi}, Monica and {Governato}, Fabio},
        title = "{On the life and death of satellite haloes}",
      journal = {\mnras},
         year = 2003,
        month = may,
       volume = {341},
       number = {2},
        pages = {434-448},
          doi = {10.1046/j.1365-8711.2003.06395.x},
archivePrefix = {arXiv},
       eprint = {astro-ph/0301271},
 primaryClass = {astro-ph},
       adsurl = {https://ui.adsabs.harvard.edu/abs/2003MNRAS.341..434T}
}

@ARTICLE{Planck2014,
       author = {{Planck Collaboration XVI}},
        title = "{Planck 2013 results. XVI. Cosmological parameters}",
      journal = {\aap},
         year = 2014,
        month = nov,
       volume = {571},
          eid = {A16},
        pages = {A16},
          doi = {10.1051/0004-6361/201321591},
archivePrefix = {arXiv},
       eprint = {1303.5076},
 primaryClass = {astro-ph.CO},
       adsurl = {https://ui.adsabs.harvard.edu/abs/2014A&A...571A..16P}
}

@ARTICLE{Marinacci2018,
       author = {{Marinacci}, Federico and {Vogelsberger}, Mark and {Pakmor}, R{\"u}diger and {Torrey}, Paul and {Springel}, Volker and {Hernquist}, Lars and {Nelson}, Dylan and {Weinberger}, Rainer and {Pillepich}, Annalisa and {Naiman}, Jill and {Genel}, Shy},
        title = "{First results from the IllustrisTNG simulations: radio haloes and magnetic fields}",
      journal = {\mnras},
         year = 2018,
        month = nov,
       volume = {480},
       number = {4},
        pages = {5113-5139},
          doi = {10.1093/mnras/sty2206},
archivePrefix = {arXiv},
       eprint = {1707.03396},
 primaryClass = {astro-ph.CO},
       adsurl = {https://ui.adsabs.harvard.edu/abs/2018MNRAS.480.5113M}
}

@ARTICLE{Naiman2018,
       author = {{Naiman}, Jill P. and {Pillepich}, Annalisa and {Springel}, Volker and {Ramirez-Ruiz}, Enrico and {Torrey}, Paul and {Vogelsberger}, Mark and {Pakmor}, R{\"u}diger and {Nelson}, Dylan and {Marinacci}, Federico and {Hernquist}, Lars and {Weinberger}, Rainer and {Genel}, Shy},
        title = "{First results from the IllustrisTNG simulations: a tale of two elements - chemical evolution of magnesium and europium}",
      journal = {\mnras},
         year = 2018,
        month = jun,
       volume = {477},
       number = {1},
        pages = {1206-1224},
          doi = {10.1093/mnras/sty618},
archivePrefix = {arXiv},
       eprint = {1707.03401},
 primaryClass = {astro-ph.GA},
       adsurl = {https://ui.adsabs.harvard.edu/abs/2018MNRAS.477.1206N}
}

@ARTICLE{Springel2018,
       author = {{Springel}, Volker and {Pakmor}, R{\"u}diger and {Pillepich}, Annalisa and {Weinberger}, Rainer and {Nelson}, Dylan and {Hernquist}, Lars and {Vogelsberger}, Mark and {Genel}, Shy and {Torrey}, Paul and {Marinacci}, Federico and {Naiman}, Jill},
        title = "{First results from the IllustrisTNG simulations: matter and galaxy clustering}",
      journal = {\mnras},
         year = 2018,
        month = mar,
       volume = {475},
       number = {1},
        pages = {676-698},
          doi = {10.1093/mnras/stx3304},
archivePrefix = {arXiv},
       eprint = {1707.03397},
 primaryClass = {astro-ph.GA},
       adsurl = {https://ui.adsabs.harvard.edu/abs/2018MNRAS.475..676S}
}

@ARTICLE{Nelson2018,
       author = {{Nelson}, Dylan and {Pillepich}, Annalisa and {Springel}, Volker and {Weinberger}, Rainer and {Hernquist}, Lars and {Pakmor}, R{\"u}diger and {Genel}, Shy and {Torrey}, Paul and {Vogelsberger}, Mark and {Kauffmann}, Guinevere and {Marinacci}, Federico and {Naiman}, Jill},
        title = "{First results from the IllustrisTNG simulations: the galaxy colour bimodality}",
      journal = {\mnras},
         year = 2018,
        month = mar,
       volume = {475},
       number = {1},
        pages = {624-647},
          doi = {10.1093/mnras/stx3040},
archivePrefix = {arXiv},
       eprint = {1707.03395},
 primaryClass = {astro-ph.GA},
       adsurl = {https://ui.adsabs.harvard.edu/abs/2018MNRAS.475..624N}
}

@ARTICLE{Barnes1986,
       author = {{Barnes}, Josh and {Hut}, Piet},
        title = "{A hierarchical O(N log N) force-calculation algorithm}",
      journal = {\nat},
         year = 1986,
        month = dec,
       volume = {324},
       number = {6096},
        pages = {446-449},
          doi = {10.1038/324446a0},
       adsurl = {https://ui.adsabs.harvard.edu/abs/1986Natur.324..446B}
}

@ARTICLE{Golden-Marx2025,
       author = {{Golden-Marx}, Jesse B. and {Zhang}, Y. and {Ogando}, R.~L.~C. and {Yanny}, B. and {da Silva Pereira}, M.~E. and {Hilton}, M. and {Aguena}, M. and {Allam}, S. and {Andrade-Oliveira}, F. and {Bacon}, D. and {Brooks}, D. and {Carnero Rosell}, A. and {Carretero}, J. and {Cheng}, T.-Y. and {da Costa}, L.~N. and {De Vicente}, J. and {Desai}, S. and {Doel}, P. and {Everett}, S. and {Ferrero}, I. and {Frieman}, J. and {Garc{\'\i}a-Bellido}, J. and {Gatti}, M. and {Giannini}, G. and {Gruen}, D. and {Gruendl}, R.~A. and {Gutierrez}, G. and {Hinton}, S.~R. and {Hollowood}, D.~L. and {Honscheid}, K. and {James}, D.~J. and {Kuehn}, K. and {Lee}, S. and {Mena-Fern{\'a}ndez}, J. and {Menanteau}, F. and {Miquel}, R. and {Mohr}, J. and {Palmese}, A. and {Pieres}, A. and {Plazas Malag{\'o}n}, A.~A. and {Samuroff}, S. and {Sanchez}, E. and {Schubnell}, M. and {Sevilla-Noarbe}, I. and {Smith}, M. and {Suchyta}, E. and {Tarle}, G. and {Vikram}, V. and {Walker}, A.~R. and {Weaverdyck}, N. and {Wiseman}, P.},
        title = "{The hierarchical growth of bright central galaxies and intracluster light as traced by the magnitude gap}",
      journal = {\mnras},
         year = 2025,
        month = apr,
       volume = {538},
       number = {2},
        pages = {622-638},
          doi = {10.1093/mnras/staf277},
archivePrefix = {arXiv},
       eprint = {2409.02184},
 primaryClass = {astro-ph.GA},
       adsurl = {https://ui.adsabs.harvard.edu/abs/2025MNRAS.538..622G}
}

@ARTICLE{Canepa2025,
       author = {{Canepa}, Louisa and {Brough}, Sarah and {Lanusse}, Francois and {Montes}, Mireia and {Hatch}, Nina},
        title = "{Measuring the Intracluster Light Fraction with Machine Learning}",
      journal = {\apj},
         year = 2025,
        month = feb,
       volume = {980},
       number = {2},
          eid = {245},
        pages = {245},
          doi = {10.3847/1538-4357/adabc7},
archivePrefix = {arXiv},
       eprint = {2501.08378},
 primaryClass = {astro-ph.GA},
       adsurl = {https://ui.adsabs.harvard.edu/abs/2025ApJ...980..245C}
}

@ARTICLE{Kimmig2025,
       author = {{Kimmig}, Lucas C. and {Brough}, Sarah and {Dolag}, Klaus and {Remus}, Rhea-Silvia and {Bah{\'e}}, Yannick M. and {Martin}, Garreth and {Pillepich}, Annalisa and {Hatch}, Nina and {Montes}, Mireia and {Lammim Ahad}, Syeda and {Bellhouse}, Callum and {Brown}, Harley J. and {Ellien}, Ama{\"e}l and {Golden-Marx}, Jesse B. and {Gonzalez}, Anthony H. and {Iodice}, Enrica and {Jim{\'e}nez-Teja}, Yolanda and {Kluge}, Matthias and {Knapen}, Johan H. and {Mihos}, J. Christopher and {Ragusa}, Rossella and {Spavone}, Marilena},
        title = "{Intra-cluster light as a dynamical clock for galaxy clusters: Insights from the MAGNETICUM, IllustrisTNG, Hydrangea, and Horizon-AGN simulations}",
      journal = {\aap},
         year = 2025,
        month = aug,
       volume = {700},
          eid = {A95},
        pages = {A95},
          doi = {10.1051/0004-6361/202554777},
archivePrefix = {arXiv},
       eprint = {2503.20857},
 primaryClass = {astro-ph.GA},
       adsurl = {https://ui.adsabs.harvard.edu/abs/2025A&A...700A..95K}
}

@ARTICLE{Dolag2010,
       author = {{Dolag}, K. and {Murante}, G. and {Borgani}, S.},
        title = "{Dynamical difference between the cD galaxy and the diffuse, stellar component in simulated galaxy clusters}",
      journal = {\mnras},
         year = 2010,
        month = jul,
       volume = {405},
       number = {3},
        pages = {1544-1559},
          doi = {10.1111/j.1365-2966.2010.16583.x},
archivePrefix = {arXiv},
       eprint = {0911.1129},
 primaryClass = {astro-ph.CO},
       adsurl = {https://ui.adsabs.harvard.edu/abs/2010MNRAS.405.1544D}
}

@ARTICLE{McLaughlin1999,
       author = {{McLaughlin}, Dean E.},
        title = "{Evidence in Virgo for the Universal Dark Matter Halo}",
      journal = {\apjl},
         year = 1999,
        month = feb,
       volume = {512},
       number = {1},
        pages = {L9-L12},
          doi = {10.1086/311860},
archivePrefix = {arXiv},
       eprint = {astro-ph/9812242},
 primaryClass = {astro-ph},
       adsurl = {https://ui.adsabs.harvard.edu/abs/1999ApJ...512L...9M}
}

@ARTICLE{Mellier2025,
       author = {{Euclid Collaboration, Mellier}, Y. and {Abdurro'uf} and {Acevedo Barroso}, J.~A. and {Ach{\'u}carro}, A. and {Adamek}, J. and {Adam}, R. and {Addison}, G.~E. and {Aghanim}, N. and {Aguena}, M. and {Ajani}, V. and {Akrami}, Y. and {Al-Bahlawan}, A. and {Alavi}, A. and {Albuquerque}, I.~S. and {Alestas}, G. and {Alguero}, G. and {Allaoui}, A. and {Allen}, S.~W. and {Allevato}, V. and {Alonso-Tetilla}, A.~V. and {Altieri}, B. and {Alvarez-Candal}, A. and {Alvi}, S. and {Amara}, A. and {Amendola}, L. and {Amiaux}, J. and {Andika}, I.~T. and {Andreon}, S. and {Andrews}, A. and {Angora}, G. and {Angulo}, R.~E. and {Annibali}, F. and {Anselmi}, A. and {Anselmi}, S. and {Arcari}, S. and {Archidiacono}, M. and {Aric{\`o}}, G. and {Arnaud}, M. and {Arnouts}, S. and {Asgari}, M. and {Asorey}, J. and {Atayde}, L. and {Atek}, H. and {Atrio-Barandela}, F. and {Aubert}, M. and {Aubourg}, E. and {Auphan}, T. and {Auricchio}, N. and {Aussel}, B. and {Aussel}, H. and {Avelino}, P.~P. and {Avgoustidis}, A. and {Avila}, S. and {Awan}, S. and {Azzollini}, R. and {Baccigalupi}, C. and {Bachelet}, E. and {Bacon}, D. and {Baes}, M. and {Bagley}, M.~B. and {Bahr-Kalus}, B. and {Balaguera-Antolinez}, A. and {Balbinot}, E. and {Balcells}, M. and {Baldi}, M. and {Baldry}, I. and {Balestra}, A. and {Ballardini}, M. and {Ballester}, O. and {Balogh}, M. and {Ba{\~n}ados}, E. and {Barbier}, R. and {Bardelli}, S. and {Baron}, M. and {Barreiro}, T. and {Barrena}, R. and {Barriere}, J.-C. and {Barros}, B.~J. and {Barthelemy}, A. and {Bartolo}, N. and {Basset}, A. and {Battaglia}, P. and {Battisti}, A.~J. and {Baugh}, C.~M. and {Baumont}, L. and {Bazzanini}, L. and {Beaulieu}, J.-P. and {Beckmann}, V. and {Belikov}, A.~N. and {Bel}, J. and {Bellagamba}, F. and {Bella}, M. and {Bellini}, E. and {Benabed}, K. and {Bender}, R. and {Benevento}, G. and {Bennett}, C.~L. and {Benson}, K. and {Bergamini}, P. and {Bermejo-Climent}, J.~R. and {Bernardeau}, F. and {Bertacca}, D. and {Berthe}, M. and {Berthier}, J. and {Bethermin}, M. and {Beutler}, F. and {Bevillon}, C. and {Bhargava}, S. and {Bhatawdekar}, R. and {Bianchi}, D. and {Bisigello}, L. and {Biviano}, A. and {Blake}, R.~P. and {Blanchard}, A. and {Blazek}, J. and {Blot}, L. and {Bosco}, A. and {Bodendorf}, C. and {Boenke}, T. and {B{\"o}hringer}, H. and {Boldrini}, P. and {Bolzonella}, M. and {Bonchi}, A. and {Bonici}, M. and {Bonino}, D. and {Bonino}, L. and {Bonvin}, C. and {Bon}, W. and {Booth}, J.~T. and {Borgani}, S. and {Borlaff}, A.~S. and {Borsato}, E. and {Bose}, B. and {Botticella}, M.~T. and {Boucaud}, A. and {Bouche}, F. and {Boucher}, J.~S. and {Boutigny}, D. and {Bouvard}, T. and {Bouwens}, R. and {Bouy}, H. and {Bowler}, R.~A.~A. and {Bozza}, V. and {Bozzo}, E. and {Branchini}, E. and {Brando}, G. and {Brau-Nogue}, S. and {Brekke}, P. and {Bremer}, M.~N. and {Brescia}, M. and {Breton}, M.-A. and {Brinchmann}, J. and {Brinckmann}, T. and {Brockley-Blatt}, C. and {Brodwin}, M. and {Brouard}, L. and {Brown}, M.~L. and {Bruton}, S. and {Bucko}, J. and {Buddelmeijer}, H. and {Buenadicha}, G. and {Buitrago}, F. and {Burger}, P. and {Burigana}, C. and {Busillo}, V. and {Busonero}, D. and {Cabanac}, R. and {Cabayol-Garcia}, L. and {Cagliari}, M.~S. and {Caillat}, A. and {Caillat}, L. and {Calabrese}, M. and {Calabro}, A. and {Calderone}, G. and {Calura}, F. and {Camacho Quevedo}, B. and {Camera}, S. and {Campos}, L. and {Ca{\~n}as-Herrera}, G. and {Candini}, G.~P. and {Cantiello}, M. and {Capobianco}, V. and {Cappellaro}, E. and {Cappelluti}, N. and {Cappi}, A. and {Caputi}, K.~I. and {Cara}, C. and {Carbone}, C. and {Cardone}, V.~F. and {Carella}, E. and {Carlberg}, R.~G. and {Carle}, M. and {Carminati}, L. and {Caro}, F. and {Carrasco}, J.~M. and {Carretero}, J. and {Carrilho}, P. and {Carron Duque}, J. and {Carry}, B.},
        title = "{Euclid: I. Overview of the Euclid mission}",
      journal = {\aap},
         year = 2025,
        month = may,
       volume = {697},
          eid = {A1},
        pages = {A1},
          doi = {10.1051/0004-6361/202450810},
archivePrefix = {arXiv},
       eprint = {2405.13491},
 primaryClass = {astro-ph.CO},
       adsurl = {https://ui.adsabs.harvard.edu/abs/2025A&A...697A...1E}
}

@ARTICLE{Valluri2007,
       author = {{Valluri}, Monica and {Vass}, Ileana M. and {Kazantzidis}, Stelios and {Kravtsov}, Andrey V. and {Bohn}, Courtlandt L.},
        title = "{On Relaxation Processes in Collisionless Mergers}",
      journal = {\apj},
         year = 2007,
        month = apr,
       volume = {658},
       number = {2},
        pages = {731-747},
          doi = {10.1086/511298},
archivePrefix = {arXiv},
       eprint = {astro-ph/0609612},
 primaryClass = {astro-ph},
       adsurl = {https://ui.adsabs.harvard.edu/abs/2007ApJ...658..731V}
}

@ARTICLE{Zhao2003,
       author = {{Zhao}, D.~H. and {Mo}, H.~J. and {Jing}, Y.~P. and {B{\"o}rner}, G.},
        title = "{The growth and structure of dark matter haloes}",
      journal = {\mnras},
         year = 2003,
        month = feb,
       volume = {339},
       number = {1},
        pages = {12-24},
          doi = {10.1046/j.1365-8711.2003.06135.x},
archivePrefix = {arXiv},
       eprint = {astro-ph/0204108},
 primaryClass = {astro-ph},
       adsurl = {https://ui.adsabs.harvard.edu/abs/2003MNRAS.339...12Z}
}

@ARTICLE{Wechsler2002,
       author = {{Wechsler}, Risa H. and {Bullock}, James S. and {Primack}, Joel R. and {Kravtsov}, Andrey V. and {Dekel}, Avishai},
        title = "{Concentrations of Dark Halos from Their Assembly Histories}",
      journal = {\apj},
         year = 2002,
        month = mar,
       volume = {568},
       number = {1},
        pages = {52-70},
          doi = {10.1086/338765},
archivePrefix = {arXiv},
       eprint = {astro-ph/0108151},
 primaryClass = {astro-ph},
       adsurl = {https://ui.adsabs.harvard.edu/abs/2002ApJ...568...52W}
}

@ARTICLE{Tormen1997,
       author = {{Tormen}, Giuseppe},
        title = "{The rise and fall of satellites in galaxy clusters}",
      journal = {\mnras},
         year = 1997,
        month = sep,
       volume = {290},
       number = {3},
        pages = {411-421},
          doi = {10.1093/mnras/290.3.411},
archivePrefix = {arXiv},
       eprint = {astro-ph/9611078},
 primaryClass = {astro-ph},
       adsurl = {https://ui.adsabs.harvard.edu/abs/1997MNRAS.290..411T}
}

@ARTICLE{Weaver2023,
       author = {{Weaver}, J.~R. and {Davidzon}, I. and {Toft}, S. and {Ilbert}, O. and {McCracken}, H.~J. and {Gould}, K.~M.~L. and {Jespersen}, C.~K. and {Steinhardt}, C. and {Lagos}, C.~D.~P. and {Capak}, P.~L. and {Casey}, C.~M. and {Chartab}, N. and {Faisst}, A.~L. and {Hayward}, C.~C. and {Kartaltepe}, J.~S. and {Kauffmann}, O.~B. and {Koekemoer}, A.~M. and {Kokorev}, V. and {Laigle}, C. and {Liu}, D. and {Long}, A. and {Magdis}, G.~E. and {McPartland}, C.~J.~R. and {Milvang-Jensen}, B. and {Mobasher}, B. and {Moneti}, A. and {Peng}, Y. and {Sanders}, D.~B. and {Shuntov}, M. and {Sneppen}, A. and {Valentino}, F. and {Zalesky}, L. and {Zamorani}, G.},
        title = "{COSMOS2020: The galaxy stellar mass function. The assembly and star formation cessation of galaxies at 0.2< z {\ensuremath{\leq}} 7.5}",
      journal = {\aap},
         year = 2023,
        month = sep,
       volume = {677},
          eid = {A184},
        pages = {A184},
          doi = {10.1051/0004-6361/202245581},
archivePrefix = {arXiv},
       eprint = {2212.02512},
 primaryClass = {astro-ph.GA},
       adsurl = {https://ui.adsabs.harvard.edu/abs/2023A&A...677A.184W}
}

@ARTICLE{Davidzon2017,
       author = {{Davidzon}, I. and {Ilbert}, O. and {Laigle}, C. and {Coupon}, J. and {McCracken}, H.~J. and {Delvecchio}, I. and {Masters}, D. and {Capak}, P. and {Hsieh}, B.~C. and {Le F{\`e}vre}, O. and {Tresse}, L. and {Bethermin}, M. and {Chang}, Y.-Y. and {Faisst}, A.~L. and {Le Floc'h}, E. and {Steinhardt}, C. and {Toft}, S. and {Aussel}, H. and {Dubois}, C. and {Hasinger}, G. and {Salvato}, M. and {Sanders}, D.~B. and {Scoville}, N. and {Silverman}, J.~D.},
        title = "{The COSMOS2015 galaxy stellar mass function . Thirteen billion years of stellar mass assembly in ten snapshots}",
      journal = {\aap},
         year = 2017,
        month = sep,
       volume = {605},
          eid = {A70},
        pages = {A70},
          doi = {10.1051/0004-6361/201730419},
archivePrefix = {arXiv},
       eprint = {1701.02734},
 primaryClass = {astro-ph.GA},
       adsurl = {https://ui.adsabs.harvard.edu/abs/2017A&A...605A..70D}
}

@ARTICLE{McLeod2021,
       author = {{McLeod}, D.~J. and {McLure}, R.~J. and {Dunlop}, J.~S. and {Cullen}, F. and {Carnall}, A.~C. and {Duncan}, K.},
        title = "{The evolution of the galaxy stellar-mass function over the last 12 billion years from a combination of ground-based and HST surveys}",
      journal = {\mnras},
         year = 2021,
        month = may,
       volume = {503},
       number = {3},
        pages = {4413-4435},
          doi = {10.1093/mnras/stab731},
archivePrefix = {arXiv},
       eprint = {2009.03176},
 primaryClass = {astro-ph.GA},
       adsurl = {https://ui.adsabs.harvard.edu/abs/2021MNRAS.503.4413M}
}

@ARTICLE{Ilbert2013,
       author = {{Ilbert}, O. and {McCracken}, H.~J. and {Le F{\`e}vre}, O. and {Capak}, P. and {Dunlop}, J. and {Karim}, A. and {Renzini}, M.~A. and {Caputi}, K. and {Boissier}, S. and {Arnouts}, S. and {Aussel}, H. and {Comparat}, J. and {Guo}, Q. and {Hudelot}, P. and {Kartaltepe}, J. and {Kneib}, J.~P. and {Krogager}, J.~K. and {Le Floc'h}, E. and {Lilly}, S. and {Mellier}, Y. and {Milvang-Jensen}, B. and {Moutard}, T. and {Onodera}, M. and {Richard}, J. and {Salvato}, M. and {Sanders}, D.~B. and {Scoville}, N. and {Silverman}, J.~D. and {Taniguchi}, Y. and {Tasca}, L. and {Thomas}, R. and {Toft}, S. and {Tresse}, L. and {Vergani}, D. and {Wolk}, M. and {Zirm}, A.},
        title = "{Mass assembly in quiescent and star-forming galaxies since z ≃ 4 from UltraVISTA}",
      journal = {\aap},
         year = 2013,
        month = aug,
       volume = {556},
          eid = {A55},
        pages = {A55},
          doi = {10.1051/0004-6361/201321100},
archivePrefix = {arXiv},
       eprint = {1301.3157},
 primaryClass = {astro-ph.CO},
       adsurl = {https://ui.adsabs.harvard.edu/abs/2013A&A...556A..55I}
}

@ARTICLE{Baldry2012,
       author = {{Baldry}, I.~K. and {Driver}, S.~P. and {Loveday}, J. and {Taylor}, E.~N. and {Kelvin}, L.~S. and {Liske}, J. and {Norberg}, P. and {Robotham}, A.~S.~G. and {Brough}, S. and {Hopkins}, A.~M. and {Bamford}, S.~P. and {Peacock}, J.~A. and {Bland-Hawthorn}, J. and {Conselice}, C.~J. and {Croom}, S.~M. and {Jones}, D.~H. and {Parkinson}, H.~R. and {Popescu}, C.~C. and {Prescott}, M. and {Sharp}, R.~G. and {Tuffs}, R.~J.},
        title = "{Galaxy And Mass Assembly (GAMA): the galaxy stellar mass function at z < 0.06}",
      journal = {\mnras},
         year = 2012,
        month = mar,
       volume = {421},
       number = {1},
        pages = {621-634},
          doi = {10.1111/j.1365-2966.2012.20340.x},
archivePrefix = {arXiv},
       eprint = {1111.5707},
 primaryClass = {astro-ph.CO},
       adsurl = {https://ui.adsabs.harvard.edu/abs/2012MNRAS.421..621B}
}

@ARTICLE{Tomczak2014,
       author = {{Tomczak}, Adam R. and {Quadri}, Ryan F. and {Tran}, Kim-Vy H. and {Labb{\'e}}, Ivo and {Straatman}, Caroline M.~S. and {Papovich}, Casey and {Glazebrook}, Karl and {Allen}, Rebecca and {Brammer}, Gabriel B. and {Kacprzak}, Glenn G. and {Kawinwanichakij}, Lalitwadee and {Kelson}, Daniel D. and {McCarthy}, Patrick J. and {Mehrtens}, Nicola and {Monson}, Andrew J. and {Persson}, S. Eric and {Spitler}, Lee R. and {Tilvi}, Vithal and {van Dokkum}, Pieter},
        title = "{Galaxy Stellar Mass Functions from ZFOURGE/CANDELS: An Excess of Low-mass Galaxies since z = 2 and the Rapid Buildup of Quiescent Galaxies}",
      journal = {\apj},
         year = 2014,
        month = mar,
       volume = {783},
       number = {2},
          eid = {85},
        pages = {85},
          doi = {10.1088/0004-637X/783/2/85},
archivePrefix = {arXiv},
       eprint = {1309.5972},
 primaryClass = {astro-ph.CO},
       adsurl = {https://ui.adsabs.harvard.edu/abs/2014ApJ...783...85T}
}

@article{Homan2014,
author = {Homan, Matthew D. and Gelman, Andrew},
title = {The No-U-turn sampler: adaptively setting path lengths in Hamiltonian Monte Carlo},
year = {2014},
issue_date = {January 2014},
publisher = {JMLR.org},
volume = {15},
number = {1},
issn = {1532-4435},
journal = {J. Mach. Learn. Res.},
month = jan,
pages = {1593–1623},
numpages = {31}
}

@ARTICLE{Neal1997,
       author = {{Neal}, Radford M.},
        title = "{Monte Carlo Implementation of Gaussian Process Models for Bayesian Regression and Classification}",
      journal = {arXiv e-prints},
         year = 1997,
        month = jan,
          eid = {physics/9701026},
        pages = {physics/9701026},
          doi = {10.48550/arXiv.physics/9701026},
archivePrefix = {arXiv},
       eprint = {physics/9701026},
 primaryClass = {physics.data-an},
       adsurl = {https://ui.adsabs.harvard.edu/abs/1997physics...1026N}
}

@ARTICLE{Goulard1992,
       author = {{Goulard}, M. and {Voltz}, M.},
        title = "{Linear coregionalization model: Tools for estimation and choice of cross-variogram matrix}",
      journal = {Mathematical Geology},
         year = 1992,
        month = apr,
       volume = {24},
       number = {3},
        pages = {269-286},
          doi = {10.1007/BF00893750},
       adsurl = {https://ui.adsabs.harvard.edu/abs/1992MatG...24..269G}
}

@article{Alvarez2012,
author = {\'{A}lvarez, Mauricio A. and Rosasco, Lorenzo and Lawrence, Neil D.},
title = {Kernels for Vector-Valued Functions: A Review},
year = {2012},
issue_date = {March 2012},
publisher = {Now Publishers Inc.},
address = {Hanover, MA, USA},
volume = {4},
number = {3},
issn = {1935-8237},
url = {https://doi.org/10.1561/2200000036},
doi = {10.1561/2200000036},
journal = {Found. Trends Mach. Learn.},
month = mar,
pages = {195–266},
numpages = {72}
}

@ARTICLE{Reina-Campos2023,
       author = {{Reina-Campos}, Marta and {Trujillo-Gomez}, Sebastian and {Pfeffer}, Joel L. and {Sills}, Alison and {Deason}, Alis J. and {Crain}, Robert A. and {Kruijssen}, J.~M. Diederik},
        title = "{Constraining the shape of dark matter haloes with globular clusters and diffuse stellar light in the E-MOSAICS simulations}",
      journal = {\mnras},
         year = 2023,
        month = jun,
       volume = {521},
       number = {4},
        pages = {6368-6382},
          doi = {10.1093/mnras/stad920},
       adsurl = {https://ui.adsabs.harvard.edu/abs/2023MNRAS.521.6368R}
}

@ARTICLE{Martin2025,
       author = {{Martin}, G. and {Watkins}, A.~E. and {Dubois}, Y. and {Devriendt}, J. and {Kaviraj}, S. and {Kim}, D. and {Kraljic}, K. and {Lazar}, I. and {Pearce}, F.~R. and {Peirani}, S. and {Pichon}, C. and {Slyz}, A. and {Yi}, S.~K.},
        title = "{Cosmic reflections I: the structural diversity of simulated and observed low-mass galaxy analogues}",
      journal = {\mnras},
         year = 2025,
        month = aug,
       volume = {541},
       number = {2},
        pages = {1831-1850},
          doi = {10.1093/mnras/staf1092},
archivePrefix = {arXiv},
       eprint = {2505.04509},
 primaryClass = {astro-ph.GA},
       adsurl = {https://ui.adsabs.harvard.edu/abs/2025MNRAS.541.1831M}
}

@ARTICLE{Schaller2015,
       author = {{Schaller}, Matthieu and {Frenk}, Carlos S. and {Bower}, Richard G. and {Theuns}, Tom and {Trayford}, James and {Crain}, Robert A. and {Furlong}, Michelle and {Schaye}, Joop and {Dalla Vecchia}, Claudio and {McCarthy}, I.~G.},
        title = "{The effect of baryons on the inner density profiles of rich clusters}",
      journal = {\mnras},
         year = 2015,
        month = sep,
       volume = {452},
       number = {1},
        pages = {343-355},
          doi = {10.1093/mnras/stv1341},
archivePrefix = {arXiv},
       eprint = {1409.8297},
 primaryClass = {astro-ph.CO},
       adsurl = {https://ui.adsabs.harvard.edu/abs/2015MNRAS.452..343S}
}

@ARTICLE{Roediger2005,
       author = {{Roediger}, E. and {Hensler}, G.},
        title = "{Ram pressure stripping of disk galaxies. From high to low density environments}",
      journal = {\aap},
         year = 2005,
        month = apr,
       volume = {433},
       number = {3},
        pages = {875-895},
          doi = {10.1051/0004-6361:20042131},
       adsurl = {https://ui.adsabs.harvard.edu/abs/2005A&A...433..875R}
}

@ARTICLE{Tonnesen2009,
       author = {{Tonnesen}, Stephanie and {Bryan}, Greg L.},
        title = "{Gas Stripping in Simulated Galaxies with a Multiphase Interstellar Medium}",
      journal = {\apj},
         year = 2009,
        month = apr,
       volume = {694},
       number = {2},
        pages = {789-804},
          doi = {10.1088/0004-637X/694/2/789},
archivePrefix = {arXiv},
       eprint = {0901.2115},
 primaryClass = {astro-ph.GA},
       adsurl = {https://ui.adsabs.harvard.edu/abs/2009ApJ...694..789T}
}

@ARTICLE{Jaffe2018,
       author = {{Jaff{\'e}}, Yara L. and {Poggianti}, Bianca M. and {Moretti}, Alessia and {Gullieuszik}, Marco and {Smith}, Rory and {Vulcani}, Benedetta and {Fasano}, Giovanni and {Fritz}, Jacopo and {Tonnesen}, Stephanie and {Bettoni}, Daniela and {Hau}, George and {Biviano}, Andrea and {Bellhouse}, Callum and {McGee}, Sean},
        title = "{GASP. IX. Jellyfish galaxies in phase-space: an orbital study of intense ram-pressure stripping in clusters}",
      journal = {\mnras},
         year = 2018,
        month = jun,
       volume = {476},
       number = {4},
        pages = {4753-4764},
          doi = {10.1093/mnras/sty500},
archivePrefix = {arXiv},
       eprint = {1802.07297},
 primaryClass = {astro-ph.GA},
       adsurl = {https://ui.adsabs.harvard.edu/abs/2018MNRAS.476.4753J}
}

@ARTICLE{Fakhouri2010,
       author = {{Fakhouri}, Onsi and {Ma}, Chung-Pei and {Boylan-Kolchin}, Michael},
        title = "{The merger rates and mass assembly histories of dark matter haloes in the two Millennium simulations}",
      journal = {\mnras},
         year = 2010,
        month = aug,
       volume = {406},
       number = {4},
        pages = {2267-2278},
          doi = {10.1111/j.1365-2966.2010.16859.x},
archivePrefix = {arXiv},
       eprint = {1001.2304},
 primaryClass = {astro-ph.CO},
       adsurl = {https://ui.adsabs.harvard.edu/abs/2010MNRAS.406.2267F}
}

@ARTICLE{Klypin2016,
       author = {{Klypin}, Anatoly and {Yepes}, Gustavo and {Gottl{\"o}ber}, Stefan and {Prada}, Francisco and {He{\ss}}, Steffen},
        title = "{MultiDark simulations: the story of dark matter halo concentrations and density profiles}",
      journal = {\mnras},
         year = 2016,
        month = apr,
       volume = {457},
       number = {4},
        pages = {4340-4359},
          doi = {10.1093/mnras/stw248},
archivePrefix = {arXiv},
       eprint = {1411.4001},
 primaryClass = {astro-ph.CO},
       adsurl = {https://ui.adsabs.harvard.edu/abs/2016MNRAS.457.4340K}
}

@ARTICLE{Hough2023,
       author = {{Hough}, Renier T. and {Rennehan}, Douglas and {Kobayashi}, Chiaki and {Loubser}, S. Ilani and {Dav{\'e}}, Romeel and {Babul}, Arif and {Cui}, Weiguang},
        title = "{SIMBA-C: an updated chemical enrichment model for galactic chemical evolution in the SIMBA simulation}",
      journal = {\mnras},
         year = 2023,
        month = oct,
       volume = {525},
       number = {1},
        pages = {1061-1076},
          doi = {10.1093/mnras/stad2394},
archivePrefix = {arXiv},
       eprint = {2308.03436},
 primaryClass = {astro-ph.GA},
       adsurl = {https://ui.adsabs.harvard.edu/abs/2023MNRAS.525.1061H}
}

@ARTICLE{Johnston1998,
       author = {{Johnston}, Kathryn V.},
        title = "{A Prescription for Building the Milky Way's Halo from Disrupted Satellites}",
      journal = {\apj},
         year = 1998,
        month = mar,
       volume = {495},
       number = {1},
        pages = {297-308},
          doi = {10.1086/305273},
archivePrefix = {arXiv},
       eprint = {astro-ph/9710007},
 primaryClass = {astro-ph},
       adsurl = {https://ui.adsabs.harvard.edu/abs/1998ApJ...495..297J}
}

@ARTICLE{Navarro1997,
       author = {{Navarro}, Julio F. and {Frenk}, Carlos S. and {White}, Simon D.~M.},
        title = "{A Universal Density Profile from Hierarchical Clustering}",
      journal = {\apj},
         year = 1997,
        month = dec,
       volume = {490},
       number = {2},
        pages = {493-508},
          doi = {10.1086/304888},
archivePrefix = {arXiv},
       eprint = {astro-ph/9611107},
 primaryClass = {astro-ph},
       adsurl = {https://ui.adsabs.harvard.edu/abs/1997ApJ...490..493N}
}

@ARTICLE{Golden-Marx2023,
       author = {{Golden-Marx}, Jesse B. and {Zhang}, Y. and {Ogando}, R.~L.~C. and {Allam}, S. and {Tucker}, D.~L. and {Miller}, C.~J. and {Hilton}, M. and {Mutlu-Pakdil}, B. and {Abbott}, T.~M.~C. and {Aguena}, M. and {Alves}, O. and {Andrade-Oliveira}, F. and {Annis}, J. and {Bacon}, D. and {Bertin}, E. and {Bocquet}, S. and {Brooks}, D. and {Burke}, D.~L. and {Carnero Rosell}, A. and {Carrasco Kind}, M. and {Castander}, F.~J. and {Conselice}, C. and {Costanzi}, M. and {da Costa}, L.~N. and {Pereira}, M.~E.~S. and {De Vicente}, J. and {Desai}, S. and {Doel}, P. and {Everett}, S. and {Ferrero}, I. and {Flaugher}, B. and {Frieman}, J. and {Garc{\'\i}a-Bellido}, J. and {Gerdes}, D.~W. and {Gruen}, D. and {Gruendl}, R.~A. and {Gutierrez}, G. and {Hinton}, S.~R. and {Hollowood}, D.~L. and {Honscheid}, K. and {James}, D.~J. and {Kuehn}, K. and {Kuropatkin}, N. and {Lahav}, O. and {Marshall}, J.~L. and {Melchior}, P. and {Mena-Fern{\'a}ndez}, J. and {Miquel}, R. and {Mohr}, J.~J. and {Palmese}, A. and {Paz-Chinch{\'o}n}, F. and {Pieres}, A. and {Plazas Malag{\'o}n}, A.~A. and {Prat}, J. and {Raveri}, M. and {Rodriguez-Monroy}, M. and {Romer}, A.~K. and {Sanchez}, E. and {Scarpine}, V. and {Sevilla-Noarbe}, I. and {Sif{\'o}n}, C. and {Smith}, M. and {Suchyta}, E. and {Swanson}, M.~E.~C. and {Tarle}, G. and {Vincenzi}, M. and {Weaverdyck}, N. and {Yanny}, B. and {DES Collaboration}},
        title = "{Characterizing the intracluster light over the redshift range 0.2 < z < 0.8 in the DES-ACT overlap}",
      journal = {\mnras},
         year = 2023,
        month = may,
       volume = {521},
       number = {1},
        pages = {478-496},
          doi = {10.1093/mnras/stad469},
archivePrefix = {arXiv},
       eprint = {2209.05519},
 primaryClass = {astro-ph.GA},
       adsurl = {https://ui.adsabs.harvard.edu/abs/2023MNRAS.521..478G}
}

@ARTICLE{Ellien2025,
       author = {{Ellien}, A. and {Montes}, M. and {Ahad}, S.~L. and {Dimauro}, P. and {Golden-Marx}, J.~B. and {Jimenez-Teja}, Y. and {Durret}, F. and {Bellhouse}, C. and {Diego}, J.~M. and {Bamford}, S.~P. and {Gonzalez}, A.~H. and {Hatch}, N.~A. and {Kluge}, M. and {Ragusa}, R. and {Slezak}, E. and {Cuillandre}, J.-C. and {Gavazzi}, R. and {Dole}, H. and {Mahler}, G. and {Congedo}, G. and {Saifollahi}, T. and {Aghanim}, N. and {Altieri}, B. and {Amara}, A. and {Andreon}, S. and {Auricchio}, N. and {Baccigalupi}, C. and {Baldi}, M. and {Balestra}, A. and {Bardelli}, S. and {Basset}, A. and {Battaglia}, P. and {Biviano}, A. and {Bonchi}, A. and {Bonino}, D. and {Branchini}, E. and {Brescia}, M. and {Brinchmann}, J. and {Caillat}, A. and {Camera}, S. and {Capobianco}, V. and {Carbone}, C. and {Cardone}, V.~F. and {Carretero}, J. and {Casas}, S. and {Castellano}, M. and {Castignani}, G. and {Cavuoti}, S. and {Cimatti}, A. and {Colodro-Conde}, C. and {Conselice}, C.~J. and {Conversi}, L. and {Copin}, Y. and {Courbin}, F. and {Courtois}, H.~M. and {Cropper}, M. and {Da Silva}, A. and {Degaudenzi}, H. and {De Lucia}, G. and {Di Giorgio}, A.~M. and {Dinis}, J. and {Dubath}, F. and {Duncan}, C.~A.~J. and {Dupac}, X. and {Dusini}, S. and {Farina}, M. and {Faustini}, F. and {Ferriol}, S. and {Fotopoulou}, S. and {Frailis}, M. and {Franceschi}, E. and {Galeotta}, S. and {George}, K. and {Gillis}, B. and {Giocoli}, C. and {G{\'o}mez-Alvarez}, P. and {Grazian}, A. and {Grupp}, F. and {Guzzo}, L. and {Haugan}, S.~V.~H. and {Hoar}, J. and {Hoekstra}, H. and {Holmes}, W. and {Hormuth}, F. and {Hornstrup}, A. and {Hudelot}, P. and {Jahnke}, K. and {Jhabvala}, M. and {Joachimi}, B. and {Keih{\"a}nen}, E. and {Kermiche}, S. and {Kiessling}, A. and {Kubik}, B. and {Kuijken}, K. and {K{\"u}mmel}, M. and {Kunz}, M. and {Kurki-Suonio}, H. and {Laureijs}, R. and {Le Mignant}, D. and {Ligori}, S. and {Lilje}, P.~B. and {Lindholm}, V. and {Lloro}, I. and {Mainetti}, G. and {Maino}, D. and {Maiorano}, E. and {Mansutti}, O. and {Marcin}, S. and {Marggraf}, O. and {Markovic}, K. and {Martinelli}, M. and {Martinet}, N. and {Marulli}, F. and {Massey}, R. and {Maurogordato}, S. and {Medinaceli}, E. and {Mei}, S. and {Melchior}, M. and {Mellier}, Y. and {Meneghetti}, M. and {Merlin}, E. and {Meylan}, G. and {Mora}, A. and {Moresco}, M. and {Moscardini}, L. and {Nakajima}, R. and {Neissner}, C. and {Nichol}, R.~C. and {Niemi}, S.-M. and {Nightingale}, J.~W. and {Padilla}, C. and {Paltani}, S. and {Pasian}, F. and {Pedersen}, K. and {Percival}, W.~J. and {Pettorino}, V. and {Pires}, S. and {Polenta}, G. and {Poncet}, M. and {Popa}, L.~A. and {Pozzetti}, L. and {Raison}, F. and {Rebolo}, R. and {Renzi}, A. and {Rhodes}, J. and {Riccio}, G. and {Romelli}, E. and {Roncarelli}, M. and {Rossetti}, E. and {Saglia}, R. and {Sakr}, Z. and {Sapone}, D. and {Sartoris}, B. and {Scaramella}, R. and {Schirmer}, M. and {Schneider}, P. and {Schrabback}, T. and {Secroun}, A. and {Sefusatti}, E. and {Seidel}, G. and {Seiffert}, M. and {Serrano}, S. and {Sirignano}, C. and {Sirri}, G. and {Stanco}, L. and {Starck}, J.-L. and {Steinwagner}, J. and {Tallada-Cresp{\'\i}}, P. and {Taylor}, A.~N. and {Teplitz}, H.~I. and {Tereno}, I. and {Toledo-Moreo}, R. and {Torradeflot}, F. and {Tsyganov}, A. and {Tutusaus}, I. and {Valenziano}, L. and {Vassallo}, T. and {Verdoes Kleijn}, G. and {Veropalumbo}, A. and {Wang}, Y. and {Weller}, J. and {Williams}, O.~R. and {Zucca}, E. and {Bolzonella}, M. and {Burigana}, C. and {Scottez}, V.},
        title = "{Euclid: Early Release Observations: The intracluster light of Abell 2390}",
      journal = {\aap},
         year = 2025,
        month = jun,
       volume = {698},
          eid = {A134},
        pages = {A134},
          doi = {10.1051/0004-6361/202554460},
archivePrefix = {arXiv},
       eprint = {2503.07484},
 primaryClass = {astro-ph.GA},
       adsurl = {https://ui.adsabs.harvard.edu/abs/2025A&A...698A.134E}
}

@ARTICLE{Giocoli2010,
       author = {{Giocoli}, Carlo and {Tormen}, Giuseppe and {Sheth}, Ravi K. and {van den Bosch}, Frank C.},
        title = "{The substructure hierarchy in dark matter haloes}",
      journal = {\mnras},
         year = 2010,
        month = may,
       volume = {404},
       number = {1},
        pages = {502-517},
          doi = {10.1111/j.1365-2966.2010.16311.x},
archivePrefix = {arXiv},
       eprint = {0911.0436},
 primaryClass = {astro-ph.CO},
       adsurl = {https://ui.adsabs.harvard.edu/abs/2010MNRAS.404..502G}
}

@ARTICLE{Watkins2025,
       author = {{Watkins}, A.~E. and {Martin}, G. and {Kaviraj}, S. and {Collins}, C. and {Dubois}, Y. and {Kraljic}, K. and {Pichon}, C. and {Yi}, S.~K.},
        title = "{2D light distributions of dwarf galaxies - key tests of the implementation of physical processes in simulations}",
      journal = {\mnras},
         year = 2025,
        month = mar,
       volume = {537},
       number = {4},
        pages = {3499-3510},
          doi = {10.1093/mnras/staf223},
archivePrefix = {arXiv},
       eprint = {2502.02632},
 primaryClass = {astro-ph.GA},
       adsurl = {https://ui.adsabs.harvard.edu/abs/2025MNRAS.537.3499W}
}

@ARTICLE{Onions2025,
       author = {{Onions}, Julian and {Pearce}, Frazer and {Knebe}, Alexander and {Gray}, Meghan and {Haggar}, Roan and {Kuchner}, Ulrike and {Contreras-Santos}, Ana and {Yepes}, Gustavo and {Cui}, Weiguang},
        title = "{The life and times of dark matter haloes: what will I be when I grow up?}",
      journal = {\mnras},
         year = 2025,
        month = sep,
       volume = {542},
       number = {2},
        pages = {1477-1485},
          doi = {10.1093/mnras/staf1293},
archivePrefix = {arXiv},
       eprint = {2508.18778},
 primaryClass = {astro-ph.GA},
       adsurl = {https://ui.adsabs.harvard.edu/abs/2025MNRAS.542.1477O}
}

@ARTICLE{Manuwal2025,
       author = {{Manuwal}, Aditya and {Avila-Reese}, Vladimir and {Montenegro-Taborda}, Daniel and {Rodriguez-Gomez}, Vicente and {Cervantes Sodi}, Bernardo},
        title = "{Inferring the dark matter distribution of massive galaxy clusters from deep optical observations: insights from the TNG300 simulation}",
      journal = {\mnras},
         year = 2025,
        month = nov,
       volume = {543},
       number = {4},
        pages = {4020-4041},
          doi = {10.1093/mnras/staf1717},
archivePrefix = {arXiv},
       eprint = {2510.03424},
 primaryClass = {astro-ph.GA},
       adsurl = {https://ui.adsabs.harvard.edu/abs/2025MNRAS.543.4020M}
}

@ARTICLE{Genel2010,
       author = {{Genel}, Shy and {Bouch{\'e}}, Nicolas and {Naab}, Thorsten and {Sternberg}, Amiel and {Genzel}, Reinhard},
        title = "{The Growth of Dark Matter Halos: Evidence for Significant Smooth Accretion}",
      journal = {\apj},
         year = 2010,
        month = aug,
       volume = {719},
       number = {1},
        pages = {229-239},
          doi = {10.1088/0004-637X/719/1/229},
archivePrefix = {arXiv},
       eprint = {1005.4058},
 primaryClass = {astro-ph.CO},
       adsurl = {https://ui.adsabs.harvard.edu/abs/2010ApJ...719..229G}
}

@ARTICLE{Alonso-Asensio2025,
       author = {{Alonso Asensio}, I. and {Contreras-Santos}, A.},
        title = "{The intracluster light as an estimator of the cluster mass profile}",
      journal = {\aap},
         year = 2025,
        month = aug,
       volume = {700},
          eid = {A205},
        pages = {A205},
          doi = {10.1051/0004-6361/202555083},
archivePrefix = {arXiv},
       eprint = {2507.05404},
 primaryClass = {astro-ph.GA},
       adsurl = {https://ui.adsabs.harvard.edu/abs/2025A&A...700A.205A}
}

@ARTICLE{Montenegro-Taborda2025,
       author = {{Montenegro-Taborda}, Daniel and {Avila-Reese}, Vladimir and {Rodriguez-Gomez}, Vicente and {Manuwal}, Aditya and {Cervantes-Sodi}, Bernardo},
        title = "{The stellar mass composition of galaxy clusters and dependencies on dark matter halo properties}",
      journal = {\mnras},
         year = 2025,
        month = mar,
       volume = {537},
       number = {4},
        pages = {3954-3975},
          doi = {10.1093/mnras/staf271},
archivePrefix = {arXiv},
       eprint = {2502.07927},
 primaryClass = {astro-ph.GA},
       adsurl = {https://ui.adsabs.harvard.edu/abs/2025MNRAS.537.3954M}
}

@ARTICLE{Jiang2015,
       author = {{Jiang}, Lilian and {Cole}, Shaun and {Sawala}, Till and {Frenk}, Carlos S.},
        title = "{Orbital parameters of infalling satellite haloes in the hierarchical {\ensuremath{\Lambda}}CDM model}",
      journal = {\mnras},
         year = 2015,
        month = apr,
       volume = {448},
       number = {2},
        pages = {1674-1686},
          doi = {10.1093/mnras/stv053},
archivePrefix = {arXiv},
       eprint = {1409.1179},
 primaryClass = {astro-ph.CO},
       adsurl = {https://ui.adsabs.harvard.edu/abs/2015MNRAS.448.1674J}
}

@ARTICLE{Fernandez2026,
       author = {{Fernandez}, Adela and {Bah{\'e}}, Yannick and {Hatch}, Nina and {Butler}, Joseph and {Kolcu}, Tutku and {Martin}, Garreth and {Montes}, Mireia},
        title = "{Intracluster light is a close tracer of the dark matter halo shape}",
      journal = {arXiv e-prints},
         year = 2026,
        month = mar,
          eid = {arXiv:2603.23158},
        pages = {arXiv:2603.23158},
          doi = {10.48550/arXiv.2603.23158},
archivePrefix = {arXiv},
       eprint = {2603.23158},
 primaryClass = {astro-ph.GA},
       adsurl = {https://ui.adsabs.harvard.edu/abs/2026arXiv260323158F}
}

@ARTICLE{Karademir2019,
       author = {{Karademir}, Geray S. and {Remus}, Rhea-Silvia and {Burkert}, Andreas and {Dolag}, Klaus and {Hoffmann}, Tadziu L. and {Moster}, Benjamin P. and {Steinwandel}, Ulrich P. and {Zhang}, Jielai},
        title = "{The outer stellar halos of galaxies: how radial merger mass deposition, shells, and streams depend on infall-orbit configurations}",
      journal = {\mnras},
         year = 2019,
        month = jul,
       volume = {487},
       number = {1},
        pages = {318-332},
          doi = {10.1093/mnras/stz1251},
archivePrefix = {arXiv},
       eprint = {1808.10454},
 primaryClass = {astro-ph.GA},
       adsurl = {https://ui.adsabs.harvard.edu/abs/2019MNRAS.487..318K}
}

@ARTICLE{Amorisco2017,
       author = {{Amorisco}, N.~C.},
        title = "{Contributions to the accreted stellar halo: an atlas of stellar deposition}",
      journal = {\mnras},
         year = 2017,
        month = jan,
       volume = {464},
       number = {3},
        pages = {2882-2895},
          doi = {10.1093/mnras/stw2229},
archivePrefix = {arXiv},
       eprint = {1511.08806},
 primaryClass = {astro-ph.GA},
       adsurl = {https://ui.adsabs.harvard.edu/abs/2017MNRAS.464.2882A}
}

@ARTICLE{Contini2020,
       author = {{Contini}, E. and {Gu}, Q.},
        title = "{On the Mass Distribution of the Intracluster Light in Galaxy Groups and Clusters}",
      journal = {\apj},
         year = 2020,
        month = oct,
       volume = {901},
       number = {2},
          eid = {128},
        pages = {128},
          doi = {10.3847/1538-4357/abb1aa},
archivePrefix = {arXiv},
       eprint = {2005.13763},
 primaryClass = {astro-ph.GA},
       adsurl = {https://ui.adsabs.harvard.edu/abs/2020ApJ...901..128C}
}

@ARTICLE{Yoo2025,
       author = {{Yoo}, Jaewon and {Shin}, Jihye and {Hwang}, Ho Seong and {Sabiu}, Cristiano G. and {Kim}, Hyowon and {Ko}, Jongwan and {Lee}, Jong Chul},
        title = "{Tracing Dark Matter in the Central Regions of Galaxy Clusters Using Galaxies, Gas, and Intracluster Light in TNG300: Connections to Cluster Dynamical State}",
      journal = {\apj},
         year = 2025,
        month = aug,
       volume = {988},
       number = {2},
          eid = {229},
        pages = {229},
          doi = {10.3847/1538-4357/ade66f},
archivePrefix = {arXiv},
       eprint = {2506.16280},
 primaryClass = {astro-ph.GA},
       adsurl = {https://ui.adsabs.harvard.edu/abs/2025ApJ...988..229Y}
}




\appendix

\section{Resolution tests}
\label{sec:resolution_test}

Resolution is a critical factor in accurately modelling the dynamical evolution of satellites and the fate of their stripped material \citep[e.g.][]{vandenBosch2018,Martin2024}. As shown in \citet{Martin2024}, DM mass resolution is the dominant factor in recovering reliable stripping rates for both stellar and DM components. Since our primary focus lies in the evolution of satellites and their stripped material, the satellites must be resolved with sufficiently low particle masses to accurately capture the internal structure of the satellite.

Similarly, the cluster---including both the DM halo and the central galaxy---must be modelled at the same numerical resolution as the satellites to prevent spurious numerical effects. Accordingly, when we generate initial conditions, we do so ensuring the cluster halo, central galaxy, and satellite share an identical mass resolution. We adopt DM particle masses ($m_{\rm DM}$) corresponding to $\gtrsim500,000$ particles for each satellite halo, with stellar particles ($m_\star$) roughly five times less massive. In Table \ref{tab:models} we summarise the $m_{\star}$ and $m_{\rm DM}$ corresponding to each of the satellite models described in Section \ref{sec:method:models:satellites}.

Given that the most radial satellite orbits reach pericentres of $\sim4$~kpc, we additionally verify via isolated simulations that the density profiles of the cluster halo and central galaxy remain stable at all radii beyond this scale at all resolution levels. Figure~\ref{fig:cluster_profile} shows the cumulative radial mass profiles of the stellar and DM components of the cluster. Purple and green filled square symbols respectively show the analytic DM and stellar profiles. Open diamond symbols show the same components measured from the \textsc{GalIC}-generated initial conditions. In order to test the stability of our cluster model, we evolve an isolated realisation of the model for 2~Gyr at each resolution level. Thin, faint lines show the result for the coarsest adopted resolution shown in Table \ref{tab:models} at evenly spaced snapshots between 0.25 and 2~Gyr. Even at this resolution, the simulation remains stable outside the central kpc where we find no significant deviation from the initial profile.

We perform further tests on the stability of each satellite model, also evolving each one in isolation for 2~Gyr. No significant deviations from the initial conditions are found, with any deviations confined to well within the stellar effective radius of the satellite.

\begin{figure}
    \centering
    \includegraphics[width=0.45\textwidth]{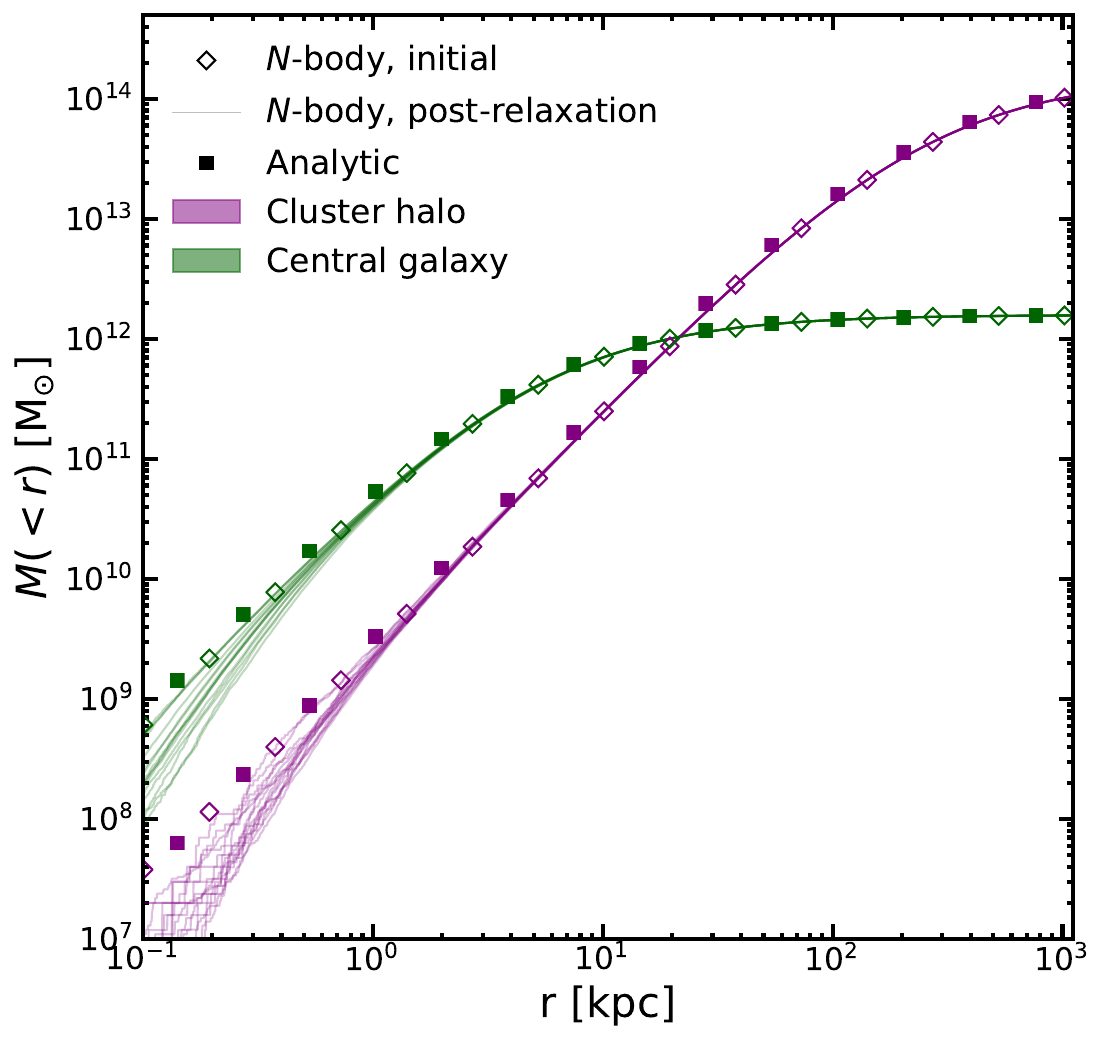}
    \caption{The DM and stellar cumulative radial mass profiles of the cluster halo and central galaxy measured out to $R_{200}$. Filled squares indicate analytic Hernquist mass profile while open diamonds indicate the mass profile measured from the \textsc{GalIC} generated initial conditions. Thin faint lines indicate the mass profile over the lowest adopted resolution shown in Table \ref{tab:models} after the isolated cluster has been evolved for between 0.25 and 2~Gyr.}
    \label{fig:cluster_profile}
\end{figure}

\section{Model predictions}
\label{app:model_predictions}

\begin{table*}
    \centering
\begin{tabular}{lllllll}
\toprule
$M_{\star}/{\rm M_{\odot}}$ & $M_{\rm h, 200}/{\rm M_{\odot}}$ & $R_{1/2}$/kpc & $c_{200}$ & $M_{\rm sat} / M_{\rm host}$ & $m_{\star}/{\rm M_{\odot}}$ & $m_{\rm DM}/{\rm M_{\odot}}$ \\
\midrule
$10^{8.5}$ & $10^{11.0}$ & $2.03$ & $6.47$ & $0.0006$ & $10^{4.3}$ & $10^{5.3}$ \\
$10^{9.9}$ & $10^{11.7}$ & $2.96$ & $5.86$ & $0.0029$ & $10^{5.3}$ & $10^{6.0}$ \\
$10^{10.5}$ & $10^{12.1}$ & $3.58$ & $5.61$ & $0.0075$ & $10^{5.8}$ & $10^{6.5}$ \\
$10^{10.9}$ & $10^{12.7}$ & $4.67$ & $5.22$ & $0.0312$ & $10^{6.3}$ & $10^{7.0}$ \\
$10^{11.1}$ & $10^{13.0}$ & $5.35$ & $5.08$ & $0.0644$ & $10^{6.3}$ & $10^{7.0}$ \\
$10^{11.3}$ & $10^{13.3}$ & $5.98$ & $4.92$ & $0.1229$ & $10^{6.3}$ & $10^{7.0}$ \\
$10^{11.5}$ & $10^{13.7}$ & $7.42$ & $4.88$ & $0.2900$ & $10^{6.3}$ & $10^{7.0}$ \\
$\mathbf{10^{12.2}}$ & $\mathbf{10^{14.2}}$ & $\mathbf{12.5}$ & $\mathbf{4.96}$ & $\mathbf{1.0000}$ & --- & --- \\
\bottomrule
\end{tabular}
    \caption{Properties and numerical resolution used for the satellite and cluster models. Columns list the stellar mass $M_\star$, halo mass $M_{\rm h,200}$, stellar half-mass radius $R_{1/2}$, halo concentration $c_{200}$, satellite--to--host mass ratio $M_{\rm sat}/M_{\rm host}$ and the stellar and DM particle masses $m_{\star}$ and $m_{\rm DM}$. Satellite model properties are assigned using scaling relations at $z=1$, while the cluster is defined at its formation redshift $z=0.6$. The cluster model is indicated in bold. The cluster model is not assigned an independent particle resolution; instead, its effective mass resolution is inherited from the satellite.}
    \label{tab:models}
\end{table*}

Figure \ref{fig:energy_am_vs_circ} presents the evolution of the specific orbital energy and angular momentum ratios between stripped stars and stripped DM as functions of the infall circularity, for the set of satellite--to--host mass ratios indicated in the legend. The top panels show the energy ratios, computed for all stripped particles (left) and for particles with $r>100$ kpc (right). We include break in the $y$-axis of the top panels because the 1:1 merger produces markedly lower ratios than the other cases. The bottom panels show the corresponding ratios for angular momentum.

\begin{figure*}
    \centering
    \includegraphics[width=0.95\textwidth]{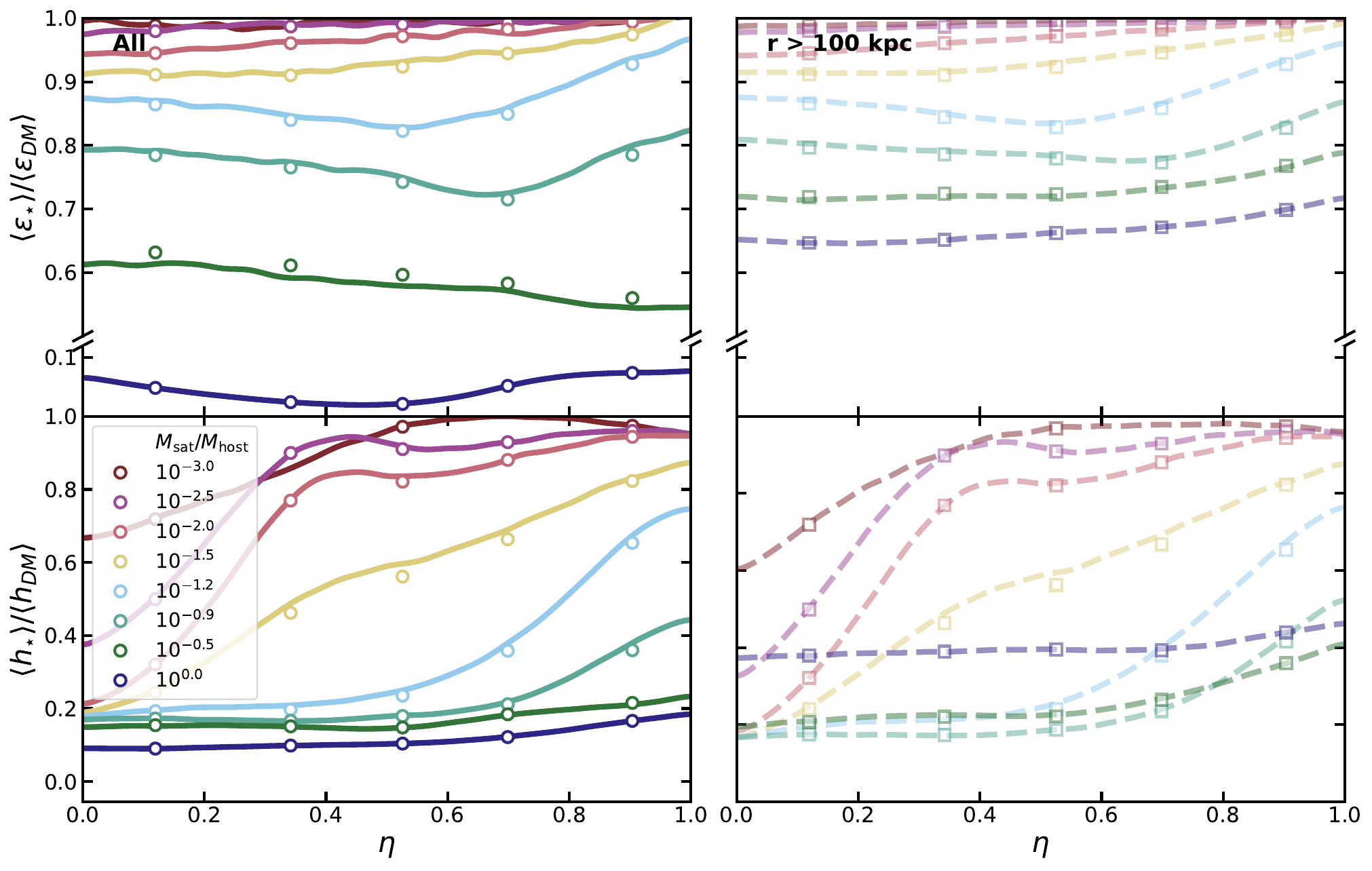}
    \caption{Average energy ratios (top) and angular momentum ratios (bottom) as a function of orbital circularity. Values drawn from each simulation in our grid are shown as open circles while the average of posterior draws from the model are shown as coloured lines. Colours correspond to satellite--to--host mass ratios indicated in the legend. Results are split between results for all stripped particles (left, solid lines) and only stripped particles with $r>100$~kpc (right, dashed lines).}
    \label{fig:energy_am_vs_circ}
\end{figure*}

For the energy ratios, most of the variation arises from the mass ratio, with only a modest dependence on circularity. The trend is monotonic: closer mass ratios produce greater stellar energy loss relative to the DM. The difference between the all particles and when only particles at $r>100$~kpc are included is smaller, since excluding the innermost region removes the lowest-energy material. At more unequal mass ratios the two selections agree much more closely, due to the fact that most stripped material remains at large radii in these cases.

The angular momentum ratios show a similar overall pattern but with a stronger dependence on the orbital angular momentum, particularly for the most radial orbits. This introduces a wider spread across circularity than is seen in the energy ratios.

To quantify the contribution of each satellite to the phase-space properties of ICL and the cluster’s DM halo, it is also necessary to measure the fraction of mass that is actually deposited into cluster the halo. Figure \ref{fig:fstripped_vs_circ} presents this for both DM and stellar components: the top panels show the stripped stellar mass as a fraction of the satellite’s infall stellar mass (left) and the fraction of that mass that ends up at $r>100$ kpc by the end of the simulation, again as a fraction of the infall stellar mass (right). The bottom panels show the same quantities for DM.

\begin{figure*}
    \centering
    \includegraphics[width=0.95\textwidth]{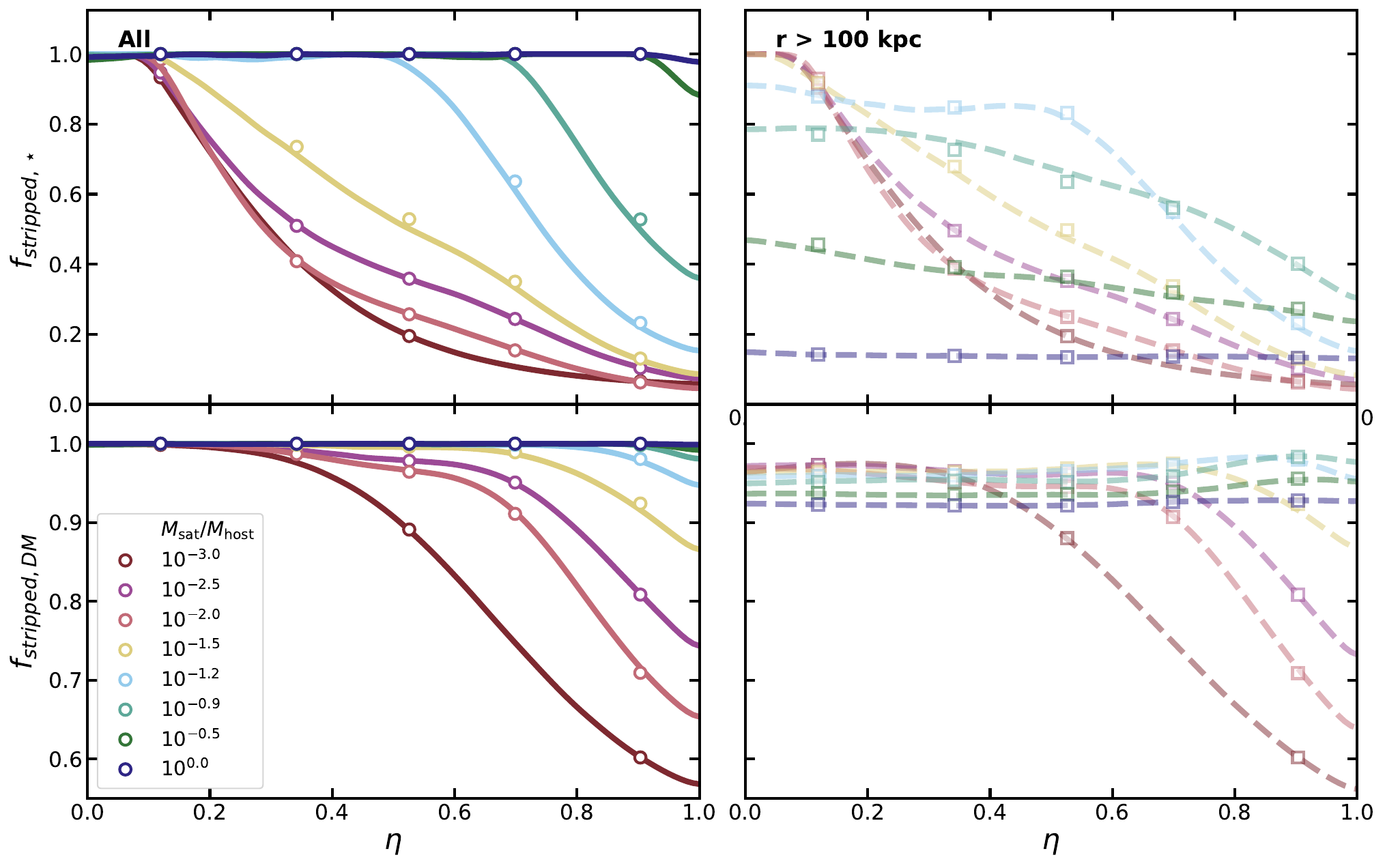}
    \caption{Stripped stellar mass fractions (top) and stripped DM mass fractions (bottom) as a function of orbital circularity. Values drawn from each simulation in our grid are shown as open circles while the average of posterior draws from the model are shown as coloured lines. Colours correspond to satellite--to--host mass ratios indicated in the legend. Results are split between results for all stripped particles (left, solid lines) and results showing the stripped mass corresponding to $r>100$~kpc as a fraction of the infalling satellite mass (right, dashed lines).}
    \label{fig:fstripped_vs_circ}
\end{figure*}

The behaviour of the stripped fractions is less straightforward than that of the energy and angular momentum ratios. Stripping efficiency increases as satellites lose orbital energy, but the material in low-mass systems is also more weakly bound resulting in an increase in stripping efficiency. Additionally, for the most equal-mass mergers, the satellite merges rapidly with the central galaxy, so a substantial proportion of the stripped stellar material merges into the central object rather than the ICL component.

These competing effects are reflected in the trends, particularly for $r>100$ kpc. Starting from equal-mass mergers and moving towards more disparate mass ratios, the stripping efficiency first decreases as dynamical friction becomes less significant, reaching a minimum at approximately 1:10. At still more disparate mass ratios, the efficiency rises again as the lower binding energies of smaller satellites make them easier to strip.

For major mergers, essentially all stellar and DM material is stripped from the satellite. However, only a small fraction of the stellar mass reaches the ICL, whereas the majority of the DM (around 90 per cent) is deposited beyond 100~kpc.

\section{Range of Schechter function parameters for infalling satellites}
\label{app:schechter_fits}

Figure~\ref{fig:mass_function_range} summarises the variation in the Schechter‐function parameters $\alpha$ (faint‑end slope) and $\log_{10} M^{\star}$ (characteristic mass) fitted to the stellar mass functions of infalling satellites across a sample of clusters drawn from four hydrodynamical simulations--\textsc{Horizon‑AGN} \citep[][]{Dubois2014, Dubois2016}, \textsc{Hydrangea} \citet{Bahe2017}, \textsc{TNG100} \citep[][]{Nelson2019} from the IllustrisTNG project, and the \textsc{GIZMO} 7K run (\citealt{Gomez2025}; Cui et al., in prep.) from \textsc{TheThreeHundred} project. The fitting procedure follows that of \citet{Brown2024}. We refer to \citet{Brown2024} for further details of the determination of infall stellar masses and the fitting method. Values for each parameter in individual clusters and their $1\sigma$ uncertainties are indicated with error bars. 

We exclude from the figure any individual-cluster Schechter fits for which the high-mass end is effectively unconstrained, corresponding to clusters where the available galaxy sample does not sample the exponential cutoff. This principally affects the \textsc{Hydrangea} set, for which only three clusters could be fit individually. This exclusion does not significantly bias the inferred parameter ranges. In all cases, the stacked infalling satellite populations of each simulation lie roughly central to the percentile regions derived from the individually fitted clusters.

For each simulation, we compute the 5th and 95th percentiles of the derived $\alpha$ and $\log_{10} M^{\star}$. These percentile intervals are shown as coloured rectangles in the figure. The four simulation suites occupy distinct regions of the $\alpha-\log_{10}M^{\star}$ plane. in particular the \textsc{Horizon-AGN} results lie systematically at shallower $\alpha$ and at lower $\log_{10}M^{\star}$ than the other suites. These offsets likely arise from differences in sub-grid prescriptions, numerical resolution, and the range of environments probed by each simulation. Given the significant differences between the four simulations, we adopt the combined parameter envelope across all four suites, enclosing the smallest minimum and largest maximum values of $\alpha$ and $\log_{10} M^{\star}$ for these regions. The black dashed rectangle defines our choice of parameter ranges, also shown in Table \ref{tab:function_params}.

\begin{figure}
    \centering
    \includegraphics[width=0.45\textwidth]{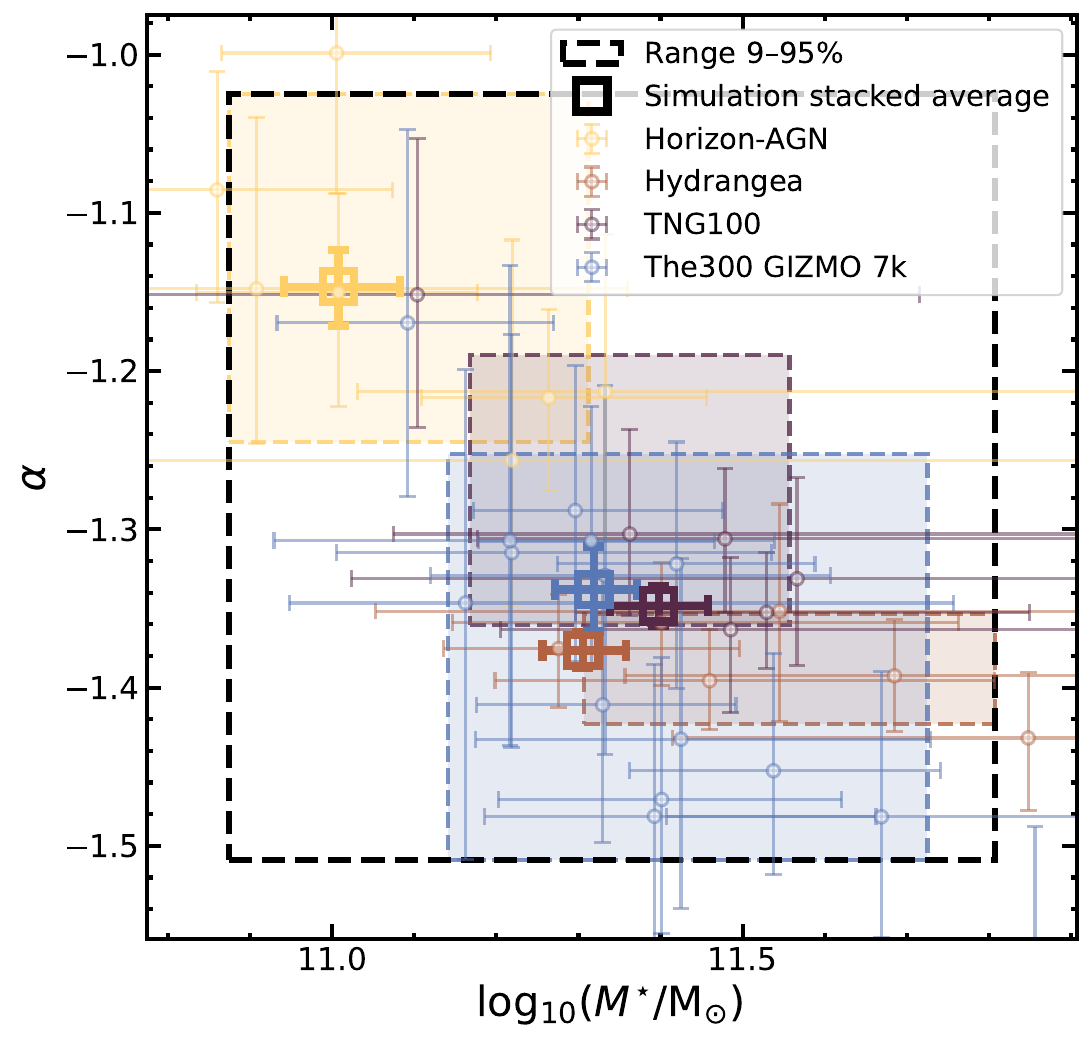}
    \caption{Stellar mass--function parameter space for infalling satellite populations across four hydrodynamical simulation suites. Points with error bars show best-fitting Schechter parameters for individual clusters, with horizontal and vertical bars indicating $1\sigma$ uncertainties on the faint-end slope $\alpha$ and characteristic mass $\log_{10}M^{\star}$, respectively. Coloured rectangles denote the 5th--95th percentile ranges of $(\alpha,\log_{10}M^{\star})$ derived from the ensemble of individually fitted clusters within each simulation. The black dashed rectangle marks the combined envelope adopted in this work, defined by the smallest minima and largest maxima across all suites, and used to set the parameter ranges listed in Table~\ref{tab:function_params}. Fits for which the exponential cutoff is unconstrained are excluded.}
    \label{fig:mass_function_range}
\end{figure}

\begin{table*}
\centering
\caption{Schechter function parameter ranges for the stellar mass functions of infalling satellites in each hydrodynamical simulation suite. Listed are the 5th--95th percentile ranges of $\log_{10}M^{\star}$ and $\alpha$ derived from individually fitted clusters, together with the parameters obtained from fitting the stacked infalling satellite population in each simulation. The final row gives the combined envelope adopted in this work.
}
\begin{tabular}{lcccccc}
\hline
Simulation & $\log_{10} M^{\star}_{\rm min}$ & $\log_{10} M^{\star}_{\rm max}$ & $\alpha_{\rm min}$ & $\alpha_{\rm max}$ & $\log_{10} M^{\star}_{\rm stack}$ & $\alpha_{\rm stack}$ \\
\hline
\textsc{Horizon-AGN} & 10.87 & 11.31 & -1.24 & -1.02 & 11.01 & -1.15 \\
\textsc{Hydrangea} & 11.31 & 11.81 & -1.42 & -1.35 & 11.31 & -1.38 \\
\textsc{TNG100} & 11.17 & 11.56 & -1.36 & -1.19 & 11.40 & -1.35 \\
\textsc{TheThreeHundred GIZMO} 7K & 11.14 & 11.73 & -1.51 & -1.25 & 11.32 & -1.34 \\
\hline
Full range & 10.87 & 11.81 & -1.51 & -1.02 & -- & -- \\
\hline
\end{tabular}
\label{tab:mf_all}
\end{table*}

\section{Sensitivity to the low-mass satellite population}
\label{sec:energy_ratio_mmin}
In this section, we explore how far down the satellite stellar mass function must be modelled in order to approximately recover the population-averaged stellar--to--DM offsets. Figure~\ref{fig:energy_ratio_vs_mmin} shows the population-averaged specific energy ratio, $\langle \varepsilon_\star / \varepsilon_{\rm DM} \rangle$ for all accreted particles, as a function of the minimum stellar mass at which we truncate the infalling satellite population. The figure samples a continuous range of characteristic stellar masses, $M^{\star}$ indicated by the colourbar, keeping the low-mass slope $\alpha$ fixed at the fiducial value.

Across these models, increasing the minimum infaller mass leads to a systematic reduction in $\langle \varepsilon_\star / \varepsilon_{\rm DM} \rangle$. This behaviour reflects the fact that low-mass satellites typically undergo stripping of stars and DM at similar orbital energies. As progressively lower-mass systems are excluded, the population-averaged energy ratio becomes increasingly dominated by a small number of intermediate- and high-mass infallers, which generate the largest stellar--DM offsets.

The dashed black line indicates the minimum infaller stellar mass required to explain 95 per cent of the DM--ICL specific energy ratio. Extending the satellite mass function to stellar masses below $\log_{10}(M_\star/M_\odot)\sim10$--$10.5$  produces little additional change in the predicted energy ratio, independent of $M^{\star}$. This indicates that the stellar--DM energy offset is insensitive to the detailed abundance of faint satellites and is instead governed primarily by the characteristic mass scale and relative frequency of intermediate- and high-mass infallers.

\begin{figure}
    \centering
    \includegraphics[width=0.45\textwidth]{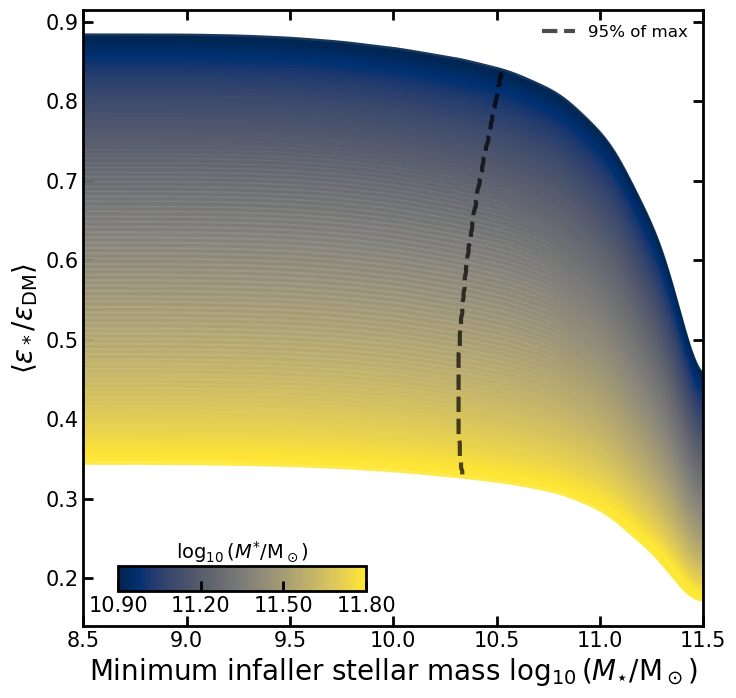}
    \caption{Population-averaged ratio of the specific orbital energy of stripped stars to that of stripped DM, $\langle \varepsilon_\star / \varepsilon_{\rm DM} \rangle$, as a function of the minimum stellar mass included in the infalling satellite population. We sample a continuous range of characteristic stellar masses $M^{\star}$ indicated by the colour bar. The dashed line indicates the infaller stellar mass above which 95 per cent of the energy ratio is explained.}
    \label{fig:energy_ratio_vs_mmin}
\end{figure}

\section{Cosmological simulations}
\label{sec:cosmo_sims}

To compare the predictions of our controlled simulations with fully cosmological environments, we make use of four independent cosmological hydrodynamical simulation suites that span a range of numerical methods, mass resolutions, and subgrid physics implementations. These simulations provide cluster and group-scale haloes with resolved satellite populations and diffuse stellar components, allowing a comparison of stellar--to--DM energy, angular momentum, and radial density offsets.

It is important to note that, at the resolutions of current large-volume cluster simulations, the tidal stripping of stellar material is not perfectly converged. In \citet{Martin2024}, we showed that limited mass and force resolution can lead to artificially enhanced stripping of stellar components, resulting in an overproduction of ICL and inaccuracies in the inferred spatial and phase-space distributions of stripped stars. More recently, \citet{Lovell2025} performed a resolution study across the IllustrisTNG suite, demonstrating that while the stripping times of DM are largely converged from TNG50-1 through to TNG300-1, the stripped mass and stripping times of stellar particles are not converged at these resolutions. This is consistent with general convergence arguments for tidal evolution. For example, \citet{Chiang2025} derived resolution criteria showing that insufficient mass and force resolution leads to systematic over-stripping and premature disruption of substructure. Together, these results imply that both the amount and detailed phase-space properties of the ICL in cosmological simulations should be interpreted with caution.

We therefore compare our model predictions to a set of complementary cosmological simulations that together sample a broad range of cluster environments, resolutions and numerical implementations. While modern high-resolution cluster simulations such as NewCluster \citep{Han2025} are now able to comfortably resolve stellar stripping in individual systems \citep{Jeon2025}, we choose to use multiple large-volume simulations here to assess the robustness of our results across a range of representative cluster populations. Below we briefly summarise the key properties and physical models of each simulation suite used in this work. Additionally, the key properties of each simulations are summarised in Table~\ref{tab:simulations}. For each simulation, we list the code, hydrodynamics scheme, subgrid physics model, simulation volume, particle mass resolutions, gravitational softening or minimum spatial resolution, and adopted cosmological parameters. All properties are expressed in physical (proper) units.

\begin{table*}
\centering
\caption{Key properties of the cosmological hydrodynamical simulations used in this work.}
\label{tab:simulations}
\begin{tabular}{lcccc}
\hline
Property & \textsc{Horizon-AGN} & \textsc{Hydrangea} & \textsc{TNG100} & \textsc{TheThreeHundred} \\
 & & & & \textsc{GIZMO 7K} \\
\hline
Code & \textsc{Ramses} & \textsc{Gadget-3} & \textsc{Arepo} & \textsc{GIZMO} \\
Hydrodynamics & AMR & SPH & Moving-mesh & Meshless-finite-mass \\
Subgrid model & \citet{Dubois2014} & \textsc{EAGLE} & \textsc{TNG} & \textsc{Simba-C} \\
\hline
Volume & $(142\,\mathrm{Mpc})^3$ & Zoom-in & $(110\,\mathrm{Mpc})^3$ & Zoom-in \\
$m_{\rm DM}$ [$\mathrm{M_\odot}$] & $8\times10^{7}$ & $9.7\times10^{6}$ & $7.5\times10^{6}$ & $2.7\times10^{8}$ \\
$m_{\star}$ [$\mathrm{M_\odot}$] & $2\times10^{6}$ & $1.8\times10^{6}$ & $1.4\times10^{6}$ & $4.4\times10^{7}$ \\
$\epsilon$ [kpc] & $1.0$ (minimum cell size) & $0.7$ & $0.7$ & $3.7$ \\
\hline
Cosmology & WMAP7 & Planck 2014 & Planck 2016 & Planck 2016 \\
\hline
$\Omega_{\rm m}$ & 0.272 & 0.307 & 0.3089 & 0.3089 \\
$\Omega_\Lambda$ & 0.728 & 0.693 & 0.6911 & 0.6911 \\
$\Omega_{\rm b}$ & 0.045 & 0.04825 & 0.0486 & 0.0486 \\
$\sigma_8$ & 0.81 & 0.8288 & 0.8159 & 0.8159 \\
$n_s$ & 0.967 & 0.9611 & 0.9667 & 0.9667 \\
$H_0$ [km s$^{-1}$ Mpc$^{-1}$] & 70.4 & 67.77 & 67.74 & 67.74 \\
\hline
\end{tabular}
\end{table*}

\subsection{\textsc{Horizon-AGN}}

\textsc{Horizon-AGN} \citep{Dubois2014, Dubois2016, Kaviraj2017} is a cosmological hydrodynamical simulation evolved with the adaptive mesh refinement code \textsc{Ramses} \citep{Teyssier2002}, run in a $(100\,h^{-1}\,\mathrm{Mpc})^3$ volume. The DM particle mass is $m_{\rm DM}=8\times10^{7}\,\mathrm{M_\odot}$ and the initial baryonic mass resolution is $m_{\rm gas}=10^{7}\,\mathrm{M_\odot}$; star particles form with a minimum mass of $m_{\star}=2\times10^{6}\,\mathrm{M_\odot}$. The maximum physical spatial resolution is $\Delta x=1\,\mathrm{kpc}$ (proper). The simulation includes radiative cooling, star formation, stellar feedback, and dual-mode AGN feedback implemented in thermal and kinetic modes. \textsc{Horizon-AGN} adopts a WMAP7-like cosmology \citep{Komatsu2011} with $(\Omega_{\rm m},\Omega_\Lambda,\Omega_{\rm b},\sigma_8,n_s,H_0)=(0.272,0.728,0.045,0.81,0.967,70.4\,\mathrm{km\,s^{-1}\,Mpc^{-1}})$. 

\subsection{\textsc{Hydrangea}}

\textsc{Hydrangea} \citep{Bahe2017} is a suite of hydrodynamical zoom simulations of galaxy clusters, built using the modified \textsc{Gadget-3} smoothed particle hydrodynamics code and the \textsc{EAGLE} subgrid model \citep{Schaye2015}, with initial conditions consistent with the \textsc{EAGLE} reference resolution: $m_{\rm DM}=9.7\times10^{6}\,\mathrm{M_\odot}$ and $m_{\rm gas}=1.8\times10^{6}\,\mathrm{M_\odot}$. The Plummer-equivalent gravitational softening length is $\epsilon=0.7\,\mathrm{kpc}$. The simulation includes radiative cooling, star formation, stellar feedback, and thermal AGN feedback following the EAGLE model. The simulations assume a $\Lambda$CDM cosmology consistent with \citet{Planck2014}, with $(\Omega_{\rm m},\Omega_\Lambda,\Omega_{\rm b},\sigma_8,n_s,H_0)=(0.307,0.693,0.04825,0.8288,0.9611,67.77\,\mathrm{km\,s^{-1}\,Mpc^{-1}})$. 

\subsection{\textsc{TNG100}}

\textsc{TNG100} \citep[][]{Marinacci2018,Nelson2018,Pillepich2018,Naiman2018,Springel2018} is a cosmological magnetohydrodynamical simulation evolving a $(110~\mathrm{Mpc})^{3}$ comoving volume using the moving-mesh code \textsc{Arepo} \citep{Springel2010}. For the \textsc{TNG100-1} run, the DM particle mass is $m_{\rm DM}=7.5\times10^{6}\,\mathrm{M_\odot}$ and the target baryonic mass resolution is $m_{\rm b}=1.4\times10^{6}\,\mathrm{M_\odot}$. The Plummer-equivalent gravitational softening length for collisionless particles is $\epsilon=0.7\,\mathrm{kpc}$ at $z=0$. The simulation includes ideal MHD, radiative cooling, star formation, kinetic stellar feedback, and dual-mode AGN feedback. A $\Lambda$CDM cosmology consistent with \citet{Planck2016} is adopted, with $(\Omega_{\rm m},\Omega_\Lambda,\Omega_{\rm b},\sigma_8,n_s,H_0)=(0.3089,0.6911,0.0486,0.8159,0.9667,67.74\,\mathrm{km\,s^{-1}\,Mpc^{-1}})$.

\subsection{\textsc{TheThreeHundred GIZMO 7K}}

\textsc{TheThreeHundred} project \citep{Cui2018} comprises 324 zoom-in simulations of a mass-complete sample of massive galaxy clusters at $z=0$, selected from a large cosmological DM only simulation, MDPL2 \citep{Klypin2016}. In this work we use the latest GIZMO 7K run (Cui et al., in prep.), which employs the meshless finite-mass hydrodynamics solver, GIZMO, coupled with an updated \textsc{Simba-C} model \citep{Hough2023} from the \textsc{Simba} galaxy formation model \citep{Dave2019}. The DM mass resolution is $m_{\rm DM}=2.7\times10^{8}\,\mathrm{M_\odot}$ and the stellar mass resolution is $m_{\star}=4.4\times10^{7}\,\mathrm{M_\odot}$ \citep{Cui2022}. The simulations include radiative cooling, star formation, stellar feedback, and multi-mode AGN feedback including kinetic jets. A $\Lambda$CDM cosmology consistent with \citet{Planck2016} is adopted, with $(\Omega_{\rm m},\Omega_\Lambda,\Omega_{\rm b},\sigma_8,n_s,H_0)=(0.3089,0.6911,0.0486,0.8159,0.9667,67.74\,\mathrm{km\,s^{-1}\,Mpc^{-1}})$.


\bsp	
\label{lastpage}
\end{document}